\documentclass[prl,aps,twocolumn,superscriptaddress,
longbibliography]{revtex4-2}

\usepackage{graphicx}
\usepackage{dcolumn}
\usepackage{bm}
\usepackage{enumitem}
\usepackage[normalem]{ulem}
\usepackage{amssymb}
\usepackage{dsfont}
\usepackage[T1]{fontenc} 
\usepackage{physics}
\usepackage{comment}
\usepackage{amsmath}
\usepackage{graphicx}
\usepackage{mathtools}
\usepackage{array}
\usepackage{makecell}
\usepackage{float}
\setcellgapes{3pt}
\usepackage[dvipsnames]{xcolor}
\usepackage{booktabs}

\usepackage[normalem]{ulem}
\usepackage{stackrel}

\usepackage{amsthm}

\usepackage{slashed}

\definecolor{azure}{rgb}{0.0, 0.5, 1.0}
\definecolor{darkblue}{rgb}{0.15,0.35,0.7}
\definecolor{reddish}{rgb}{0.65, 0.2, 0.2}
\definecolor{brandeisblue}{rgb}{0.0, 0.44, 1.0}
\definecolor{ceruleanblue}{rgb}{0.16, 0.32, 0.75}
\definecolor{indigo(dye)}{rgb}{0.0, 0.25, 0.42}
\definecolor{grey}{rgb}{0.9,0.9,0.9}
\definecolor{dgrey}{rgb}{0.3,0.3,0.3}
\definecolor{dgreen}{rgb}{0.345098, 0.596078, 0.14902}

\usepackage[linktocpage=true]{hyperref}
\hypersetup{
colorlinks=true,
citecolor=ceruleanblue,
linkcolor=ceruleanblue,
urlcolor=ceruleanblue,
pdfauthor={},
pdftitle={},
pdfsubject={}
}

\usepackage{colortbl}
\definecolor{dgreen}{rgb}{0, 0.55, 0}
\definecolor{llightyellow}{rgb}{1.0, 0.95, 0.7}
\definecolor{llightblue}{rgb}{0.7, 0.9, 1.0}
\definecolor{llightpink}{rgb}{1.0, 0.85, 0.95}
\definecolor{llightgreen}{rgb}{0.7, 1.0, 0.4}
\colorlet{lightyellow}{llightyellow!50!white}
\colorlet{lightblue}{llightblue!50!white}
\colorlet{lightgreen}{llightgreen!50!white}
\colorlet{lightpink}{llightpink!50!white}

\usepackage{cleveref}
\usepackage{bm}
\crefname{lem}{lemma}{lemmas}
\crefname{thm}{theorem}{theorems}
\crefname{cor}{corollary}{corollaries}
\crefname{rem}{remark}{remarks}
\crefname{prop}{proposition}{propositions}

\usepackage{colortbl}
\definecolor{dgreen}{rgb}{0, 0.55, 0}
\definecolor{llightyellow}{rgb}{1.0, 0.95, 0.7}
\definecolor{llightblue}{rgb}{0.7, 0.9, 1.0}
\definecolor{llightpink}{rgb}{1.0, 0.85, 0.95}
\definecolor{llightgreen}{rgb}{0.7, 1.0, 0.4}
\colorlet{lightyellow}{llightyellow!50!white}
\colorlet{lightblue}{llightblue!50!white}
\colorlet{lightgreen}{llightgreen!50!white}
\colorlet{lightpink}{llightpink!50!white}

\usepackage{tikzit}

\usepackage[framemethod=TikZ]{mdframed} 
\usepackage{tikz-cd} 
\usetikzlibrary{arrows,snakes,shapes.arrows,decorations.markings,shapes.geometric}

     \tikzset{>=triangle 90}
     \tikzstyle{bbc}=[draw,circle,fill=black,scale=.75]
     \tikzstyle{rc}=[circle,fill=red,scale=.6]
     \tikzstyle{wc}=[draw,circle,scale=.75]

\tikzset{snake it/.style={decorate, decoration=snake}}

\tikzset{
	on each segment/.style={
		decorate,
		decoration={
			show path construction,
			moveto code={},
			lineto code={
				\path [#1]
				(\tikzinputsegmentfirst) -- (\tikzinputsegmentlast);
			},
			curveto code={
				\path [#1] (\tikzinputsegmentfirst)
				.. controls
				(\tikzinputsegmentsupporta) and (\tikzinputsegmentsupportb)
				..
				(\tikzinputsegmentlast);
			},
			closepath code={
				\path [#1]
				(\tikzinputsegmentfirst) -- (\tikzinputsegmentlast);
			},
		},
	},
	mid arrow/.style={postaction={decorate,decoration={
				markings,
				mark=at position .5 with {\arrow[#1]{stealth}}
	}}},
}

\usetikzlibrary{decorations.markings}

\tikzset{line/.style={line width=0.25mm},
curve/.style={line,smooth,tension=1},
->-/.style={decoration={
  markings,
  mark=at position #1 with {\arrow[>=stealth]{>}}},postaction={decorate}},
-<-/.style={decoration={
  markings,
  mark=at position #1 with {\arrow[>=stealth]{<}}},postaction={decorate}},
}

\tikzset{bg/.style={opacity=.5}}

\tikzstyle{red dot}=[fill={rgb,255: red,240; green,165; blue,165}, draw=black, thick, shape=circle, minimum size=3mm, inner sep=0.2mm,font=\small]
\tikzstyle{green dot}=[fill={rgb,255: red,216; green,248; blue,216}, draw=black, thick, shape=circle,minimum size=3mm, inner sep=0.2mm,font=\small]
\tikzstyle{had}=[fill=yellow, draw=black, shape=rectangle]

\tikzstyle{tiny red dot}=[fill={rgb,255: red,240; green,165; blue,165}, draw=black, thick, shape=circle, minimum size=2.2mm, inner sep=0.12mm,font=\small]
\tikzstyle{tiny green dot}=[fill={rgb,255: red,216; green,248; blue,216}, draw=black, thick, shape=circle,minimum size=2.2mm, inner sep=0.12mm,font=\small]

\tikzstyle{thick line}=[fill=none, thick, draw=black, <-]
\tikzstyle{narrow}=[fill=none, thick, draw=black, ->-=0.5]

\tikzset{
    partial ellipse/.style args={#1:#2:#3}{
        insert path={+ (#1:#3) arc (#1:#2:#3)}
    }
}

\newcommand\IF{\mathbb{F}}

\newcommand\IZ{\mathbb{Z}}

\newcommand\CA{\mathcal{A}}
\newcommand\CB{\mathcal{B}}
\newcommand\CC{\mathcal{C}}
\newcommand\CD{\mathcal{D}}

\newcommand\CF{\mathcal{F}}
\newcommand\CG{\mathcal{G}}
\newcommand\CH{\mathcal{H}}

\newcommand\CJ{\mathcal{J}}
\newcommand\CK{\mathcal{K}}
\newcommand\CL{\mathcal{L}}
\newcommand\CM{\mathcal{M}}

\newcommand\CO{\mathcal{O}}
\newcommand\CP{\mathcal{P}}

\newcommand\CS{\mathcal{S}}

\newcommand\CX{\mathcal{X}}

\newcommand\CZ{\mathcal{Z}}

\newcommand\Spin{\operatorname{Spin}}

\newcommand{\dsi}{\mathds{1}}

\newcommand{\ii}{\mathsf{i}}

\newcommand\Rep{\operatorname{Rep}}
\newcommand\TY{\operatorname{TY}}

\newcommand\p{\operatorname{p}}

\newcommand{\figref}[1]{Fig.\,\ref{#1}}
\newcommand{\tabref}[1]{Tab.\,\ref{#1}}

\newcommand{\Fun}{\mathrm{Fun}}
\newcommand{\Ind}{\mathrm{Ind}}

\newcommand{\dt}{\delta t}

\newcommand{\DS}{{\CD(S_3)}}

\newcommand{\emp}{e\text{-}m}

\newcommand{\coeffa}{\lambda_1}
\newcommand{\coeffb}{\lambda_2}
\newcommand{\coeffc}{\lambda_3}

\definecolor{pink}{rgb}{1.0, 0.2, 0.6}
\definecolor{cyan}{rgb}{0.0, 0.6, 1.0}

\newcommand{\ems}{\mathbb{Z}_2^{\mathrm{em}}}
\newcommand{\pp}{\mathrm{p}}
\newcommand{\vv}{\mathrm{v}}
\newcommand{\Amin}{\CA_{\min}}
\newcommand{\zsu}{SU(2)_4\times SU(2)_{-4}}
\newcommand{\zspin}{\Spin(n)_2\times\Spin(n)_{-2}}
\newcommand{\SMref}[1]{\hyperref[#1]{SM~\ref*{#1}}}

\newcommand{\sfD}{\mathsf{D}}

\newcommand{\BZ}{{\color{Blue}Z}}

\usetikzlibrary{calc}

\begin{document}

\preprint{APS/123-QED}

\title{Self-dual $S_3$ gauge theory in 2+1d: lattice model and topological phase transitions}

\author{Da-Chuan Lu}
\affiliation{Department of Physics, Harvard University, Cambridge, MA 02138, USA}
\affiliation{Department of Physics and Center for Theory of Quantum Matter, University of Colorado, Boulder, CO 80309, USA}

\author{Chong Wang}
\affiliation{Perimeter Institute for Theoretical Physics, Waterloo, Ontario, Canada N2L 2Y5}

\author{Ashvin Vishwanath}
\affiliation{Department of Physics, Harvard University, Cambridge, MA 02138, USA}

\begin{abstract}
{Electric-magnetic self-duality of the $\mathbb{Z}_2$ gauge theory, realized microscopically as a half-lattice-translation exchanging electric charge and magnetic flux, has been an influential example of a duality symmetry with an exact lattice realization. 
We construct the first non-Abelian generalization of this construction: a lattice model of the $S_3$ quantum double $\mathcal{D}(S_3)$ {on a tensor product Hilbert space} in which the $\mathbb{Z}^{\mathrm{em}}_2$ anyon-permutation symmetry, exchanging the non-Abelian chargeon $C$ and fluxon $F$, is realized via lattice translation. 
Consequently we find that the zigzag boundary termination of the model realizes, without fine-tuning, a gapless critical edge state described by the tetracritical Ising CFT.
The bulk admits three independent $\mathbb{Z}_2^{\mathrm{em}}$-preserving bosonic perturbations, driving $\mathcal{D}(S_3)$ into  distinct gapped phases. We analyze these transitions by three independent methods: category-theoretic anyon condensation, microscopic lattice Hamiltonians, and Chern-Simons-Higgs theory which all agree yielding a unified picture. These examples motivate a \textit{minimal-condensation principle}: proliferating a bosonic anyon generically drives condensation of a minimal condensable algebra containing it, with symmetry-related condensates appearing as degenerate vacua that spontaneously break the anyon-permutation symmetry. 
Our model construction extends to an infinite family of self-dual dihedral quantum doubles $\mathcal{D}(D_{2n})$. Notably, each model is sign-problem-free, opening the door to large scale numerical exploration of the phases of non-Abelian Chern-Simons-Higgs theories. 
}
\end{abstract}

\maketitle


\textit{\textbf{Introduction.} ---}
Phase transitions out of topologically ordered states represent a central challenge in the study of quantum matter, from anyon condensation in abstract topological field theories \cite{Bais:2002pb,Bais:2008ni,Kong:2013aya,Eliens:2013epa,Cheng:2026qax} to concrete transitions out of fractional quantum Hall (FQH) states in moir\'e materials~\cite{zeng2023thermodynamic,cai2023signatures,park2023observation,xu2023observation,Shi:2025nij,Pichler:2025bda,han2025anyon}. Two complementary languages are commonly used to study such transitions. Quantum-double and string-net Hamiltonians \cite{dijkgraafTopologicalGaugeTheories1990,propitiusDiscreteGaugeTheories1996,propitiusTopologicalInteractionsBroken1995b,kitaev2003fault,Levin:2004mi}, recently realized on quantum processors \cite{song2018demonstration,Verresen:2020dmk,Semeghini:2021wls,Iqbal:2023wvm,xu2024non}, provide microscopic lattice realizations of finite-gauge and related topological orders, where anyon proliferation can be implemented by local ribbon perturbations. Continuum Chern-Simons-Higgs (CSH) theories \cite{witten1989quantum,girvin1987offFQH0,zhang1989effectiveFQH1,read1989orderFQH2,lopez1991fractionalFQH3,ezawa1991chernFQH4,halperin1993theoryFQH5,wen1992classificationFQH6,ZhouWangHe2025}, by contrast, provide compact Lagrangian descriptions of many topological orders and Higgs transitions, especially in FQH states and fractionalized Chern insulators.

Including global symmetries of a topological order further constrains the structure of its phase transitions \cite{barkeshli2019symmetry,Bombin:2010xn,Barkeshli:2012pr,Barkeshli:2013yta}. The simplest example is the $\mathbb{Z}_2$ gauge theory, which has a $\ems$ anyon-permutation symmetry exchanging the electric charge and magnetic flux, commonly referred to as self-duality. Although the lattice phase diagram of the corresponding Fradkin--Shenker model has been thoroughly studied~\cite{Fradkin:1978dv, Trebst:2006ci, Vidal:2008uy, Tupitsyn:2008ah, wu2012phase,Somoza:2020jkq}, the self-dual transition is more subtle: the $\ems$ forces charge and flux gaps to close simultaneously. Continuum descriptions of this criticality were obtained only recently using CSH theory for the self-dual toric-code transition~\cite{Ji:2026yfj}.

This example suggests a useful route from microscopic lattice models to continuum field theories. Gauging the anyon-permutation symmetry can map a finite gauge theory to a topological order (TO) with a Chern-Simons description, thereby converting lattice anyon-proliferation transitions into CSH theories in the gauged theory. The generalization of this mechanism to non-Abelian TOs remains largely unexplored, yet is particularly compelling for several reasons. Non-Abelian quantum doubles such as $\DS$ support universal topological quantum computation when braiding is supplemented by suitable fusion or charge measurements \cite{mochon2004anyonS31,Cui:2015hnw,Chen:2024xkwS33,Lo:2026oxmS34} and are connected to metaplectic modular tensor categories~\cite{Hastings:2012cd, Hastings:2013cik,Cui:2015hnw,gustafson2020metaplectic,deaton2020integral, ardonne2021classification,Shi:2026mdn}. Non-Abelian anyon-proliferation transitions are also relevant to FQH transitions \cite{Barkeshli:2010sm,Barkeshli:2010ncu,Vaezi:2014gga,Hermanns:2009bq, Zhang:2024bye,Yutushui:2025ptt}. More broadly, related exotic Chern-Simons-Higgs theories arise in proposed descriptions of bilayer FQH transitions~\cite{lopez1995fermionic,rajaraman1997generalized, simon2007pseudopotentials,barkeshli2010u,papic2010tunneling, peterson2015abelian,han2025anyon,voinea2026critical}, but typically lack sign-problem-free microscopic lattice realizations that would enable controlled Monte Carlo tests.

In this work, we construct a new lattice realization of $\DS$ in which the $\ems$ symmetry, exchanging the non-Abelian chargeon $C$ and fluxon $F$ \cite{Beigi:2010htr}, is manifested as lattice translation composed with local unitaries. The Hamiltonian is stoquastic in the computational basis, enabling sign-problem-free quantum Monte Carlo study of the full phase diagram. A local unitary transformation yields the equivalent non-Abelian Wen-plaquette form \cite{Wen:2003yv}, in which $\ems$ acts as pure lattice translation, providing a simpler geometric realization of the anyon permutation symmetry. The Wen-plaquette form and its zigzag boundary, which realizes a self-dual $\Rep(S_3)$ symmetric spin chain \cite{arkya2024sdreps3}, are discussed in the \hyperref[sec:endmatter]{End Matter}.

The self-dual $\DS$ lattice model admits three independent $\ems$-preserving perturbations, each proliferating a distinct bosonic anyon and driving the system into a different gapped phase. To obtain continuum field theories associated with these lattice transitions, we gauge the infrared $\ems$ anyon-permutation symmetry. This maps $\DS$ to $SU(2)_4\times SU(2)_{-4}$ and converts each neighboring phase into a Chern-Simons theory with a known Lagrangian. The three transitions then become tractable Chern-Simons-Higgs problems: a bi-adjoint Higgs transition, a bifundamental Higgs transition~\cite{Ji:2026yfj}, and an anyon condensation transition~\cite{Cheng:2026qax}.

The above construction generalizes to an infinite family of non-Abelian quantum doubles $\CD(D_{2n})$ for odd $n$, with $\DS$ as $n=3$. These models are sign-problem-free and their field theories are described by $\Spin(n)_2\times \Spin(n)_{-2}$; details are discussed in the \hyperref[sec:endmatter]{End Matter}.

Hamiltonian anyon proliferation is a dynamical process, whereas anyon condensation is a mathematical framework for characterizing the resulting gapped phases \cite{Kong:2013aya,kitaevModelsGappedBoundaries2012c,davydovWittGroupNondegenerate2011,neupertBosonCondensationTopologically2016a,burnellAnyonCondensationIts2018}. Motivated by the examples below, we propose the following working principle:
\begin{mdframed}[linewidth=0.8pt, innertopmargin=4pt, innerbottommargin=4pt]
\textbf{Minimal-condensation principle.}
Deforming the Hamiltonian of a topological order $\CC$ by proliferating a \underline{simple} bosonic anyon $b \in \CC$ generically drives it to a nearby gapped phase described by anyon condensation of a minimal condensable algebra $\Amin^b \ni b$ in $\CC$, i.e.\ one with the smallest $\dim(\CA^b)$ among condensable algebras containing $b$. When several such minimal algebras exist and are permuted by an anyon permutation symmetry, the system may spontaneously break that symmetry, realizing them as degenerate vacua.
\end{mdframed}
where $\dim(\CA) = \sum_{a\in \CA}n_a d_a$. For a non-simple bosonic object $b=\bigoplus_i b_i$, the dynamics may select a single component and the corresponding $\Amin^{b_i}$, while realizing them as degenerate vacua generally requires fine-tuning. We note that this principle captures the simplest scenario, in which the proliferated bosons condense without forming an additional nontrivial phase of their own; more general interactions may instead favor other phases depending on the model details \footnote{We thank Meng Cheng and Sahand Seifnashri for discussing the general case.}. We find that this principle accounts for all three transitions in the $\DS$ lattice model and their dihedral-group $D_{2n}$ generalizations.

\begin{figure}
    \centering
    \includegraphics[width=1\linewidth]{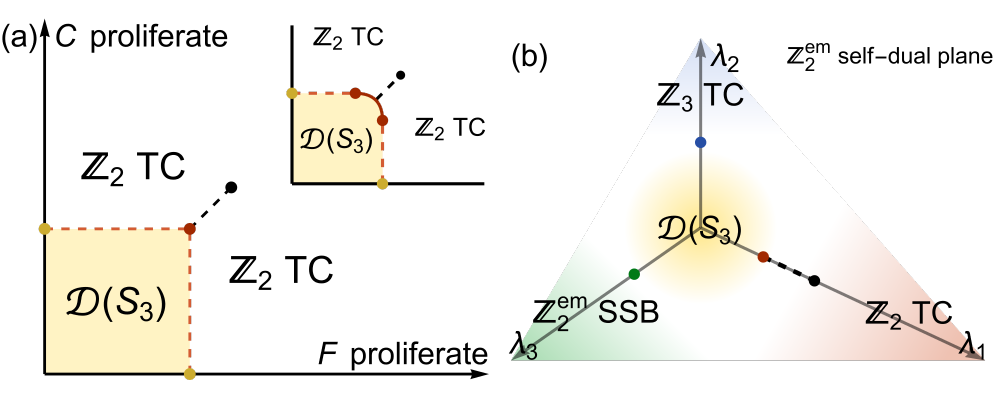}
    \caption{Phase diagrams of $\DS$. (a) Schematic $(C,F)$ plane for independent proliferation of the non-Abelian chargeon $C$ and fluxon $F$, analogous to the $e$-$m$ proliferation diagram of the $\IZ_2$ toric code. The $\DS$ phase near the origin is bordered by two $\IZ_2$ TC phases connected by Higgs-confinement continuity. Yellow dots denote 3d 3-state Potts$^*$ transitions, while the red dot is the $\ems$-symmetric multicritical point studied here. Along the self-dual line, a possible first-order line with $\ems$ SSB terminates at a 3d Ising point. The inset shows an alternative scenario where the $\ems$-odd perturbation is irrelevant, broadening the multicritical point into a critical line bounded by the red multicritical points. (b) Proposed phase diagram of the self-dual model~\eqref{eq:sds3p} in $(\coeffa,\coeffb,\coeffc)$ space. The perturbations $\coeffa,\coeffb,\coeffc$ are positive and proliferate $C+F$, $B$, and $D$, respectively, driving transitions to the $\IZ_2$ TC, the $\IZ_3$ TC, and an $\ems$-SSB phase. Along the $\coeffa$ direction, it may pass through an intermediate $\ems$-broken regime before reaching the $\ems$-symmetric $\IZ_2$ TC, where $\ems$ acts trivially.}
\label{fig:phase_diagram}
\end{figure}

\textit{\textbf{Lattice model.} ---} The $S_3$ gauge theory has several known lattice constructions, which fall into two classes. The first class, including the Kitaev quantum  double~\cite{kitaev2003fault,Brennen:2009vgn} and the gauging construction of Ref.~\cite{ruben2021DS3}, realizes $\DS$ on a tensor product Hilbert space with exact solvability, but does not make the $\mathbb{Z}_2^\mathrm{em}$ anyon permutation symmetry manifest in a simple way. The second class, exemplified by the $SU(2)_4$ string-net model \cite{Levin:2004mi,Wang:2019ydd}, realizes $\DS$ with $\mathbb{Z}_2^\mathrm{em}$ after anyon condensation of the $\IZ_2$ boson but at the cost of working on a projected Hilbert space without a tensor-product structure~\cite{Heinrich:2016wld,Cheng:2016aag,Ren:2024ayb,Fu:2025pwd}. 
Anyon proliferation transitions out of $\DS$ remain largely unexplored. Ref.~\cite{Xu:2022fnf} numerically studied its phase diagram using PEPS, but without the $\ems$ symmetry central to our work. Generalizations of quantum double and string-net models that incorporate anyon excitations can be found in \cite{Bombin:2007qv,Chen:2021gfu,Hu:2015dga,Lin:2020bak,Christian:2023ekm,Zhao:2024ilc}.

\begin{figure}
    \centering
    \includegraphics[width=0.9\linewidth]{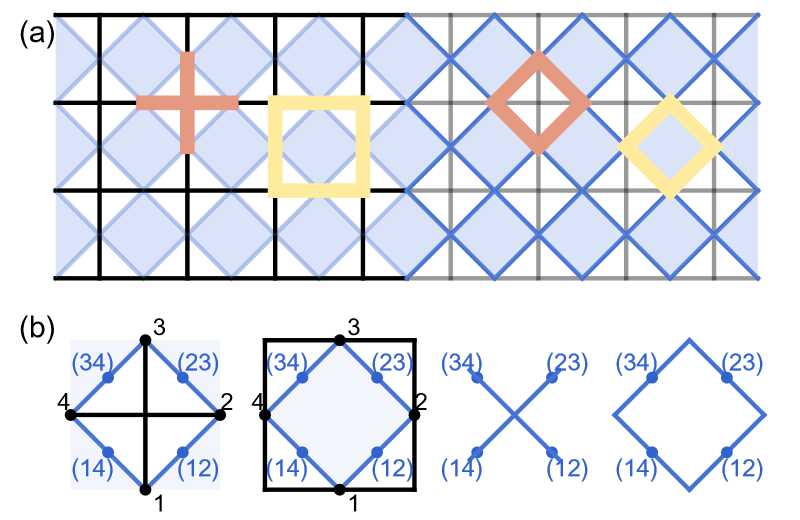}
   \caption{(a)~Lattice structure. Qutrits reside on the black links, and qubits on the blue links, primitive cell of blue lattice has half the area. Left: the quantum-double convention, with star (red) and plaquette (yellow) operators. Right: the equivalent Wen-plaquette convention, with unfilled and filled diamond plaquettes. (b)~Stabilizer definitions. From left to right: qutrit plaquette $\mathcal{B}_p$ and star $\mathcal{A}_v$, whose Pauli powers on links $1$--$4$ depend on the qubit states $(ij)$ on the adjacent blue links; qubit plaquette $B_\pp$ and star $A_\vv$, each acting on the four qubit links of the blue lattice.}
    \label{fig:latt}
\end{figure}

\textit{Sign-problem-free self-dual lattice model.} We construct the lattice model by gauging the charge conjugation symmetry that acts on the links of the $\IZ_3$ toric code, details are presented in \SMref{sm:prepare}. Similar to the gauging procedure in \cite{ruben2021DS3} but with a different translation-invariant gauging graph, we construct a lattice model of $\DS$ in which the low-energy $\mathbb{Z}_2^\mathrm{em}$ \footnote{The microscopic $\ems$ permutation symmetry acts as translation by $(\frac{1}{2},\frac{1}{2})$ composed with possible local unitaries, and thus it is squared to translation rather than an identity. However, in the infrared, the translation symmetry acts trivially, and only the $\IZ_2$ acts faithfully on the anyon content. Throughout this paper, we use $\ems$ to refer to both the microscopic and infrared symmetry, with this identification implicit.} is manifested as microscopic translation composed with local single-site unitaries. The Hamiltonian is,
\begin{equation}\label{eq:s3tc}
    H_{\DS} = -\sum_p \CB_p-\sum_{\pp} B_\pp-\sum_v \CA_v-\sum_\vv A_\vv +h.c.,
\end{equation}
where the stabilizers are,
\begin{align}
\CB^\mathrm{QD}_p &= \begin{tikzpicture}[every node/.style={outer sep=0pt, inner sep=1pt},baseline={(current bounding box.center)},line width=2pt,scale=0.9]
  \draw[gray, thick] (0,0) -- (1,0);
  \draw[gray, thick] (1,0) -- (1,1);
  \draw[gray, thick] (1,1) -- (0,1);
  \draw[gray, thick] (0,0) -- (0,1);
  \draw[RoyalBlue,opacity=0.7, thick] (0,0.5) -- (0.5,1);
  \draw[RoyalBlue,opacity=0.7, thick] (0,0.5) -- (0.5,0);
  \draw[RoyalBlue,opacity=0.7, thick] (1,0.5) -- (0.5,1);
  \draw[RoyalBlue,opacity=0.7, thick] (1,0.5) -- (0.5,0);
  \node at (0.5,-0.2) {$\CZ_1$};
  \node at (1.0,1.3) {$\CZ_3^{ {\color{Blue}-Z_{(12)}Z_{(23)}}}$};
  \node[rotate=90] at (-0.3,0.7) {$\CZ_4^{ {\color{Blue}-Z_{(14)}}}$};
  \node[rotate=90] at (1.3,0.6) {$\CZ_2^{ {\color{Blue}Z_{(12)}}}$};
\end{tikzpicture},\; \CA^\mathrm{QD}_v =\begin{tikzpicture}[every node/.style={outer sep=0pt, inner sep=1pt},baseline={(current bounding box.center)},line width=2pt,scale=0.9]
  \draw[gray, thick] (0,0) -- (0.8,0);
  \draw[gray, thick] (0,0) -- (0,0.7);
  \draw[gray, thick] (0,0) -- (0,-0.8);
  \draw[gray, thick] (0,0) -- (-0.7,0);
  \draw[RoyalBlue,opacity=0.7, thick] (0,0.5) -- (0.5,0);
  \draw[RoyalBlue,opacity=0.7, thick] (0,0.5) -- (-0.5,0);
  \draw[RoyalBlue,opacity=0.7, thick] (0,-0.5) -- (0.5,0);
  \draw[RoyalBlue,opacity=0.7, thick] (0,-0.5) -- (-0.5,0);
  \node[rotate=90] at (0.8,0.1) {$\CX_2^{ {\color{Blue}Z_{(12)}}}$};
  \node at (0.4,0.7) {$\CX_3^{ {\color{Blue}Z_{(12)}Z_{(23)}}}$};
  \node at (0,-0.7) {$\CX_1^\dagger$};
  \node[rotate=90] at (-0.7,0.2) {$\CX_4^{ {\color{Blue}-Z_{(14)}}}$};
\end{tikzpicture},\\
B_\pp &= \begin{tikzpicture}[baseline={(current bounding box.center)},line width=0pt]
  \draw[RoyalBlue,opacity=0.7, thick] (-0.5,0) -- (0,0.5);
  \draw[RoyalBlue,opacity=0.7, thick] (-0.5,0) -- (0,-0.5);
  \draw[RoyalBlue,opacity=0.7, thick] (0.5,0) -- (0,0.5);
  \draw[RoyalBlue,opacity=0.7, thick] (0.5,0) -- (0,-0.5);
  \node at (0.25,0.25) {\color{Blue}$Z$};
  \node at (0.25,-0.25) {\color{Blue}$Z$};
  \node at (-0.25,0.25) {\color{Blue}$Z$};
  \node at (-0.25,-0.25) {\color{Blue}$Z$};
\end{tikzpicture},\quad A_\vv = \begin{tikzpicture}[baseline={(current bounding box.center)},line width=0pt]
  \draw[RoyalBlue,opacity=0.7, thick] (-0.5,-0.5) -- (0.5,0.5);
  \draw[RoyalBlue,opacity=0.7, thick] (0.5,-0.5) -- (-0.5,0.5);
  \node at (0,0) {$\CC$};
  \node at (0.3,0.3) {${\color{Blue}X}$};
  \node at (0.3,-0.3) {${\color{Blue}X}$};
  \node at (-0.3,0.3) {${\color{Blue}X}$};
  \node at (-0.3,-0.3) {${\color{Blue}X}$};
\end{tikzpicture}
\end{align}
where $\CX,\CZ$ are qutrit generalized Pauli operators on the black links, while ${\color{Blue}X_{(ij)}},{\color{Blue}Z_{(ij)}}$ are qubit Pauli operators on the ${\color{RoyalBlue}blue}$ link adjacent to the black links $i,j$, as in \figref{fig:latt}. $\CO^\BZ\equiv \CO\frac{1+\BZ}{2}+\CO^\dagger\frac{1-\BZ}{2}$. The qutrit operators obey $\CZ\CX=\omega\CX\CZ$, with $\omega=e^{2\pi\ii/3}$, and the charge conjugation $\CC$ acts as $\CC\CX\CC^\dagger=\CX^\dagger$, $\CC\CZ\CC^\dagger=\CZ^\dagger$. That these stabilizers commute with each other follows from the gauging construction of the charge-conjugation symmetry of the $\IZ_3$ toric code. The $\emp$ permutation symmetry acts as translation by $(\frac{1}{2}, \frac{1}{2})$ followed by qutrit Hadamard gate $\CH,\CH^\dagger$ acting on the horizontal and vertical bonds respectively, $\CH^\dagger \CX \CH =\CZ^\dagger$: since the blue sublattice has half the primitive-cell area as the black sublattice, this exchanges the qutrit operators $\CB_p \leftrightarrow \CA_v$, while leaving the qubit operators $B_\pp$ and $A_\vv$ invariant. For comparison with existing models, e.g. quantum double, the blue lattice contains {\em twice} as many qubits as there are qutrits on the black lattice.

We consider the $\ems$ symmetric Hamiltonian with three independent perturbations,
\begin{align}\label{eq:sds3p}
    H^\mathrm{SD} &= H_{\DS}-\coeffa\sum_{\ell} (\CZ_\ell+\CX_\ell+h.c.) \nonumber\\
    &-\coeffb\sum_{\mathrm{l}} {\color{Blue}Z_\mathrm{l}} -\coeffc \sum_{\mathrm{l}}{\color{Blue}X_\mathrm{l}}.
\end{align}
where $\ell,\mathrm{l}$ refer to the links of black and blue lattices respectively. Each perturbation corresponds to the short ribbon operator for a distinct boson in $\DS$ (\tabref{tab:anyons}): $\CZ+\CX$ proliferates $C,F$ with equal strength, thus $\ems$ symmetric, while ${\color{Blue}Z}$ and ${\color{Blue}X}$ proliferate the $\ems$-invariant bosons $B$ and $D$ respectively. 
In the regime $\lambda_i \geq 0$, the above Hamiltonian is sign-problem-free in the computational basis. The full phase diagram, including transition universality classes and the multicritical point, is therefore directly accessible to large-scale quantum Monte Carlo, which we leave for future work. The quantum to classical gauge theory mapping is presented in \SMref{sm:qtocl}.

\textit{Non-Abelian Wen-plaquette and its edge.} 
In the \hyperref[sec:endmatter]{End Matter}, we obtain a non-Abelian Wen-plaquette form by onsite unitary transformation in which the $e$-$m$ permutation is pure lattice translation. To our knowledge, this provides the first non-Abelian generalization of the Wen-plaquette model. In the $\IZ_2$ case, boundary translation implements Kramers-Wannier duality and pins the edge to self-dual point realizing the Ising CFT. Likewise, on our zigzag boundary of black lattice, translation acts as generalized Kramers-Wannier duality and enforces self-duality of the $\Rep(S_3)$-symmetric chain, stabilizing the tetracritical Ising CFT $M(6,5)$ without further fine-tuning~\cite{Chang:2018iay,arkya2024sdreps3} \footnote{With only $\IZ_2$ symmetry, tetracritical Ising has three $\IZ_2$ symmetric relevant operators, $\Rep(S_3)$ forbids two and the self-duality forbids the remaining one.}. The rough or smooth blue-sublattice terminations realize the same critical CFT but connect to different gapped phases when translation is broken, as summarized in \tabref{tab:boundarys}.

\begin{table}[]
    \centering
    \begin{tabular}{c|c|c|c}
    \hline\hline
        Black bdy & Blue bdy & CFT & Gapped phases\\ \hline
        Zigzag & Smooth &  TetraIsing & $\IZ_2$ SSB v.s. $\Rep(S_3)$ SSB \\\hline 
        Zigzag & Rough & TetraIsing & $\Rep(S_3)$ sym v.s. $\frac{\Rep(S_3)}{\IZ_2}$ SSB \\\hline \hline
    \end{tabular}
    \caption{Critical black zigzag boundaries and their neighboring gapped phases for the two possible terminations of the blue sublattice.}
    \label{tab:boundarys}
\end{table}

\begin{table}
\centering
\renewcommand{\arraystretch}{1.2}
\begin{tabular}{c ccccc}
\hline\hline
Anyon & $([g],\rho)$ & $\mathbb{Z}_3$ excitations & $d_a$ & $\theta_a$ & $\mathbb{Z}_2^\mathrm{em}$ \\
\hline
$A$ & $([e],\mathbf{1})$          & vacuum                           & $1$ & $1$                    & $\checkmark$ \\
$B$ & $([e],\mathrm{sgn})$        & c.c.\ charge                     & $1$ & $1$                    & $\checkmark$ \\
$C$ & $([e],\mathbf{2})$                 & $\{e,e^2\}$                  & $2$ & $1$                    & $C\leftrightarrow F$ \\
$D$ & $([s],\mathbf{1})$          & c.c.\ defect                     & $3$ & $1$                    & $\checkmark$ \\
$E$ & $([s],-1)$                  & c.c.\ defect $\times$ c.c.\ ch. & $3$ & $-1$                   & $\checkmark$ \\
$F$ & $([r],\mathbf{1})$          & $\{m,m^2\}$                  & $2$ & $1$                    & $F\leftrightarrow C$ \\
$G$ & $([r],\omega)$            & $\{em,e^2m^2\}$          & $2$ & $e^{i\frac{2\pi}{3}}$  & $\checkmark$ \\
$H$ & $([r],\bar{\omega})$      & $\{e^2m,em^2\}$          & $2$ & $e^{-i\frac{2\pi}{3}}$ & $\checkmark$ \\
\hline\hline
\end{tabular}
\caption{Anyons of $\DS$ with representation label $([g],\rho)$ (conjugacy class and centralizer irrep for $\CD(G)$.), $\mathbb{Z}_3$ excitation content, quantum dimension $d_a$, topological spin $\theta_a$, and $\mathbb{Z}_2^\mathrm{em}$ action. $S_3 = \langle s,r \mid s^2 = r^3 = e,\, srs = r^{-1}\rangle$; $\mathrm{sgn}$ and $\mathbf{2}$ are the sign and 2d irreps of $S_3$; $\omega,\bar\omega$ are the conjugate nontrivial irreps of $\mathbb{Z}_3$. The relation between $\DS$ and charge-conjugation symmetry enriched $\IZ_3$ toric code is given in \cite{Lyons2024DS3}. Here, c.c. stands for charge-conjugation.}
\label{tab:anyons}
\end{table}

\textit{\textbf{Phase diagram.} ---}
The phase diagram of $\DS$ with $\ems$-preserving perturbations is shown in Fig.~\ref{fig:phase_diagram}. Each strong-coupling limit $\lambda_i\to\infty$ yields a distinct gapped phase, which is identified explicitly from its effective Hamiltonian. All these cases are consistent with the minimal-condensation principle.

\textit{$C$+$F$ phase ($\coeffa \to \infty$).} Proliferating either $C$ or $F$ by the terms $\CZ_\ell$ or $\CX_\ell$ individually drives $\DS$ to $\CD(\IZ_2)$; the transition is expected to be the gauged 3d 3-state Potts model, which is weakly first order~\cite{Swendsen1979threepotts,gavai1989threepotts,janke1997threepotts,Chester:2022hzt,Yang:2025wqn}. On the other hand, $\coeffa$ corresponds to simultaneous proliferation of $C+F$. For $\coeffa \to \infty$, each qutrit is polarized into the state $\propto(1+\sqrt{3},1,1)$, and projecting onto this subspace yields an effective $\IZ_2$ toric code (TC) on the blue lattice, identifying the phase as $\CD(\IZ_2)$ with $\ems$ preserving. This is consistent with the minimal-condensation principle: the minimal condensable algebras containing $C$ and $F$ are $\CA_1 = A+C \cong \Fun(S_3/\IZ_2)$ and $\CA_1' = A+F$, related by $\ems$, both giving $\CD(\IZ_2)$ upon condensation~\cite{Lan:2016rcq}. They are consistent condensable algebras \SMref{SM:anyoncond}, even though not all fusion products of $C$ or $F$ are contained. Because $C$ and $F$ map asymmetrically to a deconfined anyon and a symmetry defect, $\ems$ acts trivially on the resulting $\CD(\IZ_2)$\footnote{Due to the non-Abelian anyon condensation, the resulting $\CD(\IZ_2)$ possesses a non-invertible 0-form symmetry~\cite{Cui2018HopfMonad,KNBalasubramanian:2025vum,Eck:2026aiz,Bottini:2026edf}.}. By analogy with the $\IZ_2$ Fradkin-Shenker model, we expect an $\ems$-broken phase at intermediate value of $\coeffa$ as depicted in \figref{fig:phase_diagram}. Further numerical study is needed to determine its nature.

\textit{$B$ phase ($\coeffb \to \infty$).} For $\coeffb \to \infty$, the Hamiltonian~\eqref{eq:s3tc} reduces to the $\IZ_3$ toric code on the black lattice. This is consistent with the minimal-condensation principle: the minimal condensable algebra containing $B$ is $\CA_2 = A+B$, and whose anyon condensation yields $\CD(\IZ_3)$. The $\ems$ symmetry descends to the $\emp$ permutation $e^i \leftrightarrow m^i$ in the resulting $\CD(\IZ_3)$.

\textit{$D$ phase ($\coeffc \to \infty$).} 
For $\coeffc \to \infty$, the $-\coeffc X$ term imposes a string tension on the qubit links, driving a confinement transition in the lattice model. Projecting to the ${\color{Blue}X}=1$ subspace, the Hamiltonian~\eqref{eq:s3tc} reduces to $\IZ_3$ toric code model with charge-conjugation symmetry enforced on each bond,
\begin{equation}\label{eq:csymham}
   H_D =\sum_{{\color{Blue}s_1.. s_4}= \pm} - \begin{tikzpicture}[every node/.style={outer sep=0pt, inner sep=1pt},baseline={(current bounding box.center)},line width=2pt,scale=0.9]
  \draw[gray, thick] (0,0) -- (1,0);
  \draw[gray, thick] (1,0) -- (1,1);
  \draw[gray, thick] (1,1) -- (0,1);
  \draw[gray, thick] (0,0) -- (0,1);
  \draw[RoyalBlue,opacity=0.7, thick] (0,0.5) -- (0.5,1);
  \draw[RoyalBlue,opacity=0.7, thick] (0,0.5) -- (0.5,0);
  \draw[RoyalBlue,opacity=0.7, thick] (1,0.5) -- (0.5,1);
  \draw[RoyalBlue,opacity=0.7, thick] (1,0.5) -- (0.5,0);
  \node at (0.5,0) {$\CZ_1^{ {\color{Blue}s_1}}$};
  \node at (0.5,1) {$\CZ_3^{ {\color{Blue}s_3}}$};
  \node at (0,0.5) {$\CZ_4^{ {\color{Blue}s_4}}$};
  \node at (1.1,0.5) {$\CZ_2^{ {\color{Blue}s_2}}$};
\end{tikzpicture}- \begin{tikzpicture}[every node/.style={outer sep=0pt, inner sep=1pt},baseline={(current bounding box.center)},line width=2pt,scale=0.9]
  \draw[gray, thick] (0,0) -- (0.7,0);
  \draw[gray, thick] (0,0) -- (0,0.7);
  \draw[gray, thick] (0,0) -- (0,-0.7);
  \draw[gray, thick] (0,0) -- (-0.7,0);
  \draw[RoyalBlue,opacity=0.7, thick] (0,0.5) -- (0.5,0);
  \draw[RoyalBlue,opacity=0.7, thick] (0,0.5) -- (-0.5,0);
  \draw[RoyalBlue,opacity=0.7, thick] (0,-0.5) -- (0.5,0);
  \draw[RoyalBlue,opacity=0.7, thick] (0,-0.5) -- (-0.5,0);
  \node at (0.7,0) {$\CX_2^{ {\color{Blue}s_2}}$};
  \node at (0.0,0.5) {$\CX_3^{ {\color{Blue}s_3}}$};
  \node at (0,-0.5) {$\CX_1^{ {\color{Blue}s_1}}$};
  \node at (-0.5,0) {$\CX_4^{ {\color{Blue}s_4}}$};
\end{tikzpicture}- \begin{tikzpicture}[every node/.style={outer sep=0pt, inner sep=1pt},baseline={(current bounding box.center)},line width=2pt,scale=0.7]
  \draw[RoyalBlue,opacity=0.7, thick] (-0.5,-0.5) -- (0.5,0.5);
  \draw[RoyalBlue,opacity=0.7, thick] (0.5,-0.5) -- (-0.5,0.5);
  \node at (0,0) {$\CC$};
\end{tikzpicture}.
\end{equation}
Unlike the $\IZ_3$ toric code, the plaquette and star terms no longer commute in $H_D$. Exact diagonalization on periodic tori up to $3\times 3$ reveals two low-energy eigenstates whose small splitting decreases rapidly with system size, and the gap to higher excited states remains finite. The two symmetry-breaking combinations are exchanged by the $\ems$ action, and each is charge conjugation symmetric, consistent with spontaneous $\ems$ breaking. Details are presented in \SMref{SM:DphaseED}.

Algebraically, no nontrivial anyon braids trivially with $D$, so proliferating $D$ confines all anyons and destroys the topological order. The two minimal condensable algebras containing $D$, $\CA_3=A+F+D$ and $\CA_3'=A+C+D$, are both Lagrangian and exchanged by $\ems$. Since proliferating $D$ alone does not select between them, the corresponding condensates form two degenerate vacua that spontaneously break $\ems$. Thus, the lattice result supports the minimal-condensation principle and illustrates how symmetry-related minimal condensable algebras can lead to spontaneous symmetry breaking.

\begin{table}[t]
\centering
\renewcommand{\arraystretch}{1.3}
\begin{tabular}{c c c c}
\hline\hline
Anyon & min cond alg & Phase & $\ems$ gauged phase \\
\hline
$A$ & $A$ & $\DS^{\emp}$ & $\zsu$\\\hline
$C$+$F$ & $A+C/F$ & $\CD(\IZ_2)\times \ems$ SSB  & 
$\CD(\IZ_2)$ \\ \hline
$B$ & $A+B$ & $\CD(\IZ_3)^{\emp}$ & 
$SU(3)_1\times SU(2)_{-4}$ \\ \hline
$D$ & $A+C/F+D$ & $\ems$ SSB & trivial \\
\hline\hline
\end{tabular}
\caption{Three $\ems$-preserving transitions out of $\DS$, listing the proliferated anyon, minimal condensable algebra, condensed phase, and $\ems$-gauged phase. The $C$+$F$ row shows the possible $\ems$-broken scenario for the intermediate coupling. In all cases, the $\DS^{\emp}$ side maps to $\zsu$ under gauging $\ems$. The corresponding gauged transition theories are Eqs.~\eqref{eq:CFtrans}, 3d Ising$^*$ transition, and \eqref{eq:Dtrans}; ${\CD(G)}^{\emp}$ denotes a nontrivial $\ems$ action.}
\label{tab:phases}
\end{table}

\textit{\textbf{Field theory of the transitions.} ---}
A direct Lagrangian description of transitions out of $\DS$ enriched by $\ems$ is not available. We instead gauge the infrared $\ems$ anyon-permutation symmetry where the microscopic translational part acts trivially. It maps $\DS$ and its neighboring phases to topological orders with known Chern-Simons descriptions. This converts the corresponding transitions, when continuous, into tractable Chern-Simons-Higgs theories. 
Gauging a finite symmetry is a topological operation that commutes with the renormalization group flow~\cite{Gaiotto:2014kfa}, and therefore does not alter the dynamics of any transition. We choose the twisted $\ems$ gauging of $\DS$ \cite{barkeshli2019symmetry,Delcamp:2023kew,Lu:2024ytl,Vancraeynest-DeCuiper:2025wkh}, which maps it to $\zsu$ and admits a Lagrangian description. The corresponding gapped phases are summarized in \tabref{tab:phases} \footnote{The twisted gauging sends the trivial product state to the double-semion topological order in $2+1$d. The untwisted gauging instead maps $\DS$ to $JK_4 \boxtimes \overline{JK_4}$, which does not admit a Lagrangian description \cite{barkeshli2019symmetry,arkya2024sdreps3}.}.

\textit{$C+F$ transition.} As discussed above, the $C+F$ transition may follow the $\IZ_2$ Fradkin-Shenker scenario: at intermediate $\coeffa$, $\DS$ is driven into the $\CD(\IZ_2)$ topological phase with $\ems$ spontaneously broken.
Gauging $\ems$ maps the transition to $\zsu \to \CD(\IZ_2)$\footnote{Since $\ems$ acts trivially on $\CD(\IZ_2)$, the gauged description does not involve $SU(2)_2\times SU(2)_{-2}$.}. 
The transition is described by $SU(2)\times SU(2)$ gauge theory with Chern-Simons term and bi-adjoint scalar matter. In particular, we consider an $SU(2)_L\times SU(2)_R$ Chern-Simons-Higgs theory with a real bi-adjoint scalar $\phi^{ij}$, $i,j=1,2,3$, viewed as a real matrix transforming as $\phi\mapsto R_L\phi R_R^{\intercal}$, where $R_{L,R}\in SO(3)_{L,R}$ are the adjoint images of $U_{L,R}\in SU(2)_{L,R}$. The Lagrangian is
\begin{align}\label{eq:CFtrans}
\mathcal L_{CF}
&=\frac12(D_\mu\phi)^{ij}(D^\mu\phi)^{ij}+\sum_{\alpha=L,R}{1\over4g_\alpha^2}
\tr f_{\alpha,\mu\nu}f_\alpha^{\mu\nu} \nonumber\\
&+\mathrm{CS}[a_L]_4+\mathrm{CS}[a_R]_{-4}  -m^2\Tr\phi\phi^{\intercal} \nonumber\\
&-\kappa_3\det\phi
-\kappa_4\Tr(\phi\phi^{\intercal})^2
-\kappa_4'(\Tr\phi\phi^{\intercal})^2,
\end{align}
where $a_\alpha=a_{\alpha,\mu}^{\ell}T^\ell dx^\mu$, $f_\alpha=da_\alpha-i a_\alpha^2$, $\alpha=L,R$, $(D_\mu\phi)^{ij}=\partial_\mu\phi^{ij} +\epsilon^{i\ell m}a_{L,\mu}^{\ell}\phi^{mj} +\epsilon^{j\ell m}a_{R,\mu}^{\ell}\phi^{im}$, $\epsilon^{123}=1$ and $\mathrm{CS}[a]_k={k\over4\pi}\tr(a\,da-\frac{2i}{3}a^3)$. 
When $\phi$ is gapped, it recovers $\zsu$. When $\phi$ is Higgsed to the isotropic configuration $\phi\propto \mathds{1}$ due to the potential, leaving a residual $SU(2)_0^\mathrm{diag}\times \IZ_2$ gauge theory. Since the diagonal $SU(2)$ gauge field without Chern-Simons term confines in $2+1$d, the infrared phase is the deconfined $\IZ_2$ gauge theory $\CD(\IZ_2)$.

The Yang-Mills terms are irrelevant relative to the Chern-Simons term but control gauge fluctuations at the Higgs transition. We focus on the weak-gauge-coupling regime, $g_{L,R}^2\ll \kappa_4,\kappa_4'$ with positive $\kappa_4,\kappa_4'$, where a direct continuous transition is possible as it first approaches the scalar Wilson-Fisher fixed point before gauge fluctuations become important \cite{Ji:2026yfj}; stronger gauge fluctuations are expected to drive the transition first order.
For \(\kappa_4>0\), the quartic term favors the isotropic Higgs configuration \(\phi\phi^{\intercal}\propto \mathbf 1_3\), and the quartic potential is bounded from below when \(\kappa_4>-\kappa_4'\) and \(\kappa_4>-3\kappa_4'\).
The cubic invariant $\kappa_3$ is symmetry-allowed, with its sign selecting the orientation of the Higgs configuration. In a purely scalar Landau theory, such a cubic invariant would generically drive the transition first order. However, in the strongly coupled gauge theory this conclusion need not apply: Ref.~\cite{2018PNASchong} argues that the cubic term may be irrelevant in a similar $SU(2)$ gauge theory, based on a combination of analytic and numerical analysis, and related RG analyses suggest that the appropriate sign of the cubic term permits a continuous transition~\cite{2022JHEPsubir}.

To access the $\ems$-breaking perturbation, we ungauge $\ems$ of \eqref{eq:CFtrans}. This corresponds to gauging the $\IZ_2$ one-form symmetry of $\zsu$, yielding a bi-adjoint scalar coupled to $\frac{\zsu}{\IZ_2} \cong SO(4)_{4,-4}$. Gauging the one-form symmetry produces a dual $\IZ_2$ zero-form magnetic symmetry, which we identify with $\ems$. The symmetric mass operator is $\Tr(\phi \phi^\intercal)$, while its charged operator is the monopole $\CM$~\cite{Aharony:2013kma,Dyer:2013fja,Ji:2026yfj}. Schematically, this self-dual multicritical point is described by, 
\begin{equation}
    \CL_{CF}'=|D_b\phi|^2-m^2 \Tr(\phi \phi^\intercal) +r \CM +V(\phi)+\mathrm{CS}[b]_{4,-4}.
\end{equation}
where $b$ is the $SO(4)$ gauge field and $\mathrm{CS}[b]_{4,-4}$ describes $SO(4)_{4,-4}$. $\CM$ in the Lagrangian explicitly breaks $\ems$ with $r\ne0$. If $\CM$ is relevant, it drives the system away from the multicritical point into the individual $\CD(\IZ_2)$ phases \footnote{Only introducing the monopole operator will generally induce symmetric scalar mass term $\phi \phi^\intercal$ and drive the multicritical point to a gapped phase. The combination of monopole operator and the scalar mass term may keep the system on the nearby transition lines.}. If $\CM$ is irrelevant, the multicritical point broadens to a critical line bounded by two other multicritical points as shown in the inset of \figref{fig:phase_diagram}. $\CM$ is dangerously irrelevant and will select the $\ems$ broken phases.

\textit{$B$ transition.}
Condensing the order-two Abelian boson $B$ gives $\CD(\IZ_3)$ with the inherited electric-magnetic symmetry $\ems: e^i \leftrightarrow m^i$. Gauging $\ems$ of $\CD(\IZ_3)$ yields the Drinfeld center of the Tambara-Yamagami fusion category, $\CZ(\TY(\IZ_3,-1)) \cong SU(3)_1 \times SU(2)_{-4}$. Recall that gauging $\ems$ of $\DS$ gives $\zsu$, the $SU(2)_{-4}$ factor is unchanged across the transition, so the nontrivial part reduces to $SU(2)_4 \longrightarrow SU(2)_4/\IZ_2 \cong SO(3)_2 \cong SU(3)_1$, i.e.\ condensation of the $\IZ_2$ boson in $SU(2)_4$. The minimal critical theory is the $\IZ_2$-gauged Ising model, so the transition is expected to have 3d Ising$^*$ critical behavior.\footnote{A continuum realization is obtained from $\mathrm{CS}[a_{U(2)}]_{4,0}+\frac{1}{2\pi}c\,d\Tr(a)+|D_c\varphi|^2+V$: for massive $\varphi$, $c$ enforces $\Tr(a)=0$ and gives $SU(2)_4$; for condensed $\varphi$, the remaining $U(1)$ is confined by the allowed monopole operator, giving $SU(2)_4/\IZ_2$~\cite{Cheng:2026qax}.}

\textit{$D$ transition.} Proliferating $D$ leads to the $\ems$ SSB phase. Gauging $\ems$ maps $\ems$ SSB to a trivial phase and the transition is described by the $\zsu$ gauge theory coupled to a bifundamental scalar $\Phi$~\cite{Ji:2026yfj},
\begin{equation}\label{eq:Dtrans}
    \CL_{D}=|D_\mu\Phi|^2-m^2 \Tr \Phi^\dagger \Phi+ \mathrm{CS}[a_L]_{\scriptsize 4}+ \mathrm{CS}[a_R]_{\scriptsize-4}+\cdots 
\end{equation}
with also potential terms and $D_\mu \Phi \equiv \partial_\mu \Phi^i_{\bar j}+i (a_{L})_k^i \Phi^k_{\bar j}-i \Phi^i_{\bar k} (a_{R})_{\bar j}^{\bar k}$.  Condensing $\Phi$ Higgses the gauge group to the diagonal $SU(2)_0$, which confines in $2+1$d, leaving a trivially gapped phase. The anyon $D$ precisely corresponds to the bifundamental $(\mathbf{2},\mathbf{2})$ in $\zsu$, whose quantum dimension is $3$ and integer topological spin. Similarly, upon ungauging the $\ems$, the monopole operator of $SO(4)$ is $\ems$ odd which can drive the critical point to $\ems$ breaking.

\textit{$D_{2n}$ Generalization.} \hyperref[sec:endmatter]{End Matter} shows the generalization to the infinite family of self-dual quantum doubles $\CD(D_{2n})$ for odd $n$, $D_6=S_3$. For even $n$, $\Spin(n)_2\times \Spin(n)_{-2}$ instead describes the twisted quantum double of $D_{2n}$ with $\ems$ gauged, which is treated in the \SMref{SM:gaugeZ2em}. 

\textit{\textbf{Discussion and outlook.} ---}

\paragraph{Lattice -- Chern-Simons correspondences}
The examples above suggest a route from lattice models on a tensor-product Hilbert space to non-chiral Chern-Simons-Higgs theories: start from a finite-group gauge theory $\CD(G)$ with an anyon-permutation symmetry $K$, gauge this symmetry, and use the gauged theory as the continuum description. Formally, finite-symmetry gauging proceeds by a $K$-crossed extension followed by equivariantization~\cite{barkeshli2019symmetry,cui2016gauge}. A broader mechanism is gauging non-invertible symmetries, dual to non-Abelian anyon condensation~\cite{Kong:2013aya,Cordova:2024goh}. Relations such as $SU(N)_N/\CA \cong \Spin(N^2-1)_1$ and $SU(N)_{N\pm2}/\CA \cong SU(N(N\pm1)/2)_1$ for suitable condensable algebras $\CA$ suggest further lattice-continuum correspondences~\cite{Xu:2016roi,cordovaNonInvertibleAnyonCondensation2024}.

\paragraph{Connection to the $S[O(2)\times O(2)]$ gauge theory}
Ref.~\cite{Gao:2025ihw} gives a complementary description of dihedral quantum-double phases using a topological $S[O(2)\times O(2)]$ BF theory, equivalently a charge-conjugation gauging of the $U(1)\times U(1)$ $K$-matrix description of the $\IZ_n$ toric code. This construction reproduces the anyon data of $\CD(D_{2n})$ and its twisted generalizations. It would be interesting to add matter fields and compare the resulting critical theory with our $\Spin(n)_2\times\Spin(n)_{-2}$ Chern-Simons-Higgs description. 
The main caveat is that, in the absence of monopole perturbations, the $S[O(2)\times O(2)]$ gauge theory possesses a continuous non-invertible cosine global symmetry \cite{Chang:2020imq}. This symmetry comes from the magnetic $U(1)\times U(1)$ symmetry of the related Abelian gauge theory description and must be broken by appropriate monopole operators. For small $n$, however, these monopole perturbations are relevant; in the $S_3$ case, they may therefore drive the theory to strong coupling~\cite{Shi:2024pem}.
\footnote{Relatedly, Ref.~\cite{Dumitrescu:2026vre} studies a staggered Fradkin-Shenker variant whose continuum description is more closely related to DQCP constructions~\cite{senthil2004deconfined,Xu:2015lxa,Wang:2017txt,Jian:2017chw,Lu:2021ucu,senthil2024deconfined}. The ordinary Fradkin-Shenker problem is recovered only by adding monopole perturbations, which are expected to make the continuum description strongly coupled.}. For large enough $n$, the $S[O(2)\times O(2)]$ gauge theory becomes controlled, while the $\Spin(n)_2\times\Spin(n)_{-2}$ may flow to first-order transitions due to growing gauge fluctuations at large rank of group. The two approaches can thus be viewed as complementary. Possible level-rank relations in this setting deserve further study~\cite{Cordova:2017vab}.

\paragraph{Proliferation of non-bosonic anyons}
This work focuses on bosonic anyon proliferation. Lattice ribbon perturbations can also lower the gap of non-bosonic anyons. For example, adding $\sum_\ell \CZ_\ell\CX_\ell+h.c.$ to~\eqref{eq:s3tc} corresponding to the anyon $G$ in \tabref{tab:anyons}. Such perturbations generally spoil stoquasticity and are not ordinary anyon condensation inside $\CC$. A treatment is to stack with an auxiliary topological order $\CC'$ so that $a\boxtimes a'\in \CC\boxtimes\CC'$ is bosonic, and then apply the minimal-condensation principle to the stacked theory. This is similar to the hierarchy constructions of fractional quantum Hall states~\cite{Zhang:2024bye}. More generally, the anyon itself may form interesting states, we leave this for future exploration.

\begin{acknowledgments}
\textit{\textbf{Acknowledgments}} We thank Meng Cheng, Diego Garc\'ia-Sep\'ulveda, Chiu Fan Bowen Lo, Yuanjie Ren, Chao-Ming Jian, Leyna Shackleton, Sahand Seifnashri, Ruben Verresen, Taige Wang, Zheng Zhou for helpful discussions. A.V. and D.C.L. are supported by the Simons Collaboration on Ultra-Quantum Matter, which is a grant from the Simons Foundation (grant No. 651440, A.V.). Research at Perimeter Institute is supported in part by the Government of Canada through the Department of Innovation, Science and Industry Canada and by the Province of Ontario through the Ministry of Colleges and Universities.
\end{acknowledgments}

\bibliographystyle{apsrev4-1}
\bibliography{main}

\section*{End Matter}\label{sec:endmatter}

\textit{\textbf{Generalization to $\CD(D_{2n})$ for odd $n$.} ---} The lattice construction generalizes straightforwardly to self-dual $\CD(D_{2n})$ model for any odd integer $n$, using $D_{2n} = \IZ_n \rtimes \IZ_2$. Replacing the qutrit by a qudit of dimension $n$ on the black lattice yields a sign-problem-free lattice model of $\CD(D_{2n})$, in which $\ems$ exchanges $([e],\mathbf{2}_i)\leftrightarrow ([r^i],\mathbf{1})$ for $i=1,\ldots,(n-1)/2$. Gauging $\ems$ with a possible twist depending on $n$ maps $\CD(D_{2n})$ to $\Spin(n)_2\times\Spin(n)_{-2}$ (see \SMref{SM:gaugeZ2em}). The three $\ems$-symmetric perturbations generalize as follows:
\begin{enumerate}[nosep,wide, labelwidth=!, labelindent=0pt]
    \item $-\coeffa\sum_i(\CZ^i+\CX^i+\mathrm{h.c.})$ Individually proliferates $\sum_i([e],\mathbf{2}_i)$ or $\sum_i ([r^i],\mathbf{1})$ drives a gauged 3d XY transition to $\CD(\IZ_2)$ with trivially acting $\ems$. 
    Proliferating $\sum_i([e],\mathbf{2}_i)+([r^i],\mathbf{1})$, yields $\CD(\IZ_2)$ with trivially acting $\ems$ at strong coupling limit. The corresponding minimal condensable algebra is $\CA_1 = ([e],\mathbf{1})+\sum_i([e],\mathbf{2}_i)$ or $\CA_1' =([e],\mathbf{1})+\sum_i([r^i],\mathbf{1})$ which are related by $\ems$ symmetry. As in the $\DS$ case, a natural scenario for intermediate coupling is the transition to $\CD(\IZ_2)$ with $\ems$ SSB, and it is equivalent through gauging to $\zspin\to\CD(\IZ_2)$ which is described by coupling a bi-vector $(\mathbf{n},\mathbf{n})$ scalar to $\zspin$: condensing it Higgses the gauge group to $\Spin(n)_0\times\IZ_2$, and $\Spin(n)_0$ confines in $2+1$d, leaving $\CD(\IZ_2)$. Note that no relevant determinant term for the bi-vector with sufficiently higher $n$ compared with \eqref{eq:CFtrans}.   
    \item $-\coeffb Z$ proliferates $([e],\mathrm{sgn})$ with the corresponding condensable algebra $\CA_2 = ([e],\mathbf{1})+([e],\mathrm{sgn})$, yielding $\CD(\IZ_n)$ with inherited $\ems$. Under gauging $\ems$, the resulting TO is mapped to $\CZ(\mathrm{TY}(\IZ_n,\nu))\cong  SU(n)_1 \times \Spin(n)_{-2}$ \cite{Cordova2025dihedral}, where $\mathrm{TY}(\IZ_n,\nu)$ is the Tambara-Yamagami fusion category of $\IZ_n$ with Frobenius-Schur indicator $\nu$ for the non-invertible line. In particular, as detailed in the SM, $\nu=-1$ for $n=3,5\mod8$ and $\nu=+1$ for $n=1,7\mod8$. The transition becomes $\zspin \to SU(n)_1 \times \Spin(n)_{-2}$, which is given by similar construction for $\CD(S_3)$, as $\Spin(n)_2/\IZ_2 = SO(n)_2 \cong SU(n)_1$.
    \item $-\coeffc X$ proliferates the quantum dimension $n$ anyon $([s],\mathbf{1})$, yielding the $\ems$ SSB phase. The corresponding minimal condensable algebras are $\CA_3 =  ([e],\mathbf{1})+\sum_i([r^i],\mathbf{1}) +([s],\mathbf{1})$ or $\CA_3' =  ([e],\mathbf{1})+\sum_i([e],\mathbf{2}_i) +([s],\mathbf{1})$ related by $\ems$.
    Under gauging, the transition $\zspin\to$ trivial is described by a CSH with bi-spinor scalar in $\Spin(n)\times \Spin(n)$: condensing it with suitable potential Higgses to diagonal $\Spin(n)_0$, which confines leaving a trivially gapped phase.
\end{enumerate}
For $n=3$, $\Spin(3)_k \cong SU(2)_{2k}$, it recovers $\CD(S_3)$. Besides $\ems$, the anyon permutation symmetry of $\CD(D_{2n})$ also contains the outer automorphism of $D_{2n}$ \cite{nikshych2014categorical}, which is given by $\IZ_n^\times/\{\pm1\}$ \footnote{For odd prime $p$, $n=p, \IZ_n^\times/\{\pm1\} =\IZ_{(p-1)/2}$. For odd prime powers $p^a$, the group is $\IZ_{\varphi(p^a)/2}$, where $\varphi$ is the Euler’s totient function.}. However, all the minimal condensable algebras described above are invariant under this part, so no further SSB will occur.

This generalization does not extend straightforwardly to even $n$, where the $\ems$ ungauged theory acquires a nontrivial twist. For example, $\Spin(4)_2\times \Spin(4)_{-2}\cong SU(2)_2^{\boxtimes 2}\times SU(2)_{-2}^{\boxtimes 2}$ corresponds to the \emph{twisted} quantum double $\CD(D_8)^\gamma\cong \CZ(\Rep(H_8))$ under gauging $\ems$\footnote{Equivalently, $\CD(D_8)^\gamma \cong \CZ(\mathrm{TY} (\IZ_2\times\IZ_2,\chi_{\mathrm{diag}},+1))$. The Tambara-Yamagami fusion category of $\IZ_2\times \IZ_2$ with diagonal bicharacter is gauging equivalent to the $D_8$ with 3-cocycle twist.}, where $H_8$ is the self-dual Kac--Paljutkin Hopf algebra, see \cite{tambara1998tensor,Gelaki:2009blp,Choi:2023vgk,Diatlyk:2023fwf,Lu:2025gpt,Lu:2026rhb} for more details. By contrast, the \emph{untwisted} $\CD(D_8)$ with $\ems$ gauged yields $\CD(D_{16})$, which lacks a direct Chern-Simons description. More generally, $\Spin(n)_2/\IZ_2\cong SU(n)_1$, and for even $n=2r$, $SU(n)_1\times SU(n)_{-1}\cong \CD(\IZ_n)^\omega$, where $\omega$ corresponds to the twist $[r]\in \IZ_n \cong H^3(\IZ_n,U(1))$ and $i^*\gamma =\omega$ with $i: \IZ_n \hookrightarrow D_{2n}$. Nevertheless, the Dijkgraaf-Witten twist in \(\CD(D_{4r})^\gamma\) only changes the topological data of non-local line operators in the nearby gapped phase, compared with the untwisted quantum double. It should not affect the critical properties of local operators. 
A corresponding lattice model can be constructed by condensing $e^{2r}m^{2r}$ in the $\IZ_{4r}$ toric code and then gauging charge conjugation.

The $D_{2n}$ with odd $n$ family provides an infinite series of self-dual quantum doubles with Chern-Simons descriptions. We note a complementary generalization due to Ref.~\cite{Beigi:2010htr}, which considers gauge groups as affine linear groups $AGL(1,p) = \IF_p^+ \rtimes \IF_p^\times$, where $\IF_p$ is the finite field with prime power $p$ elements and $+$, $\times$ denote its additive and multiplicative groups respectively. The first few cases are $AGL(1,2) \cong \IZ_2$, $AGL(1,3) \cong S_3$, and $AGL(1,4) \cong A_4$. For each $p$, $\CD(AGL(1,p))$ possesses an $\ems$ permutation symmetry exchanging exactly one pair of anyons of quantum dimension $p-1$, and since $\IF_p^+$ and $\IF_p^\times$ are both abelian, $\CD(AGL(1,p))$ admits a lattice realization on a tensor product Hilbert space via gauging. A self-dual Hamiltonian and phase diagram can be constructed analogously. However, unlike the $D_{2n}$ family, this series does not always admit a simple Chern-Simons description: for instance, gauging $\ems$ in $\CD(A_4)$ maps it to $\CD(S_4)$, which is a non-Abelian finite gauge theory without a known Chern-Simons description.

\textit{\textbf{non-Abelian Wen-plaquette model and boundary Hamiltonian} ---}
Applying the qutrit Hadamard gate on every vertical bond of \eqref{eq:s3tc} yields a unitarily equivalent model whose Hamiltonian, 
\begin{equation}\label{eq:s3wen}
    H_{\DS}^\mathrm{WP} = -\sum_p \CB^\mathrm{WP}_p-\sum_{\pp} B_\pp-\sum_v \CA^\mathrm{WP}_v-\sum_\vv A_\vv+h.c.
\end{equation}
with the transformed qutrit stabilizers,
\begin{align}
     \CB_p^\mathrm{WP} &= \begin{tikzpicture}[every node/.style={outer sep=0pt, inner sep=1pt},baseline={(current bounding box.center)},line width=2pt]
  \draw[gray, thick] (0,0) -- (1,0);
  \draw[gray, thick] (1,0) -- (1,1);
  \draw[gray, thick] (1,1) -- (0,1);
  \draw[gray, thick] (0,0) -- (0,1);
  \draw[RoyalBlue,opacity=0.7, thick] (0,0.5) -- (0.5,1);
  \draw[RoyalBlue,opacity=0.7, thick] (0,0.5) -- (0.5,0);
  \draw[RoyalBlue,opacity=0.7, thick] (1,0.5) -- (0.5,1);
  \draw[RoyalBlue,opacity=0.7, thick] (1,0.5) -- (0.5,0);
  \node at (0.5,-0.2) {$\CZ_1$};
  \node at (1.0,1.25) {$\CZ_3^{ \scriptscriptstyle{\color{Blue}-Z_{(12)}Z_{(23)}}}$};
  \node[rotate=90] at (-0.3,0.7) {$\CX_4^{ {\color{Blue}-Z_{(14)}}}$};
  \node[rotate=90] at (1.3,0.7) {$\CX_2^{ {\color{Blue}Z_{(12)}}}$};
\end{tikzpicture},\;\CA_v^\mathrm{WP} = \begin{tikzpicture}[every node/.style={outer sep=0pt, inner sep=1pt},baseline={(current bounding box.center)},line width=2pt,scale=0.9]
  \draw[gray, thick] (0,0) -- (0.5,0);
  \draw[gray, thick] (0,0) -- (0,0.5);
  \draw[gray, thick] (0,0) -- (0,-0.5);
  \draw[gray, thick] (0,0) -- (-0.5,0);
  \draw[RoyalBlue,opacity=0.7, thick] (0,0.5) -- (0.5,0);
  \draw[RoyalBlue,opacity=0.7, thick] (0,0.5) -- (-0.5,0);
  \draw[RoyalBlue,opacity=0.7, thick] (0,-0.5) -- (0.5,0);
  \draw[RoyalBlue,opacity=0.7, thick] (0,-0.5) -- (-0.5,0);
  \node[rotate=90] at (0.8,0) {$\CX_2^{ {\color{Blue}Z_{(12)}}}$};
  \node at (0.4,0.7) {$\CZ_3^{\scriptscriptstyle {\color{Blue}-Z_{(12)}Z_{(23)}}}$};
  \node at (0,-0.6) {$\CZ_1$};
  \node[rotate=90] at (-0.7,0.2) {$\CX_4^{\tiny {\color{Blue}-Z_{(14)}}}$};
\end{tikzpicture},
\end{align}
$\CB^\mathrm{WP}_p$ and $\CA^\mathrm{WP}_v$ can be alternatively viewed as the plaquette operators acting on the unshaded and shaded diamonds respectively as shown in \figref{fig:latt}. The above model is a non-Abelian analogue of the Wen plaquette model. The $\emp$ permutation symmetry is implemented by translation by $(\frac12,\frac12)$, and in fact, it generalizes to any $\CD(D_{2n})$ by substituting the qutrit with qudit of dimension $n$ on the black lattice.

The two-sublattice structure with $45^\circ$ relative rotation leads to a richer boundary classification than in the ordinary Wen-plaquette or toric code models. The $45^\circ$ diagonal cut of the black lattice is dubbed as zigzag cut, then there is a choice of rough or smooth cut position of the blue lattice. Similarly a rough or smooth cut for the black lattice will lead to a zigzag cut of the blue lattice. The general boundary Hamiltonian is given by,
\begin{align}\label{eq:bdyham}
H_\mathrm{bdy} &= -J_1\sum_i \CZ_i^{-Z_{i+1/2}} \CZ_{i+1} -J_2 \sum_i \CX_i \\
&-J_3\sum_i Z_{i+1/2} -J_4 \sum_i X_{i-1/2} \CC_i  X_{i+1/2}  +h.c. \nonumber
\end{align}
where $\CZ_i^{Z_{i+1/2}}\equiv \CZ_i \frac{1+Z_{i+1/2}}{2}+\CZ^\dagger_i \frac{1-Z_{i+1/2}}{2}$. This Hamiltonian can be obtained from the $\IZ_3$ transverse-field Ising model with gauged charge conjugation symmetry. It has $\Rep(S_3)$ symmetry and describes the transition between 4 $\Rep(S_3)$ symmetric gapped phases \cite{arkya2024sdreps3}. The boundary conditions and their corresponding theories are summarized in \tabref{tab:boundary}, details can also be found in \SMref{sec:bdyHam}.

Analogously to the bulk generalization to $\CD(D_{2n})$, the boundary theory carries $\Rep(D_{2n})$ symmetry. The boundary universality class depends on $n$ through the underlying $n$-state clock model structure. For $n=4$, varying $J_3/J_4$ at fixed $J_1, J_2$ yields an Ising transition, while varying $J_1/J_2$ at fixed $J_3, J_4$ realizes the $c=1$ compact boson CFT on the orbifold branch with $r=2$ \cite{Bantay:2016nkw,Benjamin:2025knd}. This follows from the fact that the critical $4$-state clock model is described by the $\mathrm{Ising}\times \mathrm{Ising}$ CFT, which sits at $r = 1$ on the orbifold branch~\cite{Ginsparg:1987eb}, and the charge conjugation symmetry acts as permuting the two $\mathrm{Ising}$s. For $n>4$, the transition driven by $J_1/J_2$ at fixed $J_3, J_4$ is described by $c=1$ compact boson CFT on the orbifold branch, in particular, an intermediate critical phase with two end points being gauged Berezinskii-Kosterlitz-Thouless transitions.

If we take the zigzag boundary of the black lattice, and a smooth or rough boundary for the blue lattice. In the presence of translation symmetry, the boundary theory sits at a critical point with an emergent self-duality that corresponds to gauging the non-invertible $\Rep(D_{2n})$ symmetry. For $n=3$, this self-duality line operator together with $\Rep(S_3)$ generates the $JK_4$ fusion category~\cite{arkya2024sdreps3}, while for $n=4$ the extended symmetry forms a $\IZ_2$-extension of $\Rep(D_8)$~\cite{Lu:2025gpt}.

\begin{table}[]
    \centering
    \begin{tabular}{c|c|c|c}
    \hline\hline
        Black bdy & Blue bdy & \eqref{eq:bdyham} Parameters & CFT \\ \hline
        Smooth & Zigzag & $J_1=0$, vary $J_3/J_4$ & Ising \\ \hline 
        Rough & Zigzag & $J_2=0$, vary $J_3/J_4$ & Ising \\ \hline
        Zigzag & Smooth & $J_3=0$, vary $J_1/J_2$ & TetraIsing \\ \hline 
        Zigzag & Rough & $J_4=0$, vary $J_1/J_2$ & TetraIsing \\ \hline \hline
    \end{tabular}
    \caption{Boundary Hamiltonians of the self-dual $\DS$ Wen--plaquette model. The two-sublattice structure yields four boundary geometries: one sublattice has a smooth or rough gapped termination, while the other forms a zigzag boundary. The translation symmetry pins the effective spin chain at the zigzag boundary to be self-dual. The couplings $J_1,\ldots,J_4$ are defined in Eq.~\eqref{eq:bdyham}. Gapping the black-qutrit sublattice gives a $\IZ_2$ spin chain with an Ising transition. Gapping the blue-qubit sublattice gives the charge-conjugation-gauged $\IZ_3$ TFIM, whose self-dual point realizes the $c=4/5$ tetracritical Ising CFT $\mathcal{M}(6,5)$, equivalently the $\IZ_2$ charge-conjugation orbifold of the three-state Potts CFT~\cite{arkya2024sdreps3,Chang:2018iay}.}
    \label{tab:boundary}
\end{table}

\clearpage
\onecolumngrid

\setcounter{section}{0}
\setcounter{equation}{0}
\setcounter{figure}{0}
\setcounter{table}{0}

\renewcommand{\thesection}{S\arabic{section}}
\renewcommand{\theequation}{S\arabic{equation}}
\renewcommand{\thefigure}{S\arabic{figure}}
\renewcommand{\thetable}{S\arabic{table}}

\makeatletter

\begin{center}
\textbf{\large Supplemental Material}
\end{center}

\setcounter{table}{0}
\renewcommand{\thetable}{S\arabic{table}}
\setcounter{figure}{0}
\renewcommand{\thefigure}{S\arabic{figure}}
\setcounter{equation}{0}
\renewcommand{\theequation}{S\arabic{equation}}

\section{I. Review of Anyon Condensation}\label{SM:anyoncond}

Anyon condensation provides a systematic mechanism for generating new topological orders from a parent modular tensor category (MTC) $\mathcal{C}$. It plays a central role in understanding gapped phase transitions out of topologically ordered states, and furnishes the categorical counterpart to Hamiltonian anyon proliferation discussed in this work.
\newline
\paragraph{\textbf{Condensable algebras and quotient categories}}

A consistent condensation pattern in a modular tensor category \(\mathcal C\) is specified 
by a condensable algebra \(\CA\in\mathcal C\), namely a connected commutative separable 
algebra object. It comes with a multiplication and unit
\[
m:\CA\otimes \CA\to \CA,\qquad \eta:\mathbf 1\to \CA,
\]
satisfying associativity, unitality, and commutativity,
\[
m\circ c_{\CA,\CA}=m.
\]
Separability means that \(m\) admits an \(\CA\)-bimodule splitting. Equivalently, one may 
choose compatible Frobenius maps
\[
\Delta:\CA\to \CA\otimes \CA,\qquad \epsilon:\CA\to \mathbf 1.
\]
Physically, \(\CA\) specifies the mutually bosonic collection of anyons that is condensed. Given $\CA$, one constructs the category of local $\CA$-modules,
\begin{equation}
\mathcal{C}_\CA^{\mathrm{loc}},
\end{equation}
which defines the topological order after condensation. The passage
\begin{equation}
\mathcal{C} \rightarrow \mathcal{C}_\CA^{\mathrm{loc}}
\end{equation}
encodes three physical effects:
\begin{itemize}
\item Anyons in $\CA$ become part of the vacuum sector.
\item Anyons with nontrivial mutual braiding with $\CA$ are confined.
\item Remaining anyons are identified under fusion with $\CA$.
\end{itemize}

The total quantum dimension $\sfD_\CC = \sqrt{\sum_{a\in \CC}  d_a^2}$ reduces according to
\begin{equation}
\sfD_{\mathcal{C}_\CA^{\mathrm{loc}}} = \frac{\sfD_{\mathcal{C}}}{\dim(\CA)}.
\end{equation}
where $\dim(\CA) = \sum_{a\in\CA} n_a d_a$. Given $\CA= \oplus_a n_a a$, a simple consistency check is,
\begin{equation}
   \sum_c N_{ab}^c n_c\ge n_{a}n_b,\quad \forall a,b.
\end{equation}

The anyon condensation can be understood as gauging $1$-form symmetry, and the resulting theory has dual (or quantum) $0$-form symmetry. When the condensable algebra is non-abelian, then the dual $0$-form symmetry is mostly non-invertible \cite{Cui2018HopfMonad,Bottini:2026edf,Eck:2026aiz}.  
\begin{align}
    \CC \xrightarrow{\text{$\CA$ condense}} \mathcal{C}_\CA^{\mathrm{loc}}\equiv \CD \text{ with dual 0-form symmetry $\CG$} \xrightarrow{\text{Gauge $\CG$}} \CC
\end{align}

\paragraph{\textbf{Electric and magnetic condensable algebras in $\CD(G)$}}

For quantum double theories $\mathcal{C} = \CD(G)$, condensable algebras are explicitly classified and fall into two canonical families.

\paragraph{Electric-type algebras.}
For a subgroup $H \le G$,
\begin{equation}
\CA^{\mathrm{el}}_H = \mathrm{Ind}_H^G(\mathbf{1}) \cong \mathrm{Fun}(G/H),
\end{equation}
which decomposes as
\begin{equation}
\CA^{\mathrm{el}}_H = \bigoplus_{\rho} m_\rho \, ([e],\rho), \quad 
m_\rho = \frac{1}{|H|} \sum_{h \in H} \chi_\rho(h).
\end{equation}
Condensation yields
\begin{equation}
\CD(G) \;\rightarrow\; \CD(H).
\end{equation}
This is also generalized to the twisted quantum double, and $\CD^\omega(G) \;\rightarrow\; \CD^{\omega|_H}(H)$, where $\omega\in H^3(G,U(1))$ and $\omega|_H$ is the restriction to $H$.

\paragraph{Magnetic-type algebras.}
For a normal subgroup $K \triangleleft G$,
\begin{equation}
    \CA^{mg}_K = \bigoplus_{[g]\in \mathrm{Conj}(G),\, [g]\subset K} ([g],\mathbf{1}), 
\end{equation}
where $[g]$ are conjugacy classes contained in $K$. Condensation will lead to $\CD(G/K)$. This corresponds to condensing fluxes. These structures underlie all condensation transitions discussed in this work and provide the microscopic realization of the minimal-condensation principle.
\newline
\paragraph{\textbf{Minimal condensable algebra}}
Anyon condensation is a kinematic operation defined at the level of the MTC, while Hamiltonian perturbations are dynamical. The minimal-condensation principle of this work asserts their equivalence:

\begin{equation}
\text{anyon proliferation of } b \quad \Longleftrightarrow \quad \text{condensation of } \CA^b_{\min} \ni b.
\end{equation}

The lattice model of $\CD(S_3)$ provides explicit evidence for this correspondence across all three independent perturbations, establishing a direct bridge between microscopic Hamiltonians and categorical phase transitions.

Given the Hamiltonian of a topological order $\CC$, the deformation that corresponds to the proliferation of certain bosonic anyon $b$ will ``destroy'' part of the topological nature, the resulting gapped phase is proposed based on the minimal-condensation principle, and supported by the lattice construction, to be captured by the \emph{minimal} condensable algebra containing a given bosonic simple anyon $b$:
\begin{equation}
b \;\leadsto\; \CA_{\min}^b \ni b.
\end{equation}
where the superscript refers to the proliferating anyon $b$. The minimality requires $\CA$ to be the condensable algebra that contains $b$ with the least quantum dimension. Physically, this reflects the fact that lowering the gap of a single anyon necessarily induces condensation of all consistent channels.

For Abelian anyons, closure under fusion requires
\begin{equation}
\Amin^b = \bigoplus_{j=0}^{ord(b)-1} b^{\otimes j}.
\end{equation}
For non-Abelian anyons, consistency additionally requires trivial self-braiding within $\CA$ and the existence of a Frobenius algebra structure.
\newline
\paragraph{\textbf{Example: condensation in $D(S_3)$}}

The structure of $D(S_3)$ provides a concrete realization of the above framework. The anyon content and symmetry properties are summarized in \tabref{tab:anyons} of the main text. Several distinct condensable algebras arise \cite{xu2024condalg}:

\paragraph{(i) $\CA = A + B$.}
Here $B = ([e],\mathrm{sgn})$ is a bosonic charge. The corresponding algebra
\begin{equation}
\CA_2 = A + B
\end{equation}
is electric-type with $H = \mathbb{Z}_3$, yielding
\begin{equation}
\CD(S_3) \rightarrow \CD(\mathbb{Z}_3),
\end{equation}
consistent with the lattice $B$-proliferation transition.

\paragraph{(ii) $\CA = A + C$ or $A + F$.}
The anyons $C$ and $F$ are permuted by the $Z_2^\mathrm{em}$ symmetry. Each generates a minimal algebra
\begin{equation}
\CA_1 = A + C, \quad \CA'_1 = A + F,
\end{equation}
leading to
\begin{equation}
\CD(S_3) \longrightarrow \CD(\mathbb{Z}_2).
\end{equation}
This is because $C=([e],\mathbf{2})$ is electric-type with $H=\IZ_2$. And $F=([r],1)$ is the magnetic-type, will lead to $\CD(S_3/\IZ_3) \cong \CD(\IZ_2)$. The anyon condensation gives a condensation functor $\DS\rightarrow \CD(\IZ_2)\boxtimes \{X,Y\}$, following \cite{Cui2018HopfMonad},
\begin{equation}
 D_\CA:\quad A \to 1,\quad B \to e,\quad C \to 1+e,\quad D \to m+X,\quad E \to \psi+X,\quad F,G,H \to Y,
\end{equation}
where $1,e,m,\psi$ are the anyons in $\CD(\IZ_2)$ and $\{X,Y\}$ are the defects of the dual 0-form symmetry,
\begin{align}
mX &= Y, & mY &= X, & \psi Y &= X, & \psi X &= Y, & eY &= Y, \\
X^2 &= 1 + e + Y, & XY &= m + \psi + X, & Y^2 &= 1 + e + Y.
\end{align}
The $X,Y$ doesn't have an ordinary $G$-grading, due to the non-invertible nature of this dual $0$-form symmetry.

In the main text, we proliferate the non-simple bosonic anyon $C+F$. Although the strong-coupling limit has a unique $\mathbb{Z}_2^\mathrm{em}$-symmetric ground state, an intermediate-coupling regime may spontaneously break $\mathbb{Z}_2^\mathrm{em}$. In such a regime, the two symmetry-related ground states correspond to the condensable
algebras $\mathcal A_1=A+C$ and $\mathcal A_1'=A+F$, which can be regarded as a fine-tuned first-order transition point.

\paragraph{(iii) $\CA = A + C + D$ or $A + F + D$.}
For the non-Abelian anyon $D$ with $d_D = 3$, minimal algebras take the form
\begin{equation}
\CA_3 = A + F + D, \quad \CA'_3 = A + C + D.
\end{equation}
These are exchanged by $Z_2^\mathrm{em}$, and condensation yields a trivial phase with spontaneous symmetry breaking. Both of $\CA_3$ is the magnetic-type Lagrangian algebra and $\CA_3'$ is the $\ems$ dual.

\paragraph{(iv) $\CA = A + B + 2C$ or $A + B + 2F$.}
These are Lagrangian algebras, permuted by the $Z_2^\mathrm{em}$ symmetry, and $A + B + 2C$ is the electric-type. However, they do not directly arise in our Hamiltonian. For each constituent anyon $B, C, F$, there exist smaller minimal condensable algebras,
\begin{equation}
\CA_2 = A + B, \qquad \CA_1 = A + C, \qquad \CA'_1 = A + F,
\end{equation}
which have strictly smaller total quantum dimension. By the minimality principle, proliferation of a single anyon selects these smaller algebras, precluding condensation into $\CA$. Realizing $\CA$ instead requires simultaneous condensation of multiple channels, which may occur only when perturbations such as $\coeffa$ and $\coeffb$ are tuned to comparable strength.
\newline
\paragraph{\textbf{When does a single non-Abelian anyon form a condensable algebra}}

$\DS$ provides a simple example where a non-Abelian anyon ($C$ or $F$), together with the vacuum $A$, forms a consistent condensable algebra. This contrasts with the naive expectation that closure under fusion requires including all the fusion products.

\paragraph{Electric type.}
For electric condensation in $D(G)$, condensable algebras take the form
\begin{equation}
\CA = \Fun(G/H) \cong \Ind_H^G(\mathbf{1}).
\end{equation}
Requiring $\CA \cong \mathbf{1} \oplus a$ implies
\begin{equation}
\chi_{\Ind_H^G(\mathbf{1})}(g) = 1 + \chi_a(g), \quad \forall g \in G.
\end{equation}
Since $\Ind_H^G(\mathbf{1})$ is a permutation representation, its character counts fixed points and is non-negative integer-valued. Hence a necessary condition is
\begin{equation}
\chi_a(g) \in \mathbb{Z}, \quad \forall g \in G.
\end{equation}
If $a$ has non-integer character values, then $\mathbf{1}+a$ cannot arise as a condensable algebra.

The canonical examples occur for 2-transitive actions:
\begin{itemize}
\item $G=S_n$, $H=S_{n-1}$, $\Fun(G/H)=\mathbf{1}\oplus V$,
\item $G=A_n$, $H=A_{n-1}$, $\Fun(G/H)=\mathbf{1}\oplus V$,
\end{itemize}
where $V$ is the $(n-1)$-dimensional standard irrep. In these cases, proliferating a single non-Abelian anyon $V$ leads to condensation of $\mathbf{1}+V$.

 Outside these 2-transitive cases, the condition $\Ind_H^G(\mathbf 1)=\mathbf 1\oplus V$ rarely holds. For most non-Abelian groups, $\Ind_H^G(\mathbf 1)$ decomposes as $\mathbf 1\oplus V_1\oplus V_2\oplus\cdots$ with multiple nontrivial irreducible components. Hence proliferating a single non-Abelian anyon generally forces condensation of a larger condensable algebra containing additional anyons.

If the minimal condensable algebra is $\mathbf{1}+V_1+V_2$, what is wrong with $\mathbf{1}+V_1?$ We basically want to see, when $\CA = \bigoplus_\rho m_\rho \rho$ is the $\Fun(G/H)$, (1) trivial irrep appears once, (2) dimension of $\CA$ should divide $G$, (3) exist $H$, such that $m_\rho = \frac{1}{|H|}\sum_{h\in H} \chi_\rho(h)$.

Another family is $\CD(D_{2n})$:
\begin{itemize}
\item For $n$ odd,
\[
\CA_n = ([e],1) \oplus \bigoplus_{k=1}^{\frac{n-1}{2}} ([e],\rho_k),
\]
\item For $n$ even,
\[
\CA_n = ([e],++) \oplus ([e],-+) \oplus \bigoplus_{k=1}^{\frac{n}{2}-1} ([e],\rho_k).
\]
\end{itemize}
Only $n=3$ realizes $\CA = \mathbf{1}+V$. For odd prime $n$, proliferating any $([e],\rho_k)$ instead selects $\CA_n$, yielding $\CD(\IZ_2)$, consistent with minimality.

An example is $\CD(D_8)$, where $D_8$ is the order 8 dihedral group, the unique $2$d irrep $V$ of $D_8$ admits two minimal algebras
\begin{equation}
\CA_{1,2} = \mathbf{1} + s_{1,2} + V,
\end{equation}
both leading to $\CD(\IZ_2)$. So proliferating $V$ may result in spontaneous symmetry breaking.

\paragraph{Magnetic type.}
Magnetic condensation is labeled by a normal subgroup $K \triangleleft G$,
\begin{equation}
\CA_K = \bigoplus_{[g]\subset K} ([g],1).
\end{equation}
To obtain $\CA_K = \mathbf{1} \oplus a$ with non-Abelian $a$, $K$ must contain exactly one nontrivial conjugacy class. Examples include:
\begin{itemize}
\item $G=S_3$, $K=\IZ_3$,
\item $G=S_4$, $K=\IZ_2 \times \IZ_2$,
\item $G=A_4$, $K=\IZ_2 \times \IZ_2$.
\end{itemize}

In contrast, for simple groups such as $A_5$, no proper normal subgroup exists. Hence all magnetic condensable algebras are Lagrangian, and proliferating any flux drives the system directly to a trivial phase.

\section{II. Gauging $\ems$ in topological orders}\label{SM:gaugeZ2em}
We summarize the procedure for gauging the $\ems$ symmetry in a topological order $\CC$, following Ref.~\cite{barkeshli2019symmetry}. We first enrich the topological order by the symmetry data, and then gauge the symmetry. The input data consists of three pieces:
\begin{enumerate}
    \item \textit{Symmetry action:} a group homomorphism $\rho: \ems \to \mathrm{Aut}(\CC)$ specifying how $\ems$ permutes the anyons. Consistency requires that the $S$ and $T$ matrices of $\CC$ are invariant under this action.
    \item \textit{Symmetry fractionalization class:} an element of $H^2_{[\rho]}(\ems, \CA)$, where $\CA$ denotes the abelian anyons of $\CC$, classifying the projective action of $\ems$ on the anyons. A consistent fractionalization class exists only if the obstruction class in $H^3_{[\rho]}(\ems, \CA)$ vanishes, which is known as the Postnikov class.
    \item \textit{Discrete torsion:} since $H^4(\ems,U(1))$ is trivial, an element of $H^3(\ems, U(1)) \cong \IZ_2$, which classifies the choice of $\IZ_2$ SPT stacked with the theory prior to gauging. A non-trivial element corresponds to twisted gauging of $\ems$. 
\end{enumerate}
Given these data, we can construct the symmetry enriched topological order $\CC_{\ems}^\times$, and gauging $\ems$ produces a new topological order 
$(\CC_{\ems}^\times)^{\ems}$, whose anyon content is determined by the construction of Ref.~\cite{barkeshli2019symmetry}. Assuming the obstructions $H^3_{[\rho]}(G, \CA)$ and $H^4(G,U(1))$ are all trivial, we summarize the steps,
\begin{equation}
    \text{MTC }\CC \xrightarrow{\text{0-form symmetry $G$ enrich}} \CC_G^\times  \xrightarrow{\text{gauge $G$}} (\CC_G^\times )^G \text{ i.e. $G$-gauged TO}.
\end{equation}

In the tables below, the left part records the $\ems$ action on the parent anyons, while the right part lists the possible fractionalization/torsion choices and the resulting gauged topological order. Anyons fixed by $\ems$ may split into $\pm$ charge sectors after gauging, whereas nontrivial $\ems$ defects become new deconfined anyons.

We collect the data in the following, 
\paragraph{\textbf{Toric code}}
The toric code provides the simplest example and fixes our notation. The symmetry exchanges $e$ and $m$ while leaving the fermion $\psi=em$ invariant, so only the defect sector distinguishes the two choices of discrete torsion.
\begin{equation}
\begin{array}{c ccc}
\hline\hline
\text{Anyon} & d_a & \theta_a & \ems \\
\hline
1 & 1 & 1 & 1 \\
{\color{red}e} & 1 & 1 & {\color{Blue}m} \\
{\color{Blue}m} & 1 & 1 & {\color{red}e} \\
\psi & 1 & -1 & \psi \\
\hline\hline
\end{array} \xrightarrow{\text{Gauge $\ems$}} \begin{array}{c|c|c}
\hline\hline
H^2_{[\rho]}(\ems,\CA) & \nu=H^3(\ems,U(1)) & \text{Gauged TO} \\
\hline
1 & 1 & \CZ(\mathrm{TY}(\IZ_2,+1))\cong\text{Ising}\boxtimes\overline{\text{Ising}} \\
1 & -1 & \CZ(\mathrm{TY}(\IZ_2,-1))\cong SU(2)_2\times SU(2)_{-2} \\
\hline\hline
\end{array}
\end{equation}
Since the $\ems$ acts trivially on $\psi$, there are two $\ems$ defects, $\sigma^\pm$ and $\sigma^\pm \times \psi = \sigma^\mp$. The fractionalization class is trivial in this case as no abelian boson with trivial $\ems$ action. $H^3(\ems,U(1)) =\IZ_2$, denoted by $\nu=\pm$, and it corresponds to the Frobenius-Schur indicators of the defects. In particular, $\theta_{\sigma^\pm,\pm}^{SU(2)_2\times \overline{SU(2)_{-2}}}/\theta_{\sigma^\pm,\pm}^{\text{Double Ising}} = \sqrt{\nu}$, as shown in below, 
\begin{equation}
\begin{array}{c}
\text{Double Ising $\nu=+1$} \\[4pt]
\begin{array}{c ccc}
\hline\hline
\text{Anyon} & d_a & \theta_a & \text{TC Excitations} \\
\hline
(1,1) & 1 & 1 & (1,+) \\
(\psi,1) & 1 & -1 & (\psi,-) \\
(1,\psi) & 1 & -1 & (\psi,+) \\
(\psi,\psi) & 1 & 1 & (1,-) \\
(\sigma,1) & \sqrt{2} & e^{i\pi/8} & (\sigma^-,+) \\
(1,\sigma) & \sqrt{2} & e^{-i\pi/8} & (\sigma^+,+) \\
(\sigma,\psi) & \sqrt{2} & -e^{i\pi/8} & (\sigma^-,-) \\
(\psi,\sigma) & \sqrt{2} & -e^{-i\pi/8} & (\sigma^+,-) \\
(\sigma,\sigma) & 2 & 1 & e\oplus m \\
\hline\hline
\end{array}
\end{array}\quad 
\begin{array}{c}
SU(2)_2 \times SU(2)_{-2} \text{ $\nu=-1$} \\[4pt]
\begin{array}{c ccc}
\hline\hline
\text{Anyon} & d_a & \theta_a & \text{TC Excitations} \\
\hline
(0,0) & 1 & 1 & (1,+) \\
(1,0) & 1 & -1 & (\psi,-) \\
(0,1) & 1 & -1 & (\psi,+) \\
(1,1) & 1 & 1 & (1,-) \\
\left(0,\frac12\right) & \sqrt{2} & e^{-3i\pi/8} & (\sigma^-,+) \\
\left(\frac12,0\right) & \sqrt{2} & e^{3i\pi/8} & (\sigma^+,+) \\
\left(1,\frac12\right) & \sqrt{2} & -e^{-3i\pi/8} & (\sigma^-,-) \\
\left(\frac12,1\right) & \sqrt{2} & -e^{3i\pi/8} & (\sigma^+,-) \\
\left(\frac12,\frac12\right) & 2 & 1 & e\oplus m \\
\hline\hline
\end{array}
\end{array}
\end{equation}
where the TC excitations are labelled by $(a,\pm)$, where $a$ are the toric code anyons together with $\ems$ defects and $\pm$ labels the $\ems$ irreducible representations.

\paragraph{\textbf{$\IZ_3$ Toric code $\CD(\IZ_3)$}}
Similar to the $\IZ_2$ toric code, the $\ems$ on $\IZ_3$ toric code does not have non-trivial fractionalization class and the FS indicators are $\nu=\pm1$. For $\nu=-1$, the gauged TO is identified with $\CZ(\mathrm{TY}(\IZ_3,-1)) \cong SU(3)_1\times SU(2)_{-4}$ \cite{Lin:2023pgm},
\begin{equation}
\begin{array}{c ccc}
\hline\hline
\text{Anyon} & d_a & \theta_a & \ems \\
\hline
1 & 1 & 1 & 1 \\
{\color{red}e} & 1 & 1 & {\color{Blue}m} \\
{\color{red}e^2} & 1 & 1 & {\color{Blue}m^2} \\
{\color{Blue}m} & 1 & 1 & {\color{red}e} \\
{\color{Blue}m^2} & 1 & 1 & {\color{red}e^2} \\
em & 1 & \omega & em \\
e^2 m^2 & 1 & \omega & e^2 m^2 \\
{\color{red}e}{\color{Blue}m^2} & 1 & \omega^2 & {\color{red}e^2}{\color{Blue}m} \\
{\color{red}e^2}{\color{Blue}m} & 1 & \omega^2 & {\color{red}e}{\color{Blue}m^2} \\
\hline\hline
\end{array}\xrightarrow{\text{Gauge $\ems$}} \begin{array}{c|c|c}
\hline\hline
H^2_{[\rho]}(\ems,\CA) & \nu=H^3(\ems,U(1)) & \text{Gauged TO} \\
\hline
1 & 1 & \CZ(\mathrm{TY}(\IZ_3,+1)) \\
1 & -1 & \CZ(\mathrm{TY}(\IZ_3,-1))\cong SU(3)_1\times SU(2)_{-4} \\
\hline\hline
\end{array}
\end{equation}
In the next table, the last column describes how lines of the gauged theory descend from $\IZ_3$ toric-code data. Fixed anyons such as $em$ and $e^2m^2$ split into $\pm$ sectors, two-element $\ems$ orbits such as $e\oplus m$ become single orbit lines, and the three $\ems$-defect types $\sigma_i$ give the $\sqrt{3}$-dimensional lines. We can identify the anyon before and after gauging $\ems$ as,
\begin{equation}
\begin{array}{c}
SU(3)_1 \times SU(2)_{-4} \\[4pt]
\begin{array}{c ccc}
\hline\hline
\text{Anyon} & d_a & \theta_a & \IZ_3\text{ Toric Code} \\
\hline
(1,0) & 1 & 1 & (1,+)\\
(1,2) & 1 & 1 & (1,-)\\
(3,0) & 1 & \omega & (em,+)\\
(3,2) & 1 & \omega & (em,-)\\
(\bar{3},0) & 1 & \omega & (e^2m^2,+)\\
(\bar{3},2) & 1 & \omega & (e^2m^2,-)\\[2pt]

\left(1,\frac12\right) & \sqrt{3} & e^{-\mathrm{i}\pi/4} & (\sigma_0,+)\\
\left(1,\frac32\right) & \sqrt{3} & -e^{-\mathrm{i}\pi/4} & (\sigma_0,-)\\
\left(3,\frac12\right) & \sqrt{3} & \omega e^{-\mathrm{i}\pi/4} & (\sigma_1,+)\\
\left(3,\frac32\right) & \sqrt{3} & -\omega e^{-\mathrm{i}\pi/4} & (\sigma_1,-)\\
\left(\bar{3},\frac12\right) & \sqrt{3} & \omega e^{-\mathrm{i}\pi/4} & (\sigma_2,+)\\
\left(\bar{3},\frac32\right) & \sqrt{3} & -\omega e^{-\mathrm{i}\pi/4} & (\sigma_2,-)\\[2pt]

(1,1) & 2 & \omega^2 & em^2\oplus e^2 m\\
(3,1) & 2 & 1 & e\oplus m\\
(\bar{3},1) & 2 & 1 & e^2\oplus m^2\\
\hline\hline
\end{array}
\end{array}
\end{equation}

\paragraph{\textbf{$S_3$ quantum double $\DS$}}
For $\DS$, the same gauging problem is slightly richer because $\DS$ may be obtained from the $\IZ_3$ toric code by gauging charge conjugation. Thus the $\ems$-gauged theory can be compared both with $\DS$ anyons and with the $\IZ_3$ toric-code. In this case, besides the $\nu\in H^3(\ems,U(1))$, there is an abelian anyon $B$ invariant under the $\ems$ transformation, therefore, $\ems$ can fractionalize on it,
\begin{equation}
    \begin{array}{c|c|c}
\hline\hline
H^2_{[\rho]}(\ems,\CA) & \nu=H^3(\ems,U(1)) & \text{Gauged TO} \\
\hline
1 & 1 & ... \\
1 & -1 & SU(2)_4\times SU(2)_{-4} \\
B & 1 & ... \\
B & -1 & ...\\
\hline\hline
\end{array}
\end{equation}
where the $...$ refer to the other metaplectic MTCs with the same anyon types and fusion rules but differ in other topological data \cite{ardonne2021classification}.

\begin{equation}
\begin{array}{c}
SU(2)_4 \times SU(2)_{-4} \\[4pt]
\begin{array}{c c cccc}
\hline\hline
\text{Anyon} & \mathrm{Spin}(3)_2\times \mathrm{Spin}(3)_{-2} & d_a & \theta_a & S_3\text{ Quantum Double} & \IZ_3 \text{ Toric code}\\
\hline
(0,0) & (0,\bar{0}) & 1 & 1 & (A,+) & (1,+),+\\
(0,2) & (0,\bar{\eta}) & 1 & 1 & (B,-) & (1,-),+\\
(2,0) & (\eta,\bar{0}) & 1 & 1 & (B,+) & (1,+),-\\
(2,2) & (\eta,\bar{\eta}) & 1 & 1 & (A,-)& (1,-),-\\[2pt]

\left(0,\frac12\right) & (0,\bar{A}) & \sqrt{3} & e^{-\mathrm{i}\pi/4} & (\sigma_R,+) & (\sigma_0,+),+\\
\left(0,\frac32\right) & (0,\bar{B}) & \sqrt{3} & -e^{-\mathrm{i}\pi/4} & (\sigma_R',+) & (\sigma_0,-),+\\
\left(2,\frac12\right) & (\eta,\bar{A}) & \sqrt{3} & e^{-\mathrm{i}\pi/4} & (\sigma_R',-) & (\sigma_0,+),-\\
\left(2,\frac32\right) & (\eta,\bar{B}) & \sqrt{3} & -e^{-\mathrm{i}\pi/4} & (\sigma_R,-) & (\sigma_0,-),-\\
\left(\frac12,0\right) & (A,\bar{0}) & \sqrt{3} & e^{\mathrm{i}\pi/4} & (\sigma_L,+) & \sigma_C,+ \\
\left(\frac32,0\right) & (B,\bar{0}) & \sqrt{3} & -e^{\mathrm{i}\pi/4} & (\sigma_L',+)& \sigma_C',+\\
\left(\frac12,2\right) & (A,\bar{\eta}) & \sqrt{3} & e^{\mathrm{i}\pi/4} & (\sigma_L',-)& \sigma_C',-\\ 
\left(\frac32,2\right) & (B,\bar{\eta}) & \sqrt{3} & -e^{\mathrm{i}\pi/4} &(\sigma_L,-)& \sigma_C,- \\[2pt]

(0,1) & (0,\bar{\ell}_1) & 2 & \omega^2 & (H,+) & em^2\oplus e^2m,+\\
(2,1) & (\eta,\bar{\ell}_1) & 2 & \omega^2 & (H,-)&  em^2\oplus e^2m,-\\
(1,0) & (\ell_1,\bar{0}) & 2 & \omega & (G,+) & (em,+)\oplus (e^2m^2,+)\\
(1,2) & (\ell_1,\bar{\eta}) & 2 & \omega & (G,-) & (em,-)\oplus (e^2m^2,-) \\[2pt]

\left(\frac12,\frac12\right) & (A,\bar{A}) & 3 & 1 & (D,+) & \sigma_C,+ \times (\sigma_0,+),+\\
\left(\frac12,\frac32\right) & (A,\bar{B}) & 3 & -1 & (E,-) & \sigma_C,+ \times (\sigma_0,-),+\\
\left(\frac32,\frac12\right) & (B,\bar{A}) & 3 & -1 & (E,+) &\sigma_C',+ \times (\sigma_0,+),+\\
\left(\frac32,\frac32\right) & (B,\bar{B}) & 3 & 1 & (D,-)& \sigma_C',+ \times (\sigma_0,-),+\\[2pt]

\left(\frac12,1\right) & (A,\bar{\ell}_1) & 2\sqrt{3} & \omega^2 e^{\mathrm{i}\pi/4} & (\tau_L,+) & \sigma_C,+ \times em^2\oplus e^2m,+\\
\left(\frac32,1\right) & (B,\bar{\ell}_1) & 2\sqrt{3} & -\omega^2 e^{\mathrm{i}\pi/4} &(\tau_L,-)&  \sigma_C',+ \times em^2\oplus e^2m,+\\
\left(1,\frac12\right) & (\ell_1,\bar{A}) & 2\sqrt{3} & \omega e^{-\mathrm{i}\pi/4} & (\tau_R,+)&  (\sigma_1,+)\oplus (\sigma_2,+)  \\
\left(1,\frac32\right) & (\ell_1,\bar{B}) & 2\sqrt{3} & -\omega e^{-\mathrm{i}\pi/4} & (\tau_R,-)&  (\sigma_1,-)\oplus (\sigma_2,-) \\[2pt]

(1,1) & (\ell_1,\bar{\ell}_1) & 4 & 1 & C\oplus F & e\oplus m \oplus e^2\oplus m^2 \\
\hline\hline
\end{array}
\end{array}
\end{equation}
The first two columns are two common notations for the anyons of $SU(2)_4 \times SU(2)_{-4}$. The $S_3$ Quantum double column identifies the anyons after gauging $\ems$ of the $\DS$. The last column identifies the anyon after gauging $\IZ_2^C\times \ems$ in $\IZ_3$ toric code. The $\IZ_3$ toric code has charge conjugation $\IZ_2^C$, sending $e\leftrightarrow e^2$ and $m\leftrightarrow m^2$, and the electric-magnetic permutation $\ems$, sending $e^i\leftrightarrow m^i$. Gauging $\IZ_2^C$ of $\IZ_3$ toric code gives $\DS$, so gauging $\ems$ in $\DS$ is equivalently described as gauging the combined 0-form symmetry $\ems\times\IZ_2^C$ of the parent $\IZ_3$ toric code. In the table, $\sigma_C,\sigma_C'$ denote charge-conjugation defects, while $(\sigma_i,\pm)$ are the $\ems$ defects introduced in the previous paragraph. $\pm$ labels the $\IZ_2$ representation that appear upon equivariantization. Entries such as $\sigma_C\times(\sigma_i,\pm)$ therefore represent sectors carrying both charge-conjugation and electric-magnetic defect flux.

After matching the fusion data and topological spins in this way, the remaining ambiguity is the possible $H^3(\ems,U(1))$ discrete torsion in gauging $\ems$. This torsion is detected by Frobenius-Schur indicators of self-dual defect lines, which are sensitive to the choice of defect associator even when the quantum dimensions and fusion orbits are fixed. For a MTC $\CC$, the Frobenius-Schur indicator of a simple self-dual line $a\in \CC$ is determined by $\nu_a = [F_{a}^{a\bar{a} a}]_{11}/\abs{[F_{a}^{a\bar{a} a}]_{11}}$. This FS indicator is also determined by the modular data \cite{ng2007frobenius}, via,
\begin{equation}\label{eq:fsind}
    \nu_a =\frac{1}{\sfD^2_\CC} \sum_{x,y}N_{xy}^a d_x d_y \left( \frac{\theta_x}{\theta_y}\right)^2
\end{equation}
where $\sfD^2_\CC = \sum_{a\in \CC} d_a^2$ and $d_a$ is the quantum dimension for each $a$. $\theta_a$ is the topological spin of the anyon $a$. For anyons $(j_1,j_2)$ in $SU(2)_4\times SU(2)_{-4}$, the FS indicators are $\nu_{(j_1,j_2)}=(-1)^{2j_1 +2j_2}$, and the charge conjugation defects and $\ems$ defects have non-trivial FS indicators. The $SU(2)_4\times SU(2)_{-4}$ is identified with gauging the $\ems$ in $\DS$ with $\nu=-1 \in H^3(\ems,U(1))$.

\paragraph{\textbf{$\CD(D_{2n})$ for odd $n = 2p+1$.}}
As for $\DS$, gauging $\ems$ with trivial fractionalization class and a possible discrete torsion $\nu$ yields $\Spin(n)_2\times\Spin(n)_{-2}$. In the following, we will use the anyon data of $\Spin(n)_2\times\Spin(n)_{-2}$ to fix the possible discrete torsion $\nu$. The anyon content of $\Spin(2p+1)_2$, following the notation of Refs.~\cite{ardonne2021classification,Cordova2025dihedral}, is:
\begin{equation}
\begin{array}{cccc}
\hline\hline
\Spin(2p+1)_2\ \text{\cite{Cordova2025dihedral}} & 
\text{Notation in \cite{ardonne2021classification}} & d_a & \theta_a \\
\hline
0 & 1 & 1 & 1 \\
\eta & \epsilon & 1 & 1 \\
A & \psi_+ & \sqrt{2p+1} & e^{i\pi p/4} \\
B & \psi_- & \sqrt{2p+1} & -e^{i\pi p/4} \\
\ell_i\ (i=1,\ldots,p-1) & \phi_i & 2 & 
e^{i\pi i(2p+1-i)/(2p+1)} \\
\ell_p & \phi_p & 2 & e^{i\pi p(p+1)/(2p+1)} \\
\hline\hline
\end{array}
\end{equation}
Using the above spin and \eqref{eq:fsind}, the FS indicator for $A/B$ anyon is given by,
\begin{equation}\label{eq:spinfs}
    \nu_{A/B} = \left(\frac{2}{n} \right) = (-1)^{(n^2-1)/8} = \begin{cases}
        +1,& n=1,7 \mod 8\\ -1,& n=3,5 \mod 8
    \end{cases}
\end{equation}
where $\left(\frac{2}{n} \right)$ is the Jacobi symbol and it shows a $\mathrm{mod } 8$ rule. The non-zero $\nu$ also tells the $A/B$ anyons are self-dual which is consistent with the trivial $\ems$ fractionalization class, as non-trivial fractionalization class leads to $\bar{A} = B$. Then the $\Spin(n)_{\pm2}$ corresponds to a particular metaplectic MTC labeled by $(p,r,\kappa,\lambda) = (p,1,(-1)^{(n^2-1)/8},\mp1)$ in the classification of Ref.~\cite{ardonne2021classification}. The non-trivial discrete torsion $\nu = (-1)^{(n^2-1)/8} \in H^3(\ems,U(1))$ is confirmed by the Frobenius-Schur indicator. So $\Spin(n)_2\times\Spin(n)_{-2}$ with suitable $n$ corresponds to $D_6,D_{10}$ with twisted gauging of $\ems$, and $D_{14},D_{18}$ with untwisted gauging.

We use the presentation $D_{2n} = \langle r,s \mid r^n = s^2 = 1,\, srs = r^{-1}\rangle$. The conjugacy classes of $D_{2n}$ are $[e]$ (size $1$), $[s]$ (size $n$), and $[r^i]$ for $i=1,\ldots,p$ (size $2$), and the irreps are $\mathbf{1}, \mathbf{1}'$ and $\mathbf{2}_i$ for $i=1,\ldots,p$. The anyons of $\Spin(n)_2\times\Spin(n)_{-2}$ are identified with those of $\CD(D_{2n})$ after gauging $\ems$ as follows:
\begin{equation}
\begin{array}{ccc}
\hline\hline
\Spin(n)_2\times\Spin(n)_{-2} & \CD(D_{2n}) & \text{Type} \\
\hline
(0,0) & ([e],\mathbf{1}) & \text{vacuum} \\
(\eta,0) & ([e],\mathbf{1}') & \text{charge} \\
(\ell_i,\bar{\ell}_i),\ i=1,\ldots,p & 
([r^i],\mathbf{1})\oplus([e],\mathbf{2}_i) & \text{flux+charge} \\
(A,0),\,(0,\bar{A}) & - & \ems\ \text{defect} \\
(B,0),\,(0,\bar{B}) & - & \ems\ \text{defect} \\
\hline\hline
\end{array}
\end{equation}
where $(A/B,0)$ and $(0,\bar{A}/\bar{B})$ are the bare $\ems$ defects with no $\CD(D_{2n})$ counterpart, as they arise from gauging the $\ems$ symmetry. The other anyons can be similarly matched.

For odd $n$, the untwisted $\IZ_n$ toric code is equivalently written as $SU(n)_1\times SU(n)_{-1}$, which is also equivalent to the abelian MTC $\CC(\IZ_n,q_+)\times \CC(\IZ_n,q_-)$ with quadratic form $q_\pm(a) = \pm a^2/n \mod{1}$ \cite{Wang:2020nmz}. In particular the anyon $e^i m^j$ in the $\IZ_n$ toric code is equivalent to $e=a+b,m=a-b$ for $(a,b) \in \CC(\IZ_n,q_+)\times \CC(\IZ_n,q_-)$. Then the $\ems$ is given by the particle-hole transformation on $\CC(\IZ_n,q_-)$. According to \cite{ardonne2021classification}, gauging the particle-hole symmetry of $\CC(\IZ_n,q)$ with possible twist given by \eqref{eq:spinfs} for some $q$ will lead to $\Spin(n)_2$\cite{barkeshli2019symmetry}. In particular, gauging $\ems$ in the $\IZ_n$ toric code gives $SU(n)_1\times \Spin(n)_{-2}$. Further gauging the charge-conjugation gives $\Spin(n)_{2}\times \Spin(n)_{-2}$.

We also comment on the case of even $n$. For example, $n=4$, $\Spin(4)_2\times \Spin(4)_{-2}\cong SU(2)_2^{\boxtimes 2}\times SU(2)_{-2}^{\boxtimes 2}$ is equivalent to the twisted quantum double $\CD(D_8)^\gamma$ under gauging $\ems$, where the 3-cocycle $\gamma\in H^3(D_8,U(1))$ restricts to the element $2\in \IZ_4 \cong H^3(\IZ_4,U(1))$ on the subgroup $\IZ_4 \subset D_8$. Note that the $\ems$ permutes two non-abelian anyons which is different from the full $\eta_1,\eta_2$ in \cite{Lu:2025gpt}. More generally, we propose that $\Spin(n)_2\times \Spin(n)_{-2}$ with $n=2r$ is equivalent to the twisted quantum double $\CD(D_{4r})^\gamma$ under gauging $\ems$, where $\gamma$ is an order-2 element of $H^3(D_{4r},U(1))$ whose restriction to $\IZ_{2r}\subset D_{4r}$ corresponds to $r\in \IZ_{2r}\cong H^3(\IZ_{2r},U(1))$. This is because if ungauging the $\ems$ and charge conjugation, $\Spin(2r)_2\times \Spin(2r)_{-2}/(\IZ_2\times \IZ_2) \cong SU(2r)_1\times SU(2r)_{-1}$ which is equivalent to the twisted $\IZ_{2r}$ quantum double with twist $r$. To remove the twist, one can double it to be $SU(4r)_1\times SU(4r)_{-1}$ and quotient the diagonal $\IZ_2$ 1-form, such that it realizes untwisted $\IZ_{2r}$ quantum double. We leave the detailed verification of this proposal to future work.

When $n$ is odd and $n=x^2$ or $n$ is even $n=2y^2$, then the $\Spin(n)_2$ is integral, meaning that every simple line has integer quantum dimension. Moreover, the modular data of $\Spin(x^2)_2$ is equivalent to those of $\CD(D_{2x})^\omega$, since $\frac{\Spin(n^2)_2}{\IZ_2} \cong SU(n^2)_1\cong \CD^{\omega=2}(\IZ_n)$ \cite{Cordova2025dihedral}. And $\Spin(8)_2\times\Spin(8)_{-2}$ is equivalent to the twisted quantum double of the small group $G =[32, 49]$ \cite{deaton2020integral}.
\newline
\paragraph{\textbf{When does gauging $\ems$ lead to (twisted) quantum double}}
We note that $\Spin(n)_2\times\Spin(n)_{-2}$ arises from gauging the $\ems\times\IZ_2^{\CC}$ 0-form symmetry in the $\IZ_n$ toric code for odd $n$. The resulting topological order contains anyons of quantum dimension $\sqrt{n}$, which is irrational for $n>1$ and not a perfect square. Although the total quantum dimension $\sfD = \sqrt{\sum_a d_a^2}$ remains an integer, the presence of irrational individual quantum dimensions implies that $\Spin(n)_2\times\Spin(n)_{-2}$ cannot be realized as the quantum double $\CD(H)$ of any finite group or Hopf algebra. However, there are examples of such gauging result in a quantum double model. One interesting example is gauging $\IZ_3\rtimes \ems \cong S_3$ in $\IZ_2\times \IZ_2$ Toric code,
\begin{equation}
    \IZ_3 : \begin{pmatrix}
        e_1\\e_2\\m_1\\m_2
    \end{pmatrix} \mapsto \left(
\begin{array}{cccc}
 0 & 1 & 0 & 0 \\
 1 & 1 & 0 & 0 \\
 0 & 0 & 1 & 1 \\
 0 & 0 & 1 & 0 \\
\end{array}
\right) \begin{pmatrix}
        e_1\\e_2\\m_1\\m_2
    \end{pmatrix} ,\quad \ems : \begin{pmatrix}
        e_1\\e_2\\m_1\\m_2
    \end{pmatrix}\mapsto \left(
\begin{array}{cccc}
 0 & 0 & 1 & 0 \\
 0 & 0 & 0 & 1 \\
 1 & 0 & 0 & 0 \\
 0 & 1 & 0 & 0 \\
\end{array}
\right) \begin{pmatrix}
        e_1\\e_2\\m_1\\m_2
    \end{pmatrix}
\end{equation}
Gauging the $\IZ_3$ will maps $\IZ_2\times \IZ_2$ Toric code to $A_4$ quantum double, and $\ems$ will permute the quantum dimension 3 anyons in the $\CD(A_4)$. Further gauging the $\ems$ in $\CD(A_4)$ will map it to $\CD(S_4)$. Gauging another $\IZ_2^T$ anyon permutation symmetry,
\begin{equation}
    \IZ_2^T: \begin{pmatrix}
        e_1\\e_2\\m_1\\m_2
    \end{pmatrix}\mapsto \left(
\begin{array}{cccc}
 1 & 0 & 0 & 1 \\
 0 & 1 & 1 & 0 \\
 0 & 0 & 1 & 0 \\
 0 & 0 & 0 & 1 \\
\end{array}
\right) \begin{pmatrix}
        e_1\\e_2\\m_1\\m_2
    \end{pmatrix}
\end{equation}
will map $\CD(A_4)$ to the quantum double $\CD(\mathrm{SL}(2,\IZ_3))$. The group $SL(2,\mathbb Z_3)=\{A\in M_{2\times 2}(\mathbb Z_3)\mid \det A=1\}$ has order $24$ and is isomorphic to the binary tetrahedral group $2T$. It is a nontrivial double cover of $A_4$, via $1\to \mathbb Z_2\to SL(2,\mathbb Z_3)\to A_4\to 1$, and hence is one of the finite subgroups of $SU(2)$. The quantum double $\CD(SL(2,\IZ_3))$ captures the categorical symmetry of $A_4$ orbifolded chiral CFT \cite{Dijkgraaf:1989hb,Perez-Lona:2024sds}.

The condition for gauging a 0-form symmetry $G$ to yield a quantum double is the existence of a Lagrangian algebra invariant under $G$~\cite{Gelaki:2009blp,Sun:2023xxv,Lu:2025yru}. 
In the $\IZ_2\times\IZ_2$ toric code, the Lagrangian algebra $\CA = 1 + e_1m_2 + e_2m_1 + e_1e_2m_1m_2$ is invariant under both $\ems$ and $\IZ_3$, which is why the gauging $\IZ_3 \rtimes \ems $ yields the quantum double $\CD(S_4)$. Also, $\CA=1+e_1+e_2+e_1 e_2$ is invariant under $\IZ_3\times \IZ_2^T$, so gauging which yields quantum double $\CD(\mathrm{SL}(2,\IZ_3))$. Similarly, gauging $\ems\circ \mathrm{SWAP}$ in $\IZ_2\times\IZ_2$ toric code will lead to $\CD(D_8)$ or gauging $\ems$ will lead to $\CD(H_8)\cong \CD(D_8)^\gamma$ because there are symmetric Lagrangian algebras.

By contrast, in the $\IZ_n$ toric code for odd $n$, neither the electric Lagrangian algebra $\CA_e = \sum_{i=0}^{n-1} e^i$ nor the magnetic one $\CA_m = \sum_{i=0}^{n-1} m^i$ is invariant under $\ems$, which explains why gauging $\ems\times \IZ_2^\CC$ produces $\Spin(n)_2\times\Spin(n)_{-2}$ rather than a quantum double.

\section{III. Prepare the self-dual $\DS$ stabilizers}\label{sm:prepare}
In this section, we provide details on the construction of the self-dual $\DS$ stabilizers. We begin with the $\IZ_3$ toric code and then gauge its charge-conjugation symmetry. While this procedure was already described in \cite{ruben2021DS3}, the gauging graph used there is not manifestly $\ems$ symmetric. Indeed, the $\IZ_3$ toric code possesses an $\ems$ permutation symmetry implemented by a qutrit Hadamard transformation followed by a $(-\frac12,-\frac12)$ translation, whereas their gauging graph is not invariant under the translation. To make the $\ems$ symmetry manifest, we instead use the following gauging graph:
\begin{equation}\label{eq:gauginggraph}
\vcenter{\hbox{\includegraphics[width=0.3\textwidth]{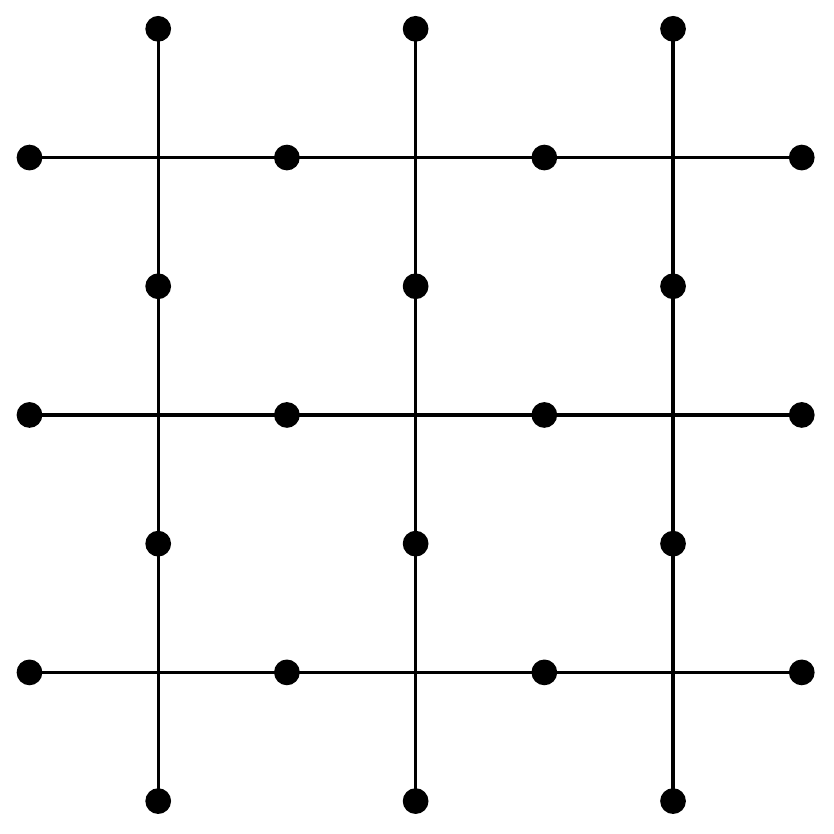}}}
\;\rightarrow\;
\vcenter{\hbox{\includegraphics[width=0.3\textwidth]{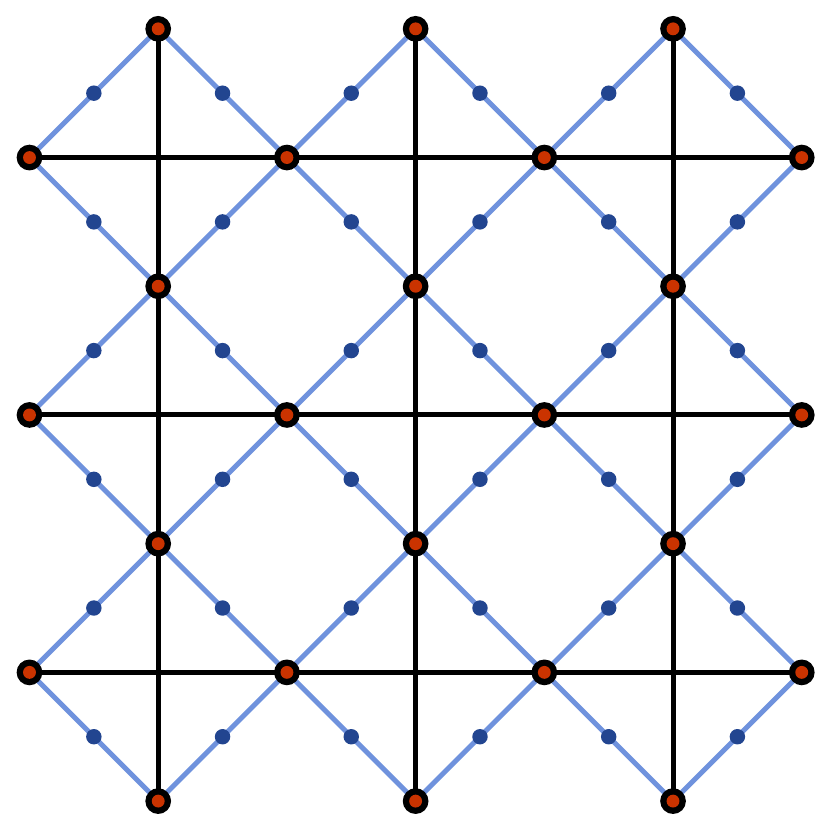}}}
\end{equation}
where the original $\IZ_3$ toric code is on the edge of the black lattice, the qutrits are labeled by the black dots {\color{black} $\bullet$}. Following the gauging procedure in \cite{ruben2021DS3}, we add additional qubit ancillas on the blue Lieb lattice, labeled by the red {\color{red} $\bullet$} and blue dots {\color{Blue} $\bullet$}.

The gauging procedure is as follows, the hermitian conjugate part is added implicitly,
\begin{enumerate}
    \item We start with the stabilizers for the $\IZ_3$ toric code, 
    \begin{equation}
        \begin{tikzpicture}[baseline={(current bounding box.center)},line width=2pt]
  \draw[gray, thick] (0,0) -- (1,0);
  \draw[gray, thick] (1,0) -- (1,1);
  \draw[gray, thick] (1,1) -- (0,1);
  \draw[gray, thick] (0,0) -- (0,1);
  \node at (0.5,0) {$\CZ$};
  \node at (0.5,1) {$\CZ^\dagger$};
  \node at (0,0.5) {$\CZ^\dagger$};
  \node at (1,0.5) {$\CZ$};
\end{tikzpicture},\quad \begin{tikzpicture}[baseline={(current bounding box.center)},line width=2pt]
  \draw[gray, thick] (0,0) -- (1,0);
  \draw[gray, thick] (0,0) -- (0,1);
  \draw[gray, thick] (0,0) -- (0,-1);
  \draw[gray, thick] (0,0) -- (-1,0);
  \node at (0.5,0) {$\CX$};
  \node at (0,0.5) {$\CX$};
  \node at (0,-0.5) {$\CX^\dagger$};
  \node at (-0.5,0) {$\CX^\dagger$};
\end{tikzpicture}
    \end{equation}
\item With the additional qubit ancillas, we have,
\begin{equation}
        \begin{tikzpicture}[baseline={(current bounding box.center)},line width=2pt]
  \draw[gray, thick] (0,0) -- (1,0);
  \draw[gray, thick] (1,0) -- (1,1);
  \draw[gray, thick] (1,1) -- (0,1);
  \draw[gray, thick] (0,0) -- (0,1);
  \draw[RoyalBlue, thick] (0,0.5) -- (0.5,1);
  \draw[RoyalBlue, thick] (0,0.5) -- (0.5,0);
  \draw[RoyalBlue, thick] (1,0.5) -- (0.5,1);
  \draw[RoyalBlue, thick] (1,0.5) -- (0.5,0);
  \node at (0.5,0) {$\CZ$};
  \node at (0.5,1) {$\CZ^\dagger$};
  \node at (0,0.5) {$\CZ^\dagger$};
  \node at (1,0.5) {$\CZ$};
\end{tikzpicture},\quad \begin{tikzpicture}[baseline={(current bounding box.center)},line width=2pt]
  \draw[gray, thick] (0,0) -- (1,0);
  \draw[gray, thick] (0,0) -- (0,1);
  \draw[gray, thick] (0,0) -- (0,-1);
  \draw[gray, thick] (0,0) -- (-1,0);
  \draw[RoyalBlue, thick] (0,0.5) -- (0.5,0);
  \draw[RoyalBlue, thick] (0,0.5) -- (-0.5,0);
  \draw[RoyalBlue, thick] (0,-0.5) -- (0.5,0);
  \draw[RoyalBlue, thick] (0,-0.5) -- (-0.5,0);
  \node at (0.5,0) {$\CX$};
  \node at (0,0.5) {$\CX$};
  \node at (0,-0.5) {$\CX^\dagger$};
  \node at (-0.5,0) {$\CX^\dagger$};
\end{tikzpicture},\quad \begin{tikzpicture}[baseline={(current bounding box.center)},line width=2pt]
  \draw[RoyalBlue, thick] (-0.5,-0.5) -- (0.5,0.5);
  \draw[RoyalBlue, thick] (0.5,-0.5) -- (-0.5,0.5);
  \node at (0,0) {\color{red}$X$};
\end{tikzpicture},\begin{tikzpicture}[baseline={(current bounding box.center)},line width=2pt]
  \draw[RoyalBlue, thick] (-0.5,-0.5) -- (0.5,0.5);
  \node at (0,0) {\color{Blue}$X$};
\end{tikzpicture},\begin{tikzpicture}[baseline={(current bounding box.center)},line width=2pt]
  \draw[RoyalBlue, thick] (0.5,-0.5) -- (-0.5,0.5);
  \node at (0,0) {\color{Blue}$X$};
\end{tikzpicture}
    \end{equation}
\item Then we use the controlled-charge conjugation from red qubit to black qutrit,
\begin{equation}
    {\color{red}C}\CC: {\color{red}X}\rightarrow {\color{red}X}\CC,\quad \CX\rightarrow \CX^{ {\color{red}Z}},\quad \CZ\rightarrow \CZ^{ {\color{red}Z}}
\end{equation}
Since the black qutrit and red qubit overlaps on the same site as in \eqref{eq:gauginggraph}, we get,
\begin{equation}
        \begin{tikzpicture}[baseline={(current bounding box.center)},line width=2pt]
  \draw[gray, thick] (0,0) -- (1,0);
  \draw[gray, thick] (1,0) -- (1,1);
  \draw[gray, thick] (1,1) -- (0,1);
  \draw[gray, thick] (0,0) -- (0,1);
  \draw[RoyalBlue, thick] (0,0.5) -- (0.5,1);
  \draw[RoyalBlue, thick] (0,0.5) -- (0.5,0);
  \draw[RoyalBlue, thick] (1,0.5) -- (0.5,1);
  \draw[RoyalBlue, thick] (1,0.5) -- (0.5,0);
  \node at (0.5,0) {$\CZ_1^{ {\color{red}Z_1}}$};
  \node at (0.5,1) {$\CZ_3^{ {\color{red}-Z_3}}$};
  \node at (0,0.5) {$\CZ_4^{ {\color{red}-Z_4}}$};
  \node at (1,0.5) {$\CZ_2^{ {\color{red}Z_2}}$};
\end{tikzpicture},\quad \begin{tikzpicture}[baseline={(current bounding box.center)},line width=2pt]
  \draw[gray, thick] (0,0) -- (1,0);
  \draw[gray, thick] (0,0) -- (0,1);
  \draw[gray, thick] (0,0) -- (0,-1);
  \draw[gray, thick] (0,0) -- (-1,0);
    \draw[RoyalBlue, thick] (0,0.5) -- (0.5,0);
  \draw[RoyalBlue, thick] (0,0.5) -- (-0.5,0);
  \draw[RoyalBlue, thick] (0,-0.5) -- (0.5,0);
  \draw[RoyalBlue, thick] (0,-0.5) -- (-0.5,0);
  \node at (0.5,0) {$\CX_2^{ {\color{red}Z_2}}$};
  \node at (0,0.5) {$\CX_3^{ {\color{red}Z_3}}$};
  \node at (0,-0.5) {$\CX_1^{ {\color{red}-Z_1}}$};
  \node at (-0.5,0) {$\CX_4^{ {\color{red}-Z_4}}$};
\end{tikzpicture},\quad \begin{tikzpicture}[baseline={(current bounding box.center)},line width=2pt]
  \draw[RoyalBlue, thick] (-0.5,-0.5) -- (0.5,0.5);
  \draw[RoyalBlue, thick] (0.5,-0.5) -- (-0.5,0.5);
  \node at (0,0) {$\CC{\color{red}X}$};
\end{tikzpicture},\begin{tikzpicture}[baseline={(current bounding box.center)},line width=2pt]
  \draw[RoyalBlue, thick] (-0.5,-0.5) -- (0.5,0.5);
  \node at (0,0) {\color{Blue}$X$};
\end{tikzpicture},\begin{tikzpicture}[baseline={(current bounding box.center)},line width=2pt]
  \draw[RoyalBlue, thick] (0.5,-0.5) -- (-0.5,0.5);
  \node at (0,0) {\color{Blue}$X$};
\end{tikzpicture}
    \end{equation}
\item Using controlled-Z to entangle the red and blue qubit to form a $\IZ_2^{(0)}\times \IZ_2^{(1)}$ cluster state,
\begin{equation}
        \begin{tikzpicture}[baseline={(current bounding box.center)},line width=2pt]
  \draw[gray, thick] (0,0) -- (1,0);
  \draw[gray, thick] (1,0) -- (1,1);
  \draw[gray, thick] (1,1) -- (0,1);
  \draw[gray, thick] (0,0) -- (0,1);
  \draw[RoyalBlue, thick] (0,0.5) -- (0.5,1);
  \draw[RoyalBlue, thick] (0,0.5) -- (0.5,0);
  \draw[RoyalBlue, thick] (1,0.5) -- (0.5,1);
  \draw[RoyalBlue, thick] (1,0.5) -- (0.5,0);
  \node at (0.5,0) {$\CZ_1^{ {\color{red}Z_1}}$};
  \node at (0.5,1) {$\CZ_3^{ {\color{red}-Z_3}}$};
  \node at (0,0.5) {$\CZ_4^{ {\color{red}-Z_4}}$};
  \node at (1,0.5) {$\CZ_2^{ {\color{red}Z_2}}$};
\end{tikzpicture},\quad \begin{tikzpicture}[baseline={(current bounding box.center)},line width=2pt]
  \draw[gray, thick] (0,0) -- (1,0);
  \draw[gray, thick] (0,0) -- (0,1);
  \draw[gray, thick] (0,0) -- (0,-1);
  \draw[gray, thick] (0,0) -- (-1,0);
    \draw[RoyalBlue, thick] (0,0.5) -- (0.5,0);
  \draw[RoyalBlue, thick] (0,0.5) -- (-0.5,0);
  \draw[RoyalBlue, thick] (0,-0.5) -- (0.5,0);
  \draw[RoyalBlue, thick] (0,-0.5) -- (-0.5,0);
  \node at (0.5,0) {$\CX_2^{ {\color{red}Z_2}}$};
  \node at (0,0.5) {$\CX_3^{ {\color{red}Z_3}}$};
  \node at (0,-0.5) {$\CX_1^{ {\color{red}-Z_1}}$};
  \node at (-0.5,0) {$\CX_4^{ {\color{red}-Z_4}}$};
\end{tikzpicture},\quad \begin{tikzpicture}[baseline={(current bounding box.center)},line width=2pt]
  \draw[RoyalBlue, thick] (-0.5,-0.5) -- (0.5,0.5);
  \draw[RoyalBlue, thick] (0.5,-0.5) -- (-0.5,0.5);
  \node at (0,0) {$\CC{\color{red}X}$};
  \node at (0.3,0.3) {${\color{Blue}Z}$};
  \node at (0.3,-0.3) {${\color{Blue}Z}$};
  \node at (-0.3,0.3) {${\color{Blue}Z}$};
  \node at (-0.3,-0.3) {${\color{Blue}Z}$};
\end{tikzpicture},\begin{tikzpicture}[baseline={(current bounding box.center)},line width=2pt]
  \draw[RoyalBlue, thick] (-0.5,-0.5) -- (0.5,0.5);
  \node at (0,0) {\color{Blue}$X$};
  \node at (-0.3,-0.3) {\color{red}$Z$};
  \node at (0.3,0.3) {\color{red}$Z$};
\end{tikzpicture},\begin{tikzpicture}[baseline={(current bounding box.center)},line width=2pt]
  \draw[RoyalBlue, thick] (0.5,-0.5) -- (-0.5,0.5);
  \node at (0,0) {\color{Blue}$X$};
  \node at (0.3,-0.3) {\color{red}$Z$};
  \node at (-0.3,0.3) {\color{red}$Z$};
\end{tikzpicture}
    \end{equation}
Although the first two terms have single ${\color{red}Z}$ on the vertices, ${\CX_i}^{\color{red}Z_i}, {\CZ_i}^{\color{red}Z_i}$ are invariant, therefore, the plaquette and star terms are invariant under the Gauss law $\begin{tikzpicture}[baseline={(current bounding box.center)},line width=2pt]
  \draw[RoyalBlue, thick] (-0.5,-0.5) -- (0.5,0.5);
  \draw[RoyalBlue, thick] (0.5,-0.5) -- (-0.5,0.5);
  \node at (0,0) {$\CC{\color{red}X}$};
  \node at (0.3,0.3) {${\color{Blue}Z}$};
  \node at (0.3,-0.3) {${\color{Blue}Z}$};
  \node at (-0.3,0.3) {${\color{Blue}Z}$};
  \node at (-0.3,-0.3) {${\color{Blue}Z}$};
\end{tikzpicture}$.

\item Then measure the red qubits in the $X$ basis. No post-selection is needed, since nontrivial measurement outcomes correspond to gauge charge excitations of the final $\DS$ state and can be paired and annihilated by local unitaries. After these corrections we work in the sector ${\color{red}X}=1$. Using the trick \footnote{Around (D.11) in \cite{ruben2021DS3}} in \cite{ruben2021DS3} to further simplify the qutrit plaquette and star terms, we basically want even number of ${\color{red}Z}$ and substitute them by ${\color{Blue}X}$, we get,
\begin{equation}
        \begin{tikzpicture}[baseline={(current bounding box.center)},line width=2pt]
  \draw[gray, thick] (0,0) -- (1,0);
  \draw[gray, thick] (1,0) -- (1,1);
  \draw[gray, thick] (1,1) -- (0,1);
  \draw[gray, thick] (0,0) -- (0,1);
  \draw[RoyalBlue, thick] (0,0.5) -- (0.5,1);
  \draw[RoyalBlue, thick] (0,0.5) -- (0.5,0);
  \draw[RoyalBlue, thick] (1,0.5) -- (0.5,1);
  \draw[RoyalBlue, thick] (1,0.5) -- (0.5,0);
  \node at (0.5,0) {$\CZ_1$};
  \node at (1.3,1) {$\CZ_3^{ {\color{Blue}-X_{(12)}X_{(23)}}}$};
  \node at (0,0.5) {$\CZ_4^{ {\color{Blue}-X_{(14)}}}$};
  \node at (1.3,0.5) {$\CZ_2^{ {\color{Blue}X_{(12)}}}$};
\end{tikzpicture},\quad \begin{tikzpicture}[baseline={(current bounding box.center)},line width=2pt]
  \draw[gray, thick] (0,0) -- (1,0);
  \draw[gray, thick] (0,0) -- (0,1);
  \draw[gray, thick] (0,0) -- (0,-1);
  \draw[gray, thick] (0,0) -- (-1,0);
    \draw[RoyalBlue, thick] (0,0.5) -- (0.5,0);
  \draw[RoyalBlue, thick] (0,0.5) -- (-0.5,0);
  \draw[RoyalBlue, thick] (0,-0.5) -- (0.5,0);
  \draw[RoyalBlue, thick] (0,-0.5) -- (-0.5,0);
  \node at (0.9,0) {$\CX_2^{ {\color{Blue}X_{(12)}}}$};
  \node at (0.6,0.7) {$\CX_3^{ {\color{Blue}X_{(12)}X_{(23)}}}$};
  \node at (0,-0.5) {$\CX_1^\dagger$};
  \node at (-0.5,0) {$\CX_4^{ {\color{Blue}-X_{(14)}}}$};
\end{tikzpicture},\quad \begin{tikzpicture}[baseline={(current bounding box.center)},line width=2pt]
  \draw[RoyalBlue, thick] (-0.5,-0.5) -- (0.5,0.5);
  \draw[RoyalBlue, thick] (0.5,-0.5) -- (-0.5,0.5);
  \node at (0,0) {$\CC$};
  \node at (0.3,0.3) {${\color{Blue}Z}$};
  \node at (0.3,-0.3) {${\color{Blue}Z}$};
  \node at (-0.3,0.3) {${\color{Blue}Z}$};
  \node at (-0.3,-0.3) {${\color{Blue}Z}$};
\end{tikzpicture},\begin{tikzpicture}[baseline={(current bounding box.center)},line width=2pt]
  \draw[RoyalBlue, thick] (-0.5,0) -- (0,0.5);
  \draw[RoyalBlue, thick] (-0.5,0) -- (0,-0.5);
  \draw[RoyalBlue, thick] (0.5,0) -- (0,0.5);
  \draw[RoyalBlue, thick] (0.5,0) -- (0,-0.5);
  \node at (0.25,0.25) {\color{Blue}$X$};
  \node at (0.25,-0.25) {\color{Blue}$X$};
  \node at (-0.25,0.25) {\color{Blue}$X$};
  \node at (-0.25,-0.25) {\color{Blue}$X$};
\end{tikzpicture}
    \end{equation}
where $\color{Blue}X_{(ij)}$ denotes the Pauli-X on blue link that connects $i,j$.
\item Finally a basis rotation of the qubit, we obtain,
\begin{equation}
       \textbf{Self-dual $S_3$ quantum double:  } \begin{tikzpicture}[baseline={(current bounding box.center)},line width=2pt]
  \draw[gray, thick] (0,0) -- (1,0);
  \draw[gray, thick] (1,0) -- (1,1);
  \draw[gray, thick] (1,1) -- (0,1);
  \draw[gray, thick] (0,0) -- (0,1);
  \draw[RoyalBlue, thick] (0,0.5) -- (0.5,1);
  \draw[RoyalBlue, thick] (0,0.5) -- (0.5,0);
  \draw[RoyalBlue, thick] (1,0.5) -- (0.5,1);
  \draw[RoyalBlue, thick] (1,0.5) -- (0.5,0);
  \node at (0.5,0) {$\CZ_1$};
  \node at (1.3,1) {$\CZ_3^{ {\color{Blue}-Z_{(12)}Z_{(23)}}}$};
  \node at (0,0.5) {$\CZ_4^{ {\color{Blue}-Z_{(14)}}}$};
  \node at (1.3,0.5) {$\CZ_2^{ {\color{Blue}Z_{(12)}}}$};
\end{tikzpicture},\quad \begin{tikzpicture}[baseline={(current bounding box.center)},line width=2pt]
  \draw[gray, thick] (0,0) -- (1,0);
  \draw[gray, thick] (0,0) -- (0,1);
  \draw[gray, thick] (0,0) -- (0,-1);
  \draw[gray, thick] (0,0) -- (-1,0);
  \draw[RoyalBlue, thick] (0,0.5) -- (0.5,0);
  \draw[RoyalBlue, thick] (0,0.5) -- (-0.5,0);
  \draw[RoyalBlue, thick] (0,-0.5) -- (0.5,0);
  \draw[RoyalBlue, thick] (0,-0.5) -- (-0.5,0);
  \node at (0.9,0) {$\CX_2^{ {\color{Blue}Z_{(12)}}}$};
  \node at (0.6,0.7) {$\CX_3^{ {\color{Blue}Z_{(12)}Z_{(23)}}}$};
  \node at (0,-0.5) {$\CX_1^\dagger$};
  \node at (-0.5,0) {$\CX_4^{ {\color{Blue}-Z_{(14)}}}$};
\end{tikzpicture},\quad \begin{tikzpicture}[baseline={(current bounding box.center)},line width=2pt]
  \draw[RoyalBlue, thick] (-0.5,-0.5) -- (0.5,0.5);
  \draw[RoyalBlue, thick] (0.5,-0.5) -- (-0.5,0.5);
  \node at (0,0) {$\CC$};
  \node at (0.3,0.3) {${\color{Blue}X}$};
  \node at (0.3,-0.3) {${\color{Blue}X}$};
  \node at (-0.3,0.3) {${\color{Blue}X}$};
  \node at (-0.3,-0.3) {${\color{Blue}X}$};
\end{tikzpicture},\begin{tikzpicture}[baseline={(current bounding box.center)},line width=2pt]
  \draw[RoyalBlue, thick] (-0.5,0) -- (0,0.5);
  \draw[RoyalBlue, thick] (-0.5,0) -- (0,-0.5);
  \draw[RoyalBlue, thick] (0.5,0) -- (0,0.5);
  \draw[RoyalBlue, thick] (0.5,0) -- (0,-0.5);
  \node at (0.25,0.25) {\color{Blue}$Z$};
  \node at (0.25,-0.25) {\color{Blue}$Z$};
  \node at (-0.25,0.25) {\color{Blue}$Z$};
  \node at (-0.25,-0.25) {\color{Blue}$Z$};
\end{tikzpicture}
    \end{equation}
\end{enumerate}

The generalization of Wen-plaquette model is obtained by applying the $\CH$ on the vertical bonds of the qutrits, 
\begin{equation}
        \textbf{$S_3$ Wen-plaquette:  }\begin{tikzpicture}[baseline={(current bounding box.center)},line width=2pt]
  \draw[gray, thick] (0,0) -- (1,0);
  \draw[gray, thick] (1,0) -- (1,1);
  \draw[gray, thick] (1,1) -- (0,1);
  \draw[gray, thick] (0,0) -- (0,1);
  \draw[RoyalBlue, thick] (0,0.5) -- (0.5,1);
  \draw[RoyalBlue, thick] (0,0.5) -- (0.5,0);
  \draw[RoyalBlue, thick] (1,0.5) -- (0.5,1);
  \draw[RoyalBlue, thick] (1,0.5) -- (0.5,0);
  \node at (0.5,0) {$\CZ_1$};
  \node at (1.3,1) {$\CZ_3^{ {\color{Blue}-Z_{(12)}Z_{(23)}}}$};
  \node at (0,0.5) {$\CX_4^{ {\color{Blue}-Z_{(14)}}}$};
  \node at (1.3,0.5) {$\CX_2^{ {\color{Blue}Z_{(12)}}}$};
\end{tikzpicture},\quad \begin{tikzpicture}[baseline={(current bounding box.center)},line width=2pt]
  \draw[gray, thick] (0,0) -- (1,0);
  \draw[gray, thick] (0,0) -- (0,1);
  \draw[gray, thick] (0,0) -- (0,-1);
  \draw[gray, thick] (0,0) -- (-1,0);
  \draw[RoyalBlue, thick] (0,0.5) -- (0.5,0);
  \draw[RoyalBlue, thick] (0,0.5) -- (-0.5,0);
  \draw[RoyalBlue, thick] (0,-0.5) -- (0.5,0);
  \draw[RoyalBlue, thick] (0,-0.5) -- (-0.5,0);
  \node at (0.9,0) {$\CX_2^{ {\color{Blue}Z_{(12)}}}$};
  \node at (0.6,0.7) {$\CZ_3^{ {\color{Blue}-Z_{(12)}Z_{(23)}}}$};
  \node at (0,-0.5) {$\CZ_1$};
  \node at (-0.5,0) {$\CX_4^{ {\color{Blue}-Z_{(14)}}}$};
\end{tikzpicture},\quad \begin{tikzpicture}[baseline={(current bounding box.center)},line width=2pt]
  \draw[RoyalBlue, thick] (-0.5,-0.5) -- (0.5,0.5);
  \draw[RoyalBlue, thick] (0.5,-0.5) -- (-0.5,0.5);
  \node at (0,0) {$\CC$};
  \node at (0.3,0.3) {${\color{Blue}X}$};
  \node at (0.3,-0.3) {${\color{Blue}X}$};
  \node at (-0.3,0.3) {${\color{Blue}X}$};
  \node at (-0.3,-0.3) {${\color{Blue}X}$};
\end{tikzpicture},\begin{tikzpicture}[baseline={(current bounding box.center)},line width=2pt]
  \draw[RoyalBlue, thick] (-0.5,0) -- (0,0.5);
  \draw[RoyalBlue, thick] (-0.5,0) -- (0,-0.5);
  \draw[RoyalBlue, thick] (0.5,0) -- (0,0.5);
  \draw[RoyalBlue, thick] (0.5,0) -- (0,-0.5);
  \node at (0.25,0.25) {\color{Blue}$Z$};
  \node at (0.25,-0.25) {\color{Blue}$Z$};
  \node at (-0.25,0.25) {\color{Blue}$Z$};
  \node at (-0.25,-0.25) {\color{Blue}$Z$};
\end{tikzpicture}
    \end{equation}
One can easily see that the first two terms are swapped by $(-\frac12,-\frac12)$-translation, while the qubit stabilizers remain unchanged, this is consistent with the fact that $\emp$ permutation only permutes the anyon $C\leftrightarrow F$ in $D(S_3)$.
\begin{equation}\label{eq:finalgraph}
    \includegraphics[width=0.3\textwidth]{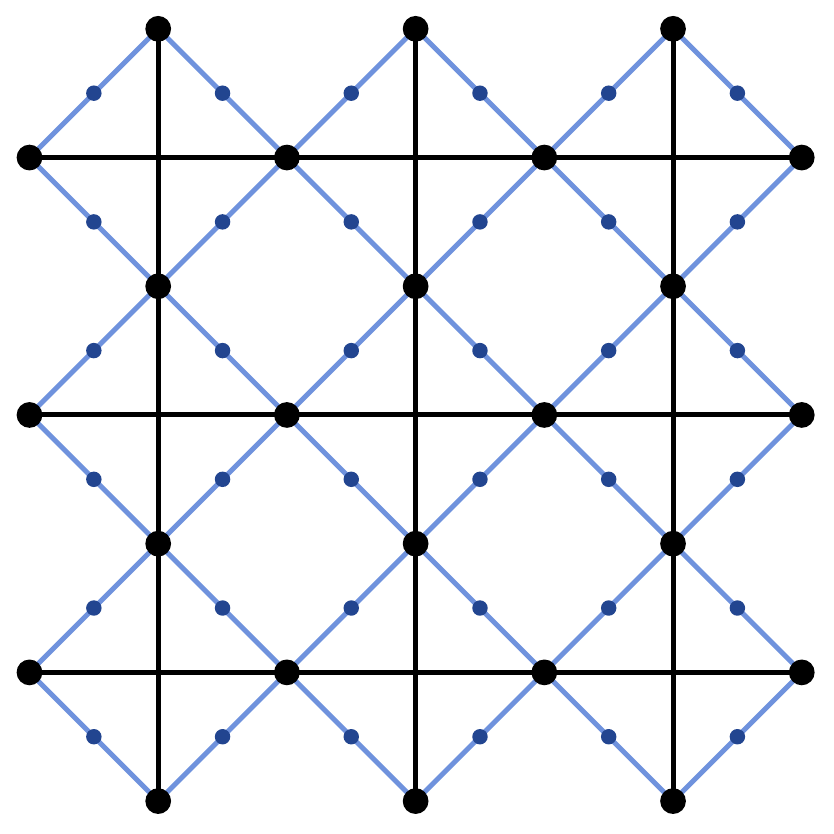}
\end{equation}
Note that due to the lattice structure \eqref{eq:finalgraph}, the number of qubits is $2\times$ the number of qutrits. The number of qubit stabilizers are also doubled compared to the qutrit stabilizers. Because of the doubled number of qubits, the $\emp$ permutation symmetry is manifested.

\paragraph{\textbf{Boundary Hamiltonian}}\label{sec:bdyHam}
Let's take the model in \eqref{eq:s3wen}, where the $\emp$ permutation symmetry is given by the translation. Since the blue lattice is 45$\deg$, there are 4 typical types of boundaries, depending on horizontal/$45\deg$ cut or smooth/rough for the black lattice. For example, we consider the following ``smooth'' boundary with $45\deg$ cut, as shown in \eqref{eq:smoothbdye}. The boundary theory realizes the $\Rep(S_3)$ symmetric model with additional self-duality. 
\begin{equation}\label{eq:smoothbdye}
    \includegraphics[width=0.3\textwidth]{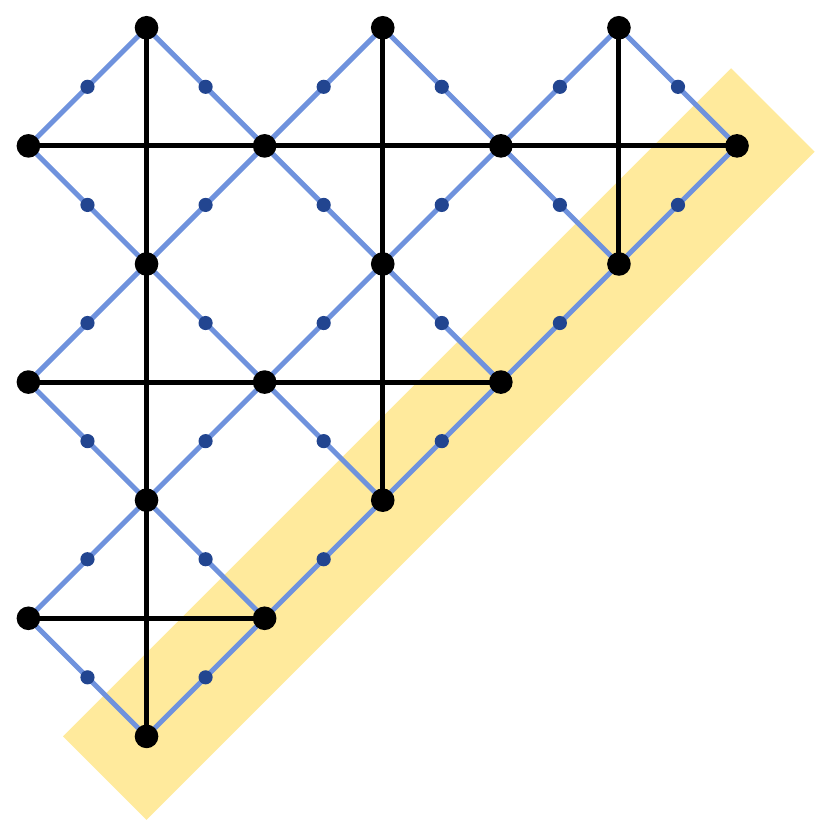}
    \end{equation}
    Recall the stabilizers for the $S_3$ Wen plaquette model,
    \begin{equation}
        \textbf{$S_3$ Wen-plaquette:  }\begin{tikzpicture}[baseline={(current bounding box.center)},line width=2pt]
  \draw[gray, thick] (0,0) -- (1,0);
  \draw[gray, thick] (1,0) -- (1,1);
  \draw[gray, thick] (1,1) -- (0,1);
  \draw[gray, thick] (0,0) -- (0,1);
  \draw[RoyalBlue, thick] (0,0.5) -- (0.5,1);
  \draw[RoyalBlue, thick] (0,0.5) -- (0.5,0);
  \draw[RoyalBlue, thick] (1,0.5) -- (0.5,1);
  \draw[RoyalBlue, thick] (1,0.5) -- (0.5,0);
  \node at (0.5,0) {$\CZ_1$};
  \node at (1.3,1) {$\CZ_3^{ {\color{Blue}-Z_{(12)}Z_{(23)}}}$};
  \node at (0,0.5) {$\CX_4^{ {\color{Blue}-Z_{(14)}}}$};
  \node at (1.3,0.5) {$\CX_2^{ {\color{Blue}Z_{(12)}}}$};
\end{tikzpicture},\quad \begin{tikzpicture}[baseline={(current bounding box.center)},line width=2pt]
  \draw[gray, thick] (0,0) -- (1,0);
  \draw[gray, thick] (0,0) -- (0,1);
  \draw[gray, thick] (0,0) -- (0,-1);
  \draw[gray, thick] (0,0) -- (-1,0);
  \draw[RoyalBlue, thick] (0,0.5) -- (0.5,0);
  \draw[RoyalBlue, thick] (0,0.5) -- (-0.5,0);
  \draw[RoyalBlue, thick] (0,-0.5) -- (0.5,0);
  \draw[RoyalBlue, thick] (0,-0.5) -- (-0.5,0);
  \node at (0.9,0) {$\CX_2^{ {\color{Blue}Z_{(12)}}}$};
  \node at (0.6,0.7) {$\CZ_3^{ {\color{Blue}-Z_{(12)}Z_{(23)}}}$};
  \node at (0,-0.5) {$\CZ_1$};
  \node at (-0.5,0) {$\CX_4^{ {\color{Blue}-Z_{(14)}}}$};
\end{tikzpicture},\quad \begin{tikzpicture}[baseline={(current bounding box.center)},line width=2pt]
  \draw[RoyalBlue, thick] (-0.5,-0.5) -- (0.5,0.5);
  \draw[RoyalBlue, thick] (0.5,-0.5) -- (-0.5,0.5);
  \node at (0,0) {$\CC$};
  \node at (0.3,0.3) {${\color{Blue}X}$};
  \node at (0.3,-0.3) {${\color{Blue}X}$};
  \node at (-0.3,0.3) {${\color{Blue}X}$};
  \node at (-0.3,-0.3) {${\color{Blue}X}$};
\end{tikzpicture},\begin{tikzpicture}[baseline={(current bounding box.center)},line width=2pt]
  \draw[RoyalBlue, thick] (-0.5,0) -- (0,0.5);
  \draw[RoyalBlue, thick] (-0.5,0) -- (0,-0.5);
  \draw[RoyalBlue, thick] (0.5,0) -- (0,0.5);
  \draw[RoyalBlue, thick] (0.5,0) -- (0,-0.5);
  \node at (0.25,0.25) {\color{Blue}$Z$};
  \node at (0.25,-0.25) {\color{Blue}$Z$};
  \node at (-0.25,0.25) {\color{Blue}$Z$};
  \node at (-0.25,-0.25) {\color{Blue}$Z$};
\end{tikzpicture}
    \end{equation}
The boundary stabilizers are,
    \begin{equation}
        \textbf{Smooth bdy:  }\begin{tikzpicture}[baseline={(current bounding box.center)},line width=2pt]
  \draw[gray, thick] (0,0) -- (0.5,0);
  \draw[gray, thick] (1,0.5) -- (1,1);
  \draw[gray, thick] (1,1) -- (0,1);
  \draw[gray, thick] (0,0) -- (0,1);
  \draw[RoyalBlue, thick] (0,0.5) -- (0.5,1);
  \draw[RoyalBlue, thick] (0,0.5) -- (0.5,0);
  \draw[RoyalBlue, thick] (1,0.5) -- (0.5,1);
  \draw[RoyalBlue, thick] (1,0.5) -- (0.5,0);
  \node at (0.5,0) {$\CX_1$};
  \node at (1.3,0.5) {$\CZ_2^{ {\color{Blue}Z_{(12)}}}$};
\end{tikzpicture}+h.c.,\quad \begin{tikzpicture}[baseline={(current bounding box.center)},line width=2pt]
  \draw[gray, thick] (0,0) -- (0.5,0);
  \draw[gray, thick] (0,0) -- (0,0.5);
  \draw[gray, thick] (0,0) -- (0,-0.5);
  \draw[gray, thick] (0,0) -- (-0.5,0);
  \draw[RoyalBlue, thick] (0,0.5) -- (0.5,0);
  \draw[RoyalBlue, thick] (0,0.5) -- (-0.5,0);
  \draw[RoyalBlue, thick] (0,-0.5) -- (0.5,0);
  \draw[RoyalBlue, thick] (0,-0.5) -- (-0.5,0);
  \node at (0.9,0) {$\CZ_2^{ {\color{Blue}Z_{(12)}}}$};
  \node at (0,-0.5) {$\CX_1$};
\end{tikzpicture}+h.c.,\quad \begin{tikzpicture}[baseline={(current bounding box.center)},line width=2pt]
  \draw[RoyalBlue, thick] (-0.5,-0.5) -- (0.5,0.5);
  \draw[RoyalBlue, thick] (0,0) -- (-0.5,0.5);
  \node at (0,0) {$\CC$};
  \node at (0.3,0.3) {${\color{Blue}X}$};
  \node at (-0.3,0.3) {${\color{Blue}X}$};
  \node at (-0.3,-0.3) {${\color{Blue}X}$};
\end{tikzpicture}
\end{equation}
The boundary algebra is an effective $\IZ_3$ spin TFIM with charge conjugation symmetry being gauged,
\begin{equation}
H_\mathrm{bdy} = -\sum_i \CZ_i^{-Z_{i+1/2}} \CZ_{i+1} -\sum_i  \CX_i -\sum_i X_{i-1/2} \CC_i  X_{i+1/2}  +h.c.  
\end{equation}
If we break the translation symmetry at boundary, then we have the general boundary Hamiltonian,
\begin{equation}
H_\mathrm{bdy} = -\sum_i \CZ_i^{-Z_{i+1/2}} \CZ_{i+1} - h\sum_i  \CX_i- \sum_i X_{i-1/2} \CC_i  X_{i+1/2}  +h.c.   
\end{equation}
The above Hamiltonian is exactly the $\Rep(S_3)$ symmetric Hamiltonian studied in \cite{arkya2024sdreps3} with $J_1 =J_4= 1,J_2=h,J_3=0$, which describes the transition between $\Rep(S_3)$ SSB to $\IZ_2$ SSB \cite{arkya2024sdreps3}. For $h=0,\infty$, the Hamiltonian is exactly solvable, and the $\Rep(S_3)$ SSB has GSD=3, $\IZ_2$ SSB has GSD=2. The transition point at $h_c=1$ due to the self-duality is described by the tetracritical Ising CFT $M(6,5)$ with central charge $4/5$, where $\Rep(S_3)$ is the subfusion category of the non-invertible symmetry generated by its topological defect lines. The tetracritical Ising CFT relates to the 3-state Potts CFT by ungauging the charge conjugation symmetry. The 3-state Potts CFT can be viewed as non-diagonal Virasoro minimal model, while tetracritical Ising is the diagonal variant \cite{Cappelli:1986hf,Chang:2018iay}.

\section{IV. Bi-adjoint Higgs transition of $SU(2)\times SU(2)$ gauge theory}\label{SM:biadjoint}

In this section, we present details of the $SU(2)_L\times SU(2)_R$ 
Chern-Simons gauge theory coupled to a bi-adjoint scalar field. The 
scalar field
\begin{equation}
    P(x) = \sum_{a,b}\phi^{ab}(x)\, T_L^a\otimes T_R^b,
\end{equation}
where $T_{L,R}^a = \sigma^a/2$ are the generators of $SU(2)_{L,R}$ 
and $\phi^{ab}$ is a real $3\times 3$ matrix, transforms in the 
$(3,3)$ bi-adjoint representation. Under $SU(2)_L\times SU(2)_R$ 
gauge transformations,
\begin{align}
    P &\mapsto (U_L\otimes U_R)\,P\,(U_L\otimes U_R)^\dagger, 
    \quad U_{L,R}\in SU(2)_{L,R}, \\
    \phi^{ab} &\mapsto (R_L)^{a}{}_{a'}(R_R)^{b}{}_{b'}\,\phi^{a'b'},
    \quad R_{L,R}\in SO(3)_{L,R},
\end{align}
where $R_{L,R}$ are the $SO(3)$ matrices. It is then clear that $(-U_L)\otimes U_R$ and $U_L\otimes (-U_R)$ have the same action as $U_L\otimes U_R$, therefore, the $\IZ_2\times \IZ_2$ center acts trivially on $P(x)$.

Denoting the $3\times 3$ real matrix $\phi^{ab}$ as $\phi$, the gauge-invariant terms in the potential are
\begin{align}
    &m^2 \Tr(\phi \phi^\intercal)+\kappa_3\det \phi +\kappa_4 \Tr((\phi \phi^\intercal)^2)+\kappa_4' (\Tr(\phi \phi^\intercal))^2\\
    &\sim 4m^2 \Tr(P^2)-\frac{8}{3}\kappa_3 \Tr(P^3)-32 \kappa_4 \Tr(P^4)+(24\kappa_4 +16\kappa_4' )\Tr(P^2)^2,\quad 
\end{align}
The generic vacuum configuration takes the form $\phi = \mathrm{diag}(v_1,v_2,v_3)$. We analyze the residual gauge group in $SO(3)_L\times SO(3)_R$ and lift it to $SU(2)_L\times SU(2)_R$ to determine the resulting phase. Starting from $SU(2)_k\times SU(2)_{-k}$ Chern-Simons theory, three distinct symmetry-breaking patterns arise:
\begin{enumerate}
    \item \textit{Generic vacuum} $v_1\ne v_2\ne v_3$: the residual gauge group in $SO(3)_L\times SO(3)_R$ is the diagonal $\IZ_2\times\IZ_2$, which lifts to $Q_8\times\IZ_2$ in $SU(2)_L\times SU(2)_R$. The resulting phase is a deconfined $Q_8\times\IZ_2$ gauge theory with a possible twist.

    \item \textit{Isotropic vacuum} $v_1=v_2=v_3=v$: favored by $\kappa_4>0$, the residual gauge group is the diagonal $SO(3)_\mathrm{diag}\subset SO(3)_L\times SO(3)_R$, lifting to $SU(2)_\mathrm{diag}\times \IZ_2$ in $SU(2)_L\times SU(2)_R$. The $SU(2)_\mathrm{diag}$ factor confines in $2+1$d, leaving a deconfined $\IZ_2$ gauge theory $\CD(\IZ_2)^k$ with twist given by $k$.

    \item \textit{Uniaxial vacuum} $v_1=v,\,v_2=v_3=0$: favored by $\kappa_4<0$, the residual gauge group is $SO(2)_L\times SO(2)_R\subset SO(3)_L\times SO(3)_R$, lifting to $(U(1)_L\times U(1)_R)\rtimes\IZ_2$ in $SU(2)_L\times SU(2)_R$, where $\IZ_2$ acts by simultaneous charge conjugation on the two $U(1)$ factors. The resulting phase is a $U(1)_{2k}\times U(1)_{-2k}$ gauge theory with charge conjugation gauged, since $SU(2)_k\to U(1)_{2k}$ from the standard embedding~\cite{Moore:1989yh}. In our case, $[(U(1)_8\times U(1)_{-8})^\times_{\IZ_2}]^{\IZ_2} \cong [(\CD(\IZ_8)^{\omega_4})^\times_{\IZ_2}]^{\IZ_2}$, the twisted quantum double of $\IZ_8$ with $\omega_4\in H^3(\IZ_8,U(1))$ and charge conjugation gauged.
\end{enumerate}
For $\kappa_4 >0$, the quartic terms favor the isotropic vacuum. The quartic potential is bounded from below if $\kappa_4>-\kappa_4',\kappa_4>-3\kappa_4'$. In this regime, the scalar condensate drives $SU(2)_4\times SU(2)_{-4}$ to $\CD(\IZ_2)$, realizing the scenario of the $C$+$F$ transition where intermediate coupling has $\ems$ SSB with a $\IZ_2$ gauge theory. The cubic term is symmetry allowed and fixes the sign of the configuration, i.e. $\mathrm{sign}(\kappa_3)=-\mathrm{sign}(v)$. At mean-field level, the cubic term may render the critical point to be first-order, although the fate in the full strongly coupled 2+1d gauge theory needs to be determined \cite{2018PNASchong,2022JHEPsubir}.

Before closing this section, we want to mention an interesting case: $SU(2)_1\times SU(2)_{-1}\cong \mathrm{DSem}$ with bi-adjoint and condense down to (1) isotropic vacuum, according to the above argument, it give raise to the $\mathrm{DSem}$ to $\CD(\IZ_2)^{k=1}\cong \mathrm{DSem}$ transition. Under gauging the $\IZ_2$ 1-form symmetry, the transition is dual to $\IZ_2$ SPT to $\IZ_2$ SPT transition in 2+1d. It looks like an unnecessary critical point, which we leave for future exploration. (2) uniaxial vacuum, the resulting theory is $[(U(1)_2\times U(1)_{-2})^\times_{\IZ_2}]^{\IZ_2} \cong \CD(\IZ_4)$. So the $\mathrm{DSem}$ transitions to $\CD(\IZ_4)$, which is related by anyon condensation of $e^2m^2$ in $\CD(\IZ_4)$. 

\section{V. Quantum to classical mapping for $\DS$}\label{sm:qtocl}
We follow the quantum-to-classical mapping of Ref.~\cite{Tupitsyn:2008ah}: the quantum partition function is evaluated in the gauge-invariant subspace, $\mathsf{Z}=\Tr(\CP e^{-\beta H_\lambda})$, where $\CP$ is the corresponding projector. Imaginary time is discretized into $N_\tau$ slices with spacing $\dt=\beta/N_\tau$. After splitting the Hamiltonian into diagonal and off-diagonal pieces, $H_\lambda=H_{\rm diag}+H_{\rm off}$, the Suzuki--Trotter step is
\begin{equation*}
\mathsf{T}=e^{-\dt H_{\rm off}}\CP e^{-\dt H_{\rm diag}},
    \qquad
    \mathsf{Z}=\lim_{N_\tau\to\infty}\Tr \mathsf{T}^{N_\tau}.
\end{equation*}
Thus $Z$-diagonal operators become spatial Boltzmann weights proportional to $\dt$, while $X$-type off-diagonal operators generate temporal bonds or temporal plaquettes after inserting complete $X$-basis states and expanding the local constraints.

We use additive notation for the microscopic variables: qubit labels $\beta,\gamma,\mu$ take values in $\IZ_2$, qutrit labels $\tau,\rho,\phi$ take values in $\IZ_3$, and all exponents are understood modulo the appropriate group order. Black-lattice links carry qutrit variables $\tau_\ell$, blue-lattice links carry qubit variables $\beta_\mathrm{l}$, red auxiliary qubits carry $\mu_\mathrm{v}$, and purple auxiliary qutrits carry $\phi_v$. The indices $\mathrm{v}$ and $v$ denote blue-lattice and black-lattice vertices; $\ell_x,\ell_y$ denote spatial black links and $\ell_t$ denotes the imaginary-time direction. The finite Fourier variables $\gamma_s^z\in\IZ_2$ and $\rho_v^z\in\IZ_3$ impose the qubit and qutrit constraints and become temporal gauge fields. In multiplicative notation we use
\begin{equation*}
    S_\mathrm{v,l_{x/y}}=(-1)^{\beta^z},\quad
    S_\mathrm{v,l_t}=(-1)^{\gamma^z},\quad
    \mu_\mathrm{v}=(-1)^{\mu^z},\quad
    \CS_{v,\ell_{x,y}}=\omega^{\tau^z},\quad
    \CS_{v,\ell_t}=\omega^{\rho^z},\quad
    \phi_v=\omega^{\phi^z},\quad
    \omega=e^{2\pi\ii/3}.
\end{equation*}
Here $S,\CS$ denote the $\IZ_2,\IZ_3$ classical gauge fields respectively and $\mu_\mathrm{v},\phi_v$ denote the corresponding matter fields. The Pauli $Z$ operator acting on the computational bases are defined by, e.g.,
\begin{equation*}
    \CZ\ket{\tau^z}=\omega^{\tau^z}\ket{\tau^z},\qquad
    Z\ket{\beta^z}=(-1)^{\beta^z}\ket{\beta^z}.
\end{equation*}
The shift operators act as $\CX\ket{\tau^z}=\ket{\tau^z+1}$ and $X\ket{\beta^z}=\ket{\beta^z+1}$, while charge conjugation acts as $\CC\ket{\tau^z}=\ket{-\tau^z}$.
We suppress overall normalization constants from basis insertions and finite Fourier transforms. The constraint projectors are represented by
\begin{equation*}
    \delta_2(n=0)\sim \sum_{\gamma\in\IZ_2}(-1)^{\gamma n},\qquad
    \delta_3(m=0)\sim \sum_{\rho\in\IZ_3}\omega^{\rho m},
\end{equation*}
and the $X$ and $Z$-basis overlaps are
\begin{equation*}
    \braket{\beta^z}{\beta^x}=(-1)^{\beta^z\beta^x},\qquad
    \braket{\tau^z}{\tau^x}=\braket{\tau^x}{\tau^z}^*=\omega^{-\tau^z\tau^x}.
\end{equation*}
where we omit the normalization factors $1/\sqrt{2}$ and $1/\sqrt{3}$. $X\ket{\beta^x}=(-1)^{\beta^x}\ket{\beta^x}$ and $\CX\ket{\tau^x}=\omega^{\tau^x}\ket{\tau^x}$. We will use the orthonormal charge-conjugation eigenbasis $\{\ket{\tau^\CC}\}$ to resolve the action of $\CC$, the eigenvectors are $(1,0,0)^\intercal$, $(0,1,1)^\intercal/\sqrt{2}$, and $(0,1,-1)^\intercal/\sqrt{2}$, with eigenvalues $1,1,-1$, respectively, and we use $(-1)^{\tau^\CC}$ to denote this eigenvalue.

Recall the Hamiltonian 
\begin{align}
    H_\DS&=- \CJ^z\begin{tikzpicture}[baseline={(current bounding box.center)},line width=2pt]
  \draw[gray, thick] (0,0) -- (1,0);
  \draw[gray, thick] (1,0) -- (1,1);
  \draw[gray, thick] (1,1) -- (0,1);
  \draw[gray, thick] (0,0) -- (0,1);
  \draw[RoyalBlue, thick] (0,0.5) -- (0.5,1);
  \draw[RoyalBlue, thick] (0,0.5) -- (0.5,0);
  \draw[RoyalBlue, thick] (1,0.5) -- (0.5,1);
  \draw[RoyalBlue, thick] (1,0.5) -- (0.5,0);
  \node at (0.5,0) {$\CZ_1$};
  \node at (1.3,1) {$\CZ_3^{ {\color{Blue}-Z_{(12)}Z_{(23)}}}$};
  \node at (0,0.5) {$\CZ_4^{ {\color{Blue}-Z_{(14)}}}$};
  \node at (1.3,0.5) {$\CZ_2^{ {\color{Blue}Z_{(12)}}}$};
\end{tikzpicture}- \CJ^x\begin{tikzpicture}[baseline={(current bounding box.center)},line width=2pt]
  \draw[gray, thick] (0,0) -- (1,0);
  \draw[gray, thick] (0,0) -- (0,1);
  \draw[gray, thick] (0,0) -- (0,-1);
  \draw[gray, thick] (0,0) -- (-1,0);
  \draw[RoyalBlue, thick] (0,0.5) -- (0.5,0);
  \draw[RoyalBlue, thick] (0,0.5) -- (-0.5,0);
  \draw[RoyalBlue, thick] (0,-0.5) -- (0.5,0);
  \draw[RoyalBlue, thick] (0,-0.5) -- (-0.5,0);
  \node at (0.9,0) {$\CX_2^{ {\color{Blue}Z_{(12)}}}$};
  \node at (0.6,0.7) {$\CX_3^{ {\color{Blue}Z_{(12)}Z_{(23)}}}$};
  \node at (0,-0.5) {$\CX_1^\dagger$};
  \node at (-0.5,0) {$\CX_4^{ {\color{Blue}-Z_{(14)}}}$};
\end{tikzpicture}- J^x\begin{tikzpicture}[baseline={(current bounding box.center)},line width=2pt]
  \draw[RoyalBlue, thick] (-0.5,-0.5) -- (0.5,0.5);
  \draw[RoyalBlue, thick] (0.5,-0.5) -- (-0.5,0.5);
  \node at (0,0) {$\CC$};
  \node at (0.3,0.3) {${\color{Blue}X}$};
  \node at (0.3,-0.3) {${\color{Blue}X}$};
  \node at (-0.3,0.3) {${\color{Blue}X}$};
  \node at (-0.3,-0.3) {${\color{Blue}X}$};
\end{tikzpicture}-J^z\begin{tikzpicture}[baseline={(current bounding box.center)},line width=2pt]
  \draw[RoyalBlue, thick] (-0.5,0) -- (0,0.5);
  \draw[RoyalBlue, thick] (-0.5,0) -- (0,-0.5);
  \draw[RoyalBlue, thick] (0.5,0) -- (0,0.5);
  \draw[RoyalBlue, thick] (0.5,0) -- (0,-0.5);
  \node at (0.25,0.25) {\color{Blue}$Z$};
  \node at (0.25,-0.25) {\color{Blue}$Z$};
  \node at (-0.25,0.25) {\color{Blue}$Z$};
  \node at (-0.25,-0.25) {\color{Blue}$Z$};
\end{tikzpicture}+h.c. \nonumber \\
H_\lambda &= H_\DS - \coeffa (\CZ+\CX+h.c.) -\coeffb Z -\coeffc X .
\end{align}
The Hamiltonian is not gauge-invariant, but it can be made gauge-invariant by introducing the dummy spin variables for qubits and qutrits as matter fields. The gauge transformations for qubits and qutrits are distinct and coupled together. We introduce the qubits at sites of blue lattice (which is equivalently on the link of the black lattice) with ${\color{red} X} = 1$, and qutrit ${\color{purple} \CX}=1$ at vertices of the black lattice. After a unitary transformation, the Hamiltonian becomes,
\begin{align}
 &H_\DS=- \CJ^z\begin{tikzpicture}[baseline={(current bounding box.center)},line width=2pt]
  \draw[gray, thick] (0,0) -- (1,0);
  \draw[gray, thick] (1,0) -- (1,1);
  \draw[gray, thick] (1,1) -- (0,1);
  \draw[gray, thick] (0,0) -- (0,1);
  \draw[RoyalBlue, thick] (0,0.5) -- (0.5,1);
  \draw[RoyalBlue, thick] (0,0.5) -- (0.5,0);
  \draw[RoyalBlue, thick] (1,0.5) -- (0.5,1);
  \draw[RoyalBlue, thick] (1,0.5) -- (0.5,0);
  \node at (0.5,0) {$\CZ_1^{{\color{red}Z_1}}$};
  \node at (0.6,1) {$\CZ_3^{ {\color{red}-Z_3}}$};
  \node at (0,0.5) {$\CZ_4^{ {\color{red}-Z_4}}$};
  \node at (1.3,0.5) {$\CZ_2^{ {\color{red}Z_2}}$};
\end{tikzpicture}- \CJ^x\begin{tikzpicture}[baseline={(current bounding box.center)},line width=2pt]
  \draw[gray, thick] (0,0) -- (1,0);
  \draw[gray, thick] (0,0) -- (0,1);
  \draw[gray, thick] (0,0) -- (0,-1);
  \draw[gray, thick] (0,0) -- (-1,0);
  \draw[RoyalBlue, thick] (0,0.5) -- (0.5,0);
  \draw[RoyalBlue, thick] (0,0.5) -- (-0.5,0);
  \draw[RoyalBlue, thick] (0,-0.5) -- (0.5,0);
  \draw[RoyalBlue, thick] (0,-0.5) -- (-0.5,0);
  \node at (0,0) {\color{purple} $\CX$};
\end{tikzpicture}- J^x\begin{tikzpicture}[baseline={(current bounding box.center)},line width=2pt]
  \draw[RoyalBlue, thick] (-0.5,-0.5) -- (0.5,0.5);
  \draw[RoyalBlue, thick] (0.5,-0.5) -- (-0.5,0.5);
  \node at (0,0) {${\color{red}X}$};
\end{tikzpicture}-J^z\begin{tikzpicture}[baseline={(current bounding box.center)},line width=2pt]
  \draw[RoyalBlue, thick] (-0.5,0) -- (0,0.5);
  \draw[RoyalBlue, thick] (-0.5,0) -- (0,-0.5);
  \draw[RoyalBlue, thick] (0.5,0) -- (0,0.5);
  \draw[RoyalBlue, thick] (0.5,0) -- (0,-0.5);
  \node at (0.25,0.25) {\color{Blue}$Z$};
  \node at (0.25,-0.25) {\color{Blue}$Z$};
  \node at (-0.25,0.25) {\color{Blue}$Z$};
  \node at (-0.25,-0.25) {\color{Blue}$Z$};
\end{tikzpicture}+h.c. \nonumber \\
&H_\lambda= H_\DS - \coeffa ({\color{purple}\CZ^\dagger}\CZ^{\color{red}Z}{\color{purple}\CZ}+\CX^{\color{red}Z}+h.c.) -\coeffb {\color{red}Z}{\color{Blue}Z_\mathrm{l}}{\color{red}Z} -\coeffc {\color{Blue}X_\mathrm{l}} .
\end{align}
with the local constraints,
\begin{equation}\label{eq:gausslawp}
    \begin{tikzpicture}[baseline={(current bounding box.center)},line width=2pt]
  \draw[RoyalBlue, thick] (-0.5,-0.5) -- (0.5,0.5);
  \draw[RoyalBlue, thick] (0.5,-0.5) -- (-0.5,0.5);
  \node at (0,0) {$\CC{\color{red}X}$};
  \node at (0.3,0.3) {${\color{Blue}X}$};
  \node at (0.3,-0.3) {${\color{Blue}X}$};
  \node at (-0.3,0.3) {${\color{Blue}X}$};
  \node at (-0.3,-0.3) {${\color{Blue}X}$};
\end{tikzpicture}=1,\quad \begin{tikzpicture}[baseline={(current bounding box.center)},line width=2pt]
  \draw[gray, thick] (0,0) -- (1,0);
  \draw[gray, thick] (0,0) -- (0,1);
  \draw[gray, thick] (0,0) -- (0,-1);
  \draw[gray, thick] (0,0) -- (-1,0);
  \draw[RoyalBlue, thick] (0,0.5) -- (0.5,0);
  \draw[RoyalBlue, thick] (0,0.5) -- (-0.5,0);
  \draw[RoyalBlue, thick] (0,-0.5) -- (0.5,0);
  \draw[RoyalBlue, thick] (0,-0.5) -- (-0.5,0);
  \node at (0.9,0) {$\CX_2^{ {\color{red}Z_2}}$};
  \node at (0.0,0.7) {$\CX_3^{ {\color{red}Z_3}}$};
  \node at (0,-0.5) {$\CX_1^{{\color{red}-Z_1}}$};
  \node at (0,0) {\color{purple} $\CX$};
  \node at (-0.5,0) {$\CX_4^{ {\color{red}-Z_4}}$};
\end{tikzpicture}=1
\end{equation}
Note that because of the addition of the matter fields, ${\color{Blue}Z_{(ij)}}$ should be replaced by $\color{red}Z_i Z_j$. 

We now apply the constrained Suzuki-Trotter construction. A one-step transfer matrix is $\mathsf{T}=e^{-\dt H_{\rm off}}\CP e^{-\dt H_{\rm diag}}$, with $\CP$ enforcing the two constraints above at each imaginary-time slice. We first read off the diagonal $Z$-basis matrix elements as spatial interactions, and then evaluate the off-diagonal terms by inserting complete $X$-eigenbases and rewriting the constraints with the Lagrange multipliers $\gamma_s^z$ and $\rho_v^z$.
We first calculate the qubit part. The diagonal operators are directly mapped to the spatial operators in the classical theory,
\begin{equation}
    (\coeffb \dt )\mu_\mathrm{v}S_\mathrm{v,l}\mu_\mathrm{v+l},\quad (J^z \dt) S_\mathrm{v,l_x}S_\mathrm{v+l_x,l_y}S_\mathrm{v+l_y,l_x}S_\mathrm{v,l_y}
\end{equation}
where $\dt = \beta/N_\tau$, with $N_\tau$ the number of imaginary-time slices. $S_\mathrm{v,l}=(-1)^{\beta^z}$ is the classical $\IZ_2$ gauge field on the link starting from $\mathrm{v}$ and pointing along $\mathrm{l}$, and $\mu_\mathrm{v}=(-1)^{\mu^z}$ is the $\IZ_2$ matter field. For the off-diagonal terms, we insert complete basis to the matrix element of a time step,
\begin{align}
        &\bra{\beta^z_{t+\dt}}\bra{\mu_{t+\dt}^z} e^{\dt J^x {\color{red}X}}e^{\dt\coeffc {\color{Blue}X}}\delta(\begin{tikzpicture}[baseline={(current bounding box.center)},line width=2pt]
  \draw[RoyalBlue, thick] (-0.5,-0.5) -- (0.5,0.5);
  \draw[RoyalBlue, thick] (0.5,-0.5) -- (-0.5,0.5);
  \node at (0,0) {$\CC{\color{red}X}$};
  \node at (0.3,0.3) {${\color{Blue}X}$};
  \node at (0.3,-0.3) {${\color{Blue}X}$};
  \node at (-0.3,0.3) {${\color{Blue}X}$};
  \node at (-0.3,-0.3) {${\color{Blue}X}$};
\end{tikzpicture}=1)\ket{\mu^z_{t}}\ket{\beta^z_t}\\
=& \sum_{\beta^x,\mu^x,\tau^\CC} \ket{\tau^\CC}\bra{\beta^z_{t+\dt}}\ket{\beta^x}\bra{\mu_{t+\dt}^z}\ket{\mu^x} e^{\dt J^x (-1)^{\mu^x}}e^{\dt\coeffc (-1)^{\beta^x}} \delta(\mu^x+\tau^\CC+\sum_{l\ni s} \beta^x_l=0)\bra{\mu^x} \ket{\mu^z_{t}}\bra{\beta^x}\ket{\beta^z_t}\bra{\tau^\CC}
\end{align}
where the complete bases of qutrit operator $\CC$ is needed to resolve its action $\CC\ket{\tau^\CC} = (-1)^{\tau^\CC}\ket{\tau^\CC}$. And we introduce the $\IZ_2$ variable $\gamma_s^z$ at sites of blue lattice as the Lagrange multiplier to rewrite the Kronecker delta,
\begin{align}
&=\sum_{\beta^x,\mu^x,\gamma_s^z,\tau^\CC} (-1)^{(\beta^z_{t+\dt}-\beta^z_{t})\beta^x}(-1)^{(\mu^z_{t+\dt}-\mu^z_{t})\mu^x}e^{\dt J^x (-1)^{\mu^x}}e^{\dt\coeffc (-1)^{\beta^x}}(-1)^{\gamma_s^z (\tau^\CC+\mu^x +\sum_{l\ni s} \beta^x_l )}\ket{\tau^\CC}\bra{\tau^\CC}\\
&=\sum_{\beta^x,\mu^x,\gamma_s^z,\tau^\CC} (-1)^{(\beta^z_{t+\dt}+\gamma_s^z+\gamma^z_{s+\mathrm{l}}-\beta^z_{t})\beta^x_\mathrm{l}}(-1)^{(\mu^z_{t+\dt}+\gamma_s^z-\mu^z_{t})\mu^x_s}e^{\dt J^x (-1)^{\mu^x}}e^{\dt\coeffc (-1)^{\beta^x}}(-1)^{\gamma_s^z \tau^\CC}\ket{\tau^\CC}\bra{\tau^\CC}\\
&=\sum_{\gamma_s^z,\tau^\CC} e^{K_3^t \sum_p
(\beta^z_{l,t+\dt}\beta^z_{l,t}
\gamma_s^z \gamma^z_{s+l})}e^{J^t \sum_s
(\mu^z_{s,t+\dt}\gamma^z_{s}\mu^z_{s,t})}(-1)^{\gamma_s^z \tau^\CC}\ket{\tau^\CC}\bra{\tau^\CC}
\end{align}
where $K_3^t = \tanh^{-1} e^{-2\dt \coeffc}$ and $J^t = \tanh^{-1} e^{-2\dt J^x}$. And $\gamma^z$ is the $\IZ_2$ gauge field along the temporal direction. We can rewrite these operators as in the classical $\IZ_2$ gauge theory,
\begin{equation}
(J^t)\mu_\mathrm{v}S_\mathrm{v,l_t}\mu_\mathrm{v+l_t},\quad (K_3^t)S_\mathrm{v,l}S_\mathrm{v+l,l_t}S_\mathrm{v+l_t,l}S_\mathrm{v,l_t}
\end{equation}
where $S_\mathrm{v,l_t}=(-1)^{\gamma^z}$ is the $\IZ_2$ gauge field on temporal blue links.

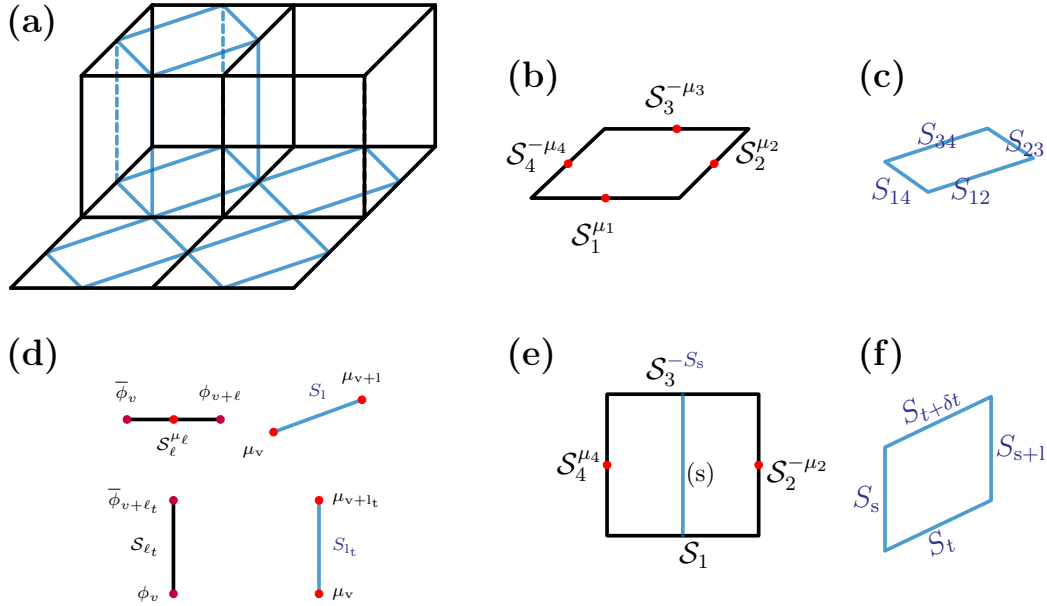
\begin{figure}[hbtp]
    \centering
    \begingroup
    \colorlet{gaugeblueedge}{RoyalBlue}
    \colorlet{gaugebluetext}{Blue}
    \resizebox{0.8\linewidth}{!}{%
    \begin{tikzpicture}[
        x=0.007cm,
        y=-0.007cm,
        line cap=round,
        line join=round,
        blackedge/.style={black,line width=1.45pt},
        blueedge/.style={gaugeblueedge,draw opacity=0.7,line width=1.45pt},
        reddot/.style={circle,fill=red,draw=red,inner sep=0pt,minimum size=3pt},
        purpledot/.style={circle,fill=purple,draw=purple,inner sep=0pt,minimum size=3pt},
        hidden/.style={densely dashed},
        oplabel/.style={font=\large,inner sep=1pt},
        smalllabel/.style={font=\normalsize,inner sep=1pt},
        bondlabel/.style={font=\scriptsize,inner sep=1pt},
        bluelabel/.style={font=\large,text=gaugebluetext,inner sep=1pt},
        paneltag/.style={font=\bfseries\Large,inner sep=1pt,fill=white,fill opacity=0.85,text opacity=1}
    ]
    \path[use as bounding box] (-35,0) rectangle (2110,1263);
    \node[paneltag,anchor=west] at (0,35) {(a)};
    \node[paneltag,anchor=west] at (990,145) {(b)};
    \node[paneltag,anchor=west] at (1685,145) {(c)};
    \node[paneltag,anchor=west] at (0,695) {(d)};
    \node[paneltag,anchor=west] at (990,695) {(e)};
    \node[paneltag,anchor=west] at (1685,695) {(f)};

    \begin{scope}
        \coordinate (O) at (15,560);
        \coordinate (ex) at (280,0);
        \coordinate (ey) at (140,-140);
        \coordinate (ez) at (0,-280);

        \foreach \xa/\ya/\xb/\yb/\xc/\yc/\xd/\yd in {
            0.5/0/1/0.5/0.5/1/0/0.5,
            1.5/0/2/0.5/1.5/1/1/0.5,
            0.5/1/1/1.5/0.5/2/0/1.5,
            1.5/1/2/1.5/1.5/2/1/1.5
        }{
            \draw[blueedge]
                ($(O)+\xa*(ex)+\ya*(ey)$)
                -- ($(O)+\xb*(ex)+\yb*(ey)$)
                -- ($(O)+\xc*(ex)+\yc*(ey)$)
                -- ($(O)+\xd*(ex)+\yd*(ey)$)
                -- cycle;
        }

        \coordinate (Yb) at ($(O)+0.5*(ex)+1*(ey)$);
        \coordinate (Yr) at ($(O)+1*(ex)+1.5*(ey)$);
        \coordinate (Yt) at ($(O)+0.5*(ex)+2*(ey)$);
        \coordinate (Yl) at ($(O)+0*(ex)+1.5*(ey)$);
        \draw[blueedge] ($(Yb)+(ez)$) -- ($(Yr)+(ez)$)
            -- ($(Yt)+(ez)$) -- ($(Yl)+(ez)$) -- cycle;
        \draw[blueedge] (Yb) -- ($(Yb)+(ez)$);
        \draw[blueedge] (Yr) -- ($(Yr)+(ez)$);
        \draw[blueedge,hidden] (Yt) -- ($(Yt)+(ez)$);
        \draw[blueedge,hidden] (Yl) -- ($(Yl)+(ez)$);

        \foreach \j in {0,1,2}{
            \draw[blackedge] ($(O)+\j*(ey)$) -- ($(O)+2*(ex)+\j*(ey)$);
        }
        \foreach \i in {0,1,2}{
            \draw[blackedge] ($(O)+\i*(ex)$) -- ($(O)+\i*(ex)+2*(ey)$);
        }
        \foreach \j in {1,2}{
            \draw[blackedge] ($(O)+\j*(ey)+1*(ez)$) -- ($(O)+2*(ex)+\j*(ey)+1*(ez)$);
        }
        \foreach \i in {0,1,2}{
            \draw[blackedge] ($(O)+\i*(ex)+1*(ey)+1*(ez)$) -- ($(O)+\i*(ex)+2*(ey)+1*(ez)$);
        }
        \foreach \i in {0,1,2}{
            \foreach \j in {1,2}{
                \draw[blackedge] ($(O)+\i*(ex)+\j*(ey)$) -- ($(O)+\i*(ex)+\j*(ey)+1*(ez)$);
            }
        }
        \draw[blackedge,hidden] ($(O)+1*(ex)+1*(ey)$) -- ($(O)+1*(ex)+1*(ey)+1*(ez)$);
        \draw[blackedge,hidden] ($(O)+2*(ex)+1*(ey)$) -- ($(O)+2*(ex)+1*(ey)+1*(ez)$);
    \end{scope}

    \begin{scope}
        \coordinate (HqL) at (245,820);
        \coordinate (HqR) at (430,820);
        \coordinate (HbL) at (535,845);
        \coordinate (HbR) at (710,781);
        \coordinate (VqB) at (337,980);
        \coordinate (VqT) at (337,1165);
        \coordinate (VbB) at (625,980);
        \coordinate (VbT) at (625,1165);
        \coordinate (HqM) at ($(HqL)!0.5!(HqR)$);
        \coordinate (HbM) at ($(HbL)!0.5!(HbR)$);
        \coordinate (VqM) at ($(VqB)!0.5!(VqT)$);
        \coordinate (VbM) at ($(VbB)!0.5!(VbT)$);
        \draw[blackedge] (HqL) -- (HqR);
        \draw[blueedge] (HbL) -- (HbR);
        \draw[blackedge] (VqB) -- (VqT);
        \draw[blueedge] (VbB) -- (VbT);
        \node[reddot] at (HqM) {};
        \foreach \p in {HqL,HqR,VqB,VqT}{
            \node[purpledot] at (\p) {};
        }
        \foreach \p in {HbL,HbR,VbB,VbT}{
            \node[reddot] at (\p) {};
        }
        \node[bondlabel,anchor=south] at ($(HqL)+(0,-28)$) {$\overline{\phi}_{v}$};
        \node[bondlabel,anchor=south] at ($(HqR)+(0,-28)$) {$\phi_{v+\ell}$};
        \node[bondlabel,anchor=north] at ($(HqM)+(0,22)$) {$\CS_\ell^{\mu_\ell}$};
        \node[bondlabel,anchor=north east] at ($(HbL)+(-10,22)$) {$\mu_{\mathrm{v}}$};
        \node[bondlabel,anchor=south] at ($(HbR)+(0,-24)$) {$\mu_{\mathrm{v+l}}$};
        \node[bondlabel,text=gaugebluetext,anchor=south] at ($(HbM)+(0,-28)$) {$S_\mathrm{l}$};
        \node[bondlabel,anchor=east] at ($(VqB)+(-20,0)$) {$\overline{\phi}_{v+\ell_t}$};
        \node[bondlabel,anchor=east] at ($(VqT)+(-20,0)$) {$\phi_v$};
        \node[bondlabel,anchor=east] at ($(VqM)+(-20,0)$) {$\CS_{\ell_t}$};
        \node[bondlabel,anchor=west] at ($(VbB)+(20,0)$) {$\mu_{\mathrm{v+l_t}}$};
        \node[bondlabel,anchor=west] at ($(VbT)+(20,0)$) {$\mu_{\mathrm{v}}$};
        \node[bondlabel,text=gaugebluetext,anchor=west] at ($(VbM)+(20,0)$) {$S_\mathrm{l_t}$};
    \end{scope}

    \begin{scope}
        \coordinate (A) at (1045,382);
        \coordinate (B) at (1190,245);
        \coordinate (C) at (1475,245);
        \coordinate (D) at (1338,382);
        \coordinate (P14) at (1118,314);
        \coordinate (P12) at (1192,382);
        \coordinate (P23) at (1407,314);
        \coordinate (P34) at (1333,245);
        \draw[blackedge] (A) -- (B) -- (C) -- (D) -- cycle;
        \foreach \p in {P14,P12,P23,P34}{
            \node[reddot] at (\p) {};
        }
        \node[oplabel,anchor=east] at (1125,296)
            {$\CS_4^{-\mu_4}$};
        \node[oplabel,anchor=north] at (1170,423) {$\CS_1^{\mu_1}$};
        \node[oplabel,anchor=west] at (1444,296)
            {$\CS_2^{\mu_2}$};
        \node[oplabel,anchor=south] at (1332,222)
            {$\CS_3^{-\mu_3}$};
    \end{scope}

    \begin{scope}
        \coordinate (Q14) at (1745,310);
        \coordinate (Q34) at (1948,244);
        \coordinate (Q23) at (2038,303);
        \coordinate (Q12) at (1830,370);
        \draw[blueedge] (Q14) -- (Q34) -- (Q23) -- (Q12) -- cycle;
        \node[bluelabel] at (1848,258) {$S_{34}$};
        \node[bluelabel] at (2022,273) {$S_{23}$};
        \node[bluelabel] at (1915,368) {$S_{12}$};
        \node[bluelabel] at (1758,368) {$S_{14}$};
    \end{scope}

    \begin{scope}
        \coordinate (A) at (1055,1190);
        \coordinate (B) at (1195,1050);
        \coordinate (C) at (1495,1050);
        \coordinate (D) at (1355,1190);
        \coordinate (Bt) at (1195,770);
        \coordinate (Ct) at (1495,770);
        \coordinate (Ct) at (1495,770);
        \coordinate (Bm) at (1195,910);
        \coordinate (Cm) at (1495,910);
        \coordinate (M) at (1345,1050);
        \coordinate (Mt) at (1345,770);
        \coordinate (Lmid) at (1128,1120);
        \coordinate (Rmid) at (1425,1120);
        \draw[blackedge] (B) -- (Bt) -- (Ct) -- (C) -- cycle;
        \draw[blueedge] (M) -- (Mt);
        \foreach \p in {Bm,Cm}{
            \node[reddot] at (\p) {};
        }
        \node[oplabel,anchor=east] at (1192,912)
            {$\CS_4^{\mu_4}$};
        \node[oplabel,anchor=south] at (1332,760)
            {$\CS_3^{\textcolor{gaugebluetext}{-S_\mathrm{s}}}$};
        \node[oplabel,anchor=west] at (1505,925)
            {$\CS_2^{-\mu_2}$};
        \node[oplabel,anchor=west] at (1332,1088) {$\CS_1$};
        \node[smalllabel] at (1380,930) {$(\mathrm{s})$};
    \end{scope}

    \begin{scope}
        \coordinate (A) at (1745,1080);
        \coordinate (B) at (1745,875);
        \coordinate (C) at (1955,775);
        \coordinate (D) at (1955,980);
        \draw[blueedge] (A) -- node[bluelabel,anchor=east,pos=0.50] {$S_\mathrm{s}$}
            (B) -- node[bluelabel,sloped,above,pos=0.50] {$S_{t+\delta t}$}
            (C) -- node[bluelabel,anchor=west,pos=0.50] {$S_\mathrm{s+l}$}
            (D) -- node[bluelabel,sloped,below,pos=0.55] {$S_t$}
            cycle;
    \end{scope}
    \end{tikzpicture}%
    }
    \endgroup
    \caption{Plaquette and bond terms in the $\DS$ classical gauge theory. (a) Spacetime lattice. Black links carry qutrit variables and blue links carry qubit variables. The purple and red dots at the end of black and blue edges are omitted. (b) Spatial qutrit plaquette with coupling $\CK^{xy}=\CJ^z \dt$. (c) Spatial qubit plaquette with coupling $K^{xy}=J^z \dt$. (d) Bond operators for black qutrit and blue qubit links. (e) Temporal qutrit plaquette in the $x$ direction with coupling $\CK_1^{xt}=\CF_{\IZ_3}(\lambda_1^x \dt)$, generated by the $\lambda_1^x$ deformation. (f) Temporal qubit plaquette with coupling $K_3^t=\CF_{\IZ_2}(\coeffc \dt)$, generated by the $\coeffc$ deformation.}
    \label{fig:clasgauge}
\end{figure}

Next, we similarly derive the qutrit terms. The diagonal $\DS$ bond and plaquette term contributes
\begin{equation}
   (\lambda_1^z\dt)\overline{\phi}_v\CS_{\ell}^{{\mu_\ell}}\phi_{v+\ell}+h.c.,\quad  (\CJ^z\dt)\CS_1^{{\mu_1}} \CS_2^{{\mu_2}}\CS_3^{{-\mu_3}}\CS_4^{{-\mu_4}}+h.c.\;.
\end{equation}
where $\CS_i=\omega^{\tau^z}$ are the spatial $\IZ_3$ gauge field on link $i$ and $\mu_i$ are the spatial $\IZ_2$ matter field on black link $i$. For the off-diagonal qutrit matrix element, $\tau_\ell^z\in\IZ_3$ labels the black-link qutrit, while $\phi_v^z\in\IZ_3$ labels the auxiliary purple qutrit matter field introduced at the black vertex $v$; $\tau^x$ and $\phi^x$ are the corresponding Fourier-dual $X$-basis labels.
Next, for the off-diagonal terms,
\begin{align}
   &\sum_{\gamma_s^z,\tau^\CC} \bra{\tau^z_{t+\dt}} (-1)^{\gamma_s^z \tau^\CC}\ket{\tau^\CC}\bra{\tau^\CC}\bra{\phi_{t+\dt}^z} e^{\dt \CJ^x {\color{purple}\CX}}e^{\dt\lambda_1^x \CX^{\color{red}Z}}\delta(\begin{tikzpicture}[baseline={(current bounding box.center)},line width=2pt]
  \draw[gray, thick] (0,0) -- (1,0);
  \draw[gray, thick] (0,0) -- (0,1);
  \draw[gray, thick] (0,0) -- (0,-1);
  \draw[gray, thick] (0,0) -- (-1,0);
  \draw[RoyalBlue, thick] (0,0.5) -- (0.5,0);
  \draw[RoyalBlue, thick] (0,0.5) -- (-0.5,0);
  \draw[RoyalBlue, thick] (0,-0.5) -- (0.5,0);
  \draw[RoyalBlue, thick] (0,-0.5) -- (-0.5,0);
  \node at (0.9,0) {$\CX_2^{ {\color{red}Z_2}}$};
  \node at (0.0,0.7) {$\CX_3^{ {\color{red}Z_3}}$};
  \node at (0,-0.5) {$\CX_1^{{\color{red}-Z_1}}$};
  \node at (0,0) {\color{purple} $\CX$};
  \node at (-0.5,0) {$\CX_4^{ {\color{red}-Z_4}}$};
\end{tikzpicture}=1) \ket{\tau^z_{t}}\ket{\phi_{t}^z}\\
=&\sum_{\gamma_s^z,\tau^\CC,\tau^x,\phi^x} \bra{\tau^z_{t+\dt}} (-1)^{\gamma_s^z \tau^\CC}\ket{\tau^\CC}\bra{\tau^\CC}\ket{\tau^x}\bra{\phi_{t+\dt}^z}\ket{\phi^x} e^{\dt \CJ^x \omega^{\phi^x}}e^{\dt\lambda_1^x \omega^{(-1)^{\mu^z}\tau^x}} \nonumber\\
&\delta\left(\phi^x_v -(-1)^{\mu_1}\tau_1^x+(-1)^{\mu_2}\tau_2^x+(-1)^{\mu_3}\tau_3^x-(-1)^{\mu_4}\tau_4^x=0\right) \bra{\tau^x}\ket{\tau^z_{t}}\bra{\phi^x}\ket{\phi_{t}^z} \nonumber\\
=&\sum_{\gamma_s^z,\tau^\CC,\tau^x,\phi^x,\rho_v^z} \bra{\tau^z_{t+\dt}} (-1)^{\gamma_s^z \tau^\CC}\ket{\tau^\CC}\bra{\tau^\CC}\ket{\tau^x}\bra{\tau^x}\ket{\tau^z_{t}}\omega^{\phi^x (\phi_{t}^z-\phi_{t+\dt}^z)} e^{\dt \CJ^x \omega^{\phi^x}}e^{\dt\lambda_1^x \omega^{(-1)^{\mu^z}\tau^x}} \nonumber\\
&\omega^{\rho_v^z\left(\phi^x_v -(-1)^{\mu_1}\tau_1^x+(-1)^{\mu_2}\tau_2^x+(-1)^{\mu_3}\tau_3^x-(-1)^{\mu_4}\tau_4^x\right)}
\end{align}
we sum over $\tau^\CC$, notice that $\sum_{\tau^\CC} (-1)^{\sum_s \gamma_s^z\tau_s^\CC}\braket{\tau^z_{t+\dt}}{\tau^\CC}\braket{\tau^\CC}{\tau^x} = \omega^{-(-1)^{\gamma^z_s}\tau^z_{t+\dt} \tau^x}$,
\begin{align}
=&\sum_{\gamma_s^z,\tau^x,\phi^x,\rho_v^z}\omega^{(\tau^z_t-(-1)^{\gamma^z_s}\tau^z_{t+\dt} )\tau^x}\omega^{\phi^x (\phi_{t}^z-\phi_{t+\dt}^z)} e^{\dt \CJ^x \omega^{\phi^x}}e^{\dt\lambda_1^x \omega^{(-1)^{\mu^z}\tau^x}} \nonumber\\
&\omega^{\rho_v^z\left(\phi^x_v -(-1)^{\mu_1}\tau_1^x+(-1)^{\mu_2}\tau_2^x+(-1)^{\mu_3}\tau_3^x-(-1)^{\mu_4}\tau_4^x\right)} \\
=&\sum_{\gamma_s^z,\tau^x,\phi^x,\rho_v^z}
e^{\dt \CJ^x \omega^{\phi^x}}\omega^{\phi^x (\phi_{t}^z+\rho_v^z-\phi_{t+\dt}^z)}\nonumber\\
&e^{\dt\lambda_1^x \omega^{(-1)^{\mu^z}\tau^x}}\omega^{(\tau^z_t-(-1)^{\gamma^z_s}\tau^z_{t+\dt} )\tau^x}\omega^{\rho_v^z\left(-(-1)^{\mu_1}\tau_1^x+(-1)^{\mu_2}\tau_2^x+(-1)^{\mu_3}\tau_3^x-(-1)^{\mu_4}\tau_4^x\right)}
\end{align}
The $\rho_v^z$ is thought of as the $\IZ_3$ classical gauge field along the temporal direction. The first line will give the bond term $\overline{\phi}_v \CS_{\ell_t} \phi_{v+\ell_t}$ with coefficient $\CJ^{t} = \frac{1}{3} \log \left(\frac{2 e^{-\dt \CJ^x}+e^{2 \dt \CJ^x}}{e^{2 \dt \CJ^x}-e^{-\dt \CJ^x}}\right)$, while the last two lines will give the plaquette terms. We can rearrange the last summation from summing over vertices to summing over links,
\begin{align}
    \sum_{\tau^x}\sum_{\gamma^z,\rho^z}   
   & e^{\dt \lambda_1^x \sum_\ell(-1)^{\mu^z}\omega^{\tau^x_\ell}+h.c.}\omega^{\sum_\ell (\tau^z_{\ell,t}-(-1)^{\gamma_s^z}\tau^z_{\ell,t+\dt}) \tau_\ell^x} \nonumber\\
&\omega^{\sum_{\ell_x} \tau_{\ell_x}^x \left((-1)^{\mu_2}\rho^z_{v}-(-1)^{\mu_4}\rho_{v+\ell_x}^z\right)}
\omega^{\sum_{\ell_y} \tau_{\ell_y}^x \left((-1)^{\mu_3}\rho^z_{v}-(-1)^{\mu_1}\rho_{v+\ell_y}^z\right)}
\end{align}
We separate the $x$ and $y$ directions, the resulting terms are,
\begin{align}
    \sum_{\tau^x}\sum_{\gamma^z,\rho^z}   
&e^{\dt \lambda_1^x \sum_{\ell_x}(-1)^{\mu^z}\omega^{\tau^x_{\ell_x}}+h.c.}\omega^{\sum_{\ell_x} \tau_{\ell_x}^x \left(\tau^z_{\ell_x,t}-(-1)^{\gamma_s^z}\tau^z_{\ell_x,t+\dt}+(-1)^{\mu_2}\rho^z_{v}-(-1)^{\mu_4}\rho_{v+\ell_x}^z\right)}\nonumber\\
&e^{\dt \lambda_1^x \sum_{\ell_y}(-1)^{\mu^z}\omega^{\tau^x_{\ell_y}}+h.c.}\omega^{\sum_{\ell_y} \tau_{\ell_y}^x \left(\tau^z_{\ell_y,t}-(-1)^{\gamma_s^z}\tau^z_{\ell_y,t+\dt}+(-1)^{\mu_3}\rho^z_{v}-(-1)^{\mu_1}\rho_{v+\ell_y}^z\right)}
\end{align}
For each horizontal and vertical link the phase variables being summed over are
\begin{align*}
\CO_{\ell_x}
&=\omega^{\tau^z_{\ell_x,t}-(-1)^{\gamma_s^z}\tau^z_{\ell_x,t+\dt}+(-1)^{\mu_2}\rho^z_{v}-(-1)^{\mu_4}\rho_{v+\ell_x}^z},\\
\CO_{\ell_y}
&=\omega^{\tau^z_{\ell_y,t}-(-1)^{\gamma_s^z}\tau^z_{\ell_y,t+\dt}+(-1)^{\mu_3}\rho^z_{v}-(-1)^{\mu_1}\rho_{v+\ell_y}^z} .
\end{align*}
In the final classical plaquette terms we relabel the four qutrit variables around the corresponding spacetime plaquette by $\CS_1,\ldots,\CS_4$: $\CS_1$ and $\CS_3 \sim \omega^{\tau^z}$ are the qutrit fields on the original black link at times $t$ and $t+\dt$, while $\CS_2,\CS_4\sim\omega^{\rho^z}$ are the temporal qutrit fields coming from the adjacent $\rho^z$ variables.  The qubit signs appearing as exponents are $\mu_\mathrm{v}=(-1)^{\mu_\mathrm{v}^z}$ on vertices of blue lattice or equivalently links of the black lattice. And $S_\mathrm{s}=(-1)^{\gamma_\mathrm{s}^z}$ on the temporal blue link. And $\CS_i^{S_{(ij)}} = \omega^{(-1)^{\beta_{(ij)}^z}\tau^z}$. The single-link $\ell_x$ sum gives $e^{2\dt\lambda_1^x}+e^{-\dt\lambda_1^x}(\CO_{\ell_x}+\CO_{\ell_x}^\dagger)$, i.e.
\begin{equation}
    e^{2\dt \lambda_1^x }+ e^{-\dt \lambda_1^x }
    \CS_{1}(\CS_2)^{-\mu_2}(\CS_3)^{-S_{s}}(\CS_4)^{\mu_4}
    +h.c.
\end{equation}
Note that $e^{b (\CO+\CO^\dagger)} = \frac{e^{2b}+2e^{-b}}{3}\dsi +\frac{e^{2b}-e^{-b}}{3}(\CO+\CO^\dagger)=\alpha \dsi +\beta (\CO+\CO^\dagger)$, and $b= \frac{1}{3}\log(\frac{\alpha+2\beta}{\alpha-\beta})$, it gives the term,
\begin{equation}
    e^{\CK_1^{xt} (\CS_{1}(\CS_2)^{-\mu_2}(\CS_3)^{-S_{s}}(\CS_4)^{\mu_4}+h.c.)}
\end{equation}
where $\CK_1^{xt} = \frac{1}{3} \log \left(\frac{2 e^{-\dt \lambda_1^x}+e^{2 \dt \lambda_1^x}}{e^{2 \dt \lambda_1^x}-e^{-\dt \lambda_1^x}}\right)$. Similarly, we get,
\begin{equation}
    e^{\CK_1^{yt} (\CS_{1}(\CS_2)^{-\mu_3}(\CS_3)^{-S_{s}}(\CS_4)^{\mu_1}+h.c.)}
\end{equation}
with the same inversion giving $\CK_1^{yt}=\CK_1^{xt}$ for this isotropic $\lambda_1$ deformation.
The local plaquette and bond terms in the $\DS$ gauge theory are listed in \figref{fig:clasgauge}. All of the operators and their mapping are summarized in \tabref{tab:opsmap}.

\begin{table}[t]
\centering
\setlength{\tabcolsep}{1pt}
\renewcommand{\arraystretch}{1.2}
\begin{tabular}{c c c}
\hline\hline
Quantum operator & Classical term & Classical coupling\\
\hline
qutrit plaquette $\CB_p$ & $\CS_1^{\mu_1}\CS_2^{\mu_2}\CS_3^{-\mu_3}\CS_4^{-\mu_4}+h.c.$ & $\CK^{xy}=\CJ^z\dt$\\
qutrit star $\CA_v$ & $\overline{\phi}_v\CS_{\ell_t}\phi_{v+\ell_t}+h.c.$ & $\CJ^t=\CF_{\IZ_3}(\dt\CJ^x)$\\
qutrit single-$X$ on $\ell_x,\ell_y$ & $\CS_{1}(\CS_2)^{-\mu_2}(\CS_3)^{-S_{\mathrm{s}}}(\CS_4)^{\mu_4}+h.c.$ & $\CK_1^{xt}=\CK_1^{yt}=\CF_{\IZ_3}(\dt\lambda_1^x)$\\
qutrit single-$Z$ ${\color{purple}\CZ^\dagger}\CZ^{\color{red}Z}{\color{purple}\CZ}$ & $\overline{\phi}_v\CS_{\ell}^{\mu_\ell}\phi_{v+\ell}+h.c.$ & $\CJ_1^{xy}=\lambda_1^z\dt$\\
qubit plaquette $B_\pp$ & $S_{\mathrm{v},\mathrm{l_x}}S_{\mathrm{v}+\mathrm{l_x},\mathrm{l_y}}S_{\mathrm{v}+\mathrm{l_y},\mathrm{l_x}}S_{\mathrm{v},\mathrm{l_y}}$ & $K^{xy}=J^z\dt$\\
qubit star $A_\vv$ & $\mu_{\mathrm{v}}S_{\mathrm{v},\mathrm{l_t}}\mu_{\mathrm{v}+\mathrm{l_t}}$ & $J^t=\CF_{\IZ_2}(\dt J^x)$\\
qubit single-$X$ ${\color{Blue}X_l}$ & $S_{\mathrm{v},\mathrm{l}}S_{\mathrm{v}+\mathrm{l},\mathrm{l_t}}S_{\mathrm{v}+\mathrm{l_t},\mathrm{l}}S_{\mathrm{v},\mathrm{l_t}}$ & $K_3^t=\CF_{\IZ_2}(\dt\coeffc)$\\
qubit single-$Z$ ${\color{red}Z}{\color{Blue}Z_l}{\color{red}Z}$ & $\mu_{\mathrm{v}}S_{\mathrm{v},\mathrm{l}}\mu_{\mathrm{v}+\mathrm{l}}$ & $J_2^{xy}=\coeffb\dt$\\
\hline\hline
\end{tabular}
\caption{Classical terms and couplings generated by the local operators in $\DS$. $\CJ,J$ denote the couplings for bond operators in $\IZ_2$ and $\IZ_3$ gauge theories, $\CK,K$ denote the couplings for the plaquette operators.}
\label{tab:opsmap}
\end{table}

The quantum-to-classical coupling map used in \tabref{tab:opsmap} is written as a function of the dimensionless coupling $g$,
\begin{equation}
    \CF_{\IZ_2}(g)=\tanh^{-1}\!\left(e^{-2g}\right),\qquad
    \CF_{\IZ_3}(g)=\frac{1}{3}\log\left(
    \frac{2e^{-g}+e^{2g}}
         {e^{2g}-e^{-g}}
    \right).
\end{equation}
Thus diagonal operators give spatial couplings proportional to $\dt$, while the off-diagonal $\IZ_2$ and $\IZ_3$ operators give temporal couplings through $\CF_{\IZ_2}$ and $\CF_{\IZ_3}$, respectively. Both maps are involutions, $\CF_{\IZ_N}(\CF_{\IZ_N}(g))=g$ for $N=2,3$, equivalently $\sinh(2g)\sinh(2\CF_{\IZ_2}(g))=1$ and $(e^{3g}-1)(e^{3\CF_{\IZ_3}(g)}-1)=3$. These are the standard Kramers-Wannier self-duality relations for the $\IZ_2$ Ising and $\IZ_3$ three-state Potts models \cite{Tupitsyn:2008ah,Wu:1982potts}.

From this mapping, it is clear to see that the quantum $\emp$ permutation symmetry permutes the plaquette and bond operator in the classical gauge theory. The generic self-dual $\DS$ model is mapped to anisotropic classical gauge theory as the single qubit terms can be tuned independently for $\coeffb,\coeffc$.

Since the $\ems$ in $\DS$ only fixes $\CJ^x=\CJ^z$ and $\lambda_1^x=\lambda_1^z$, it is tempting to study the model with only the qutrit part being isotropic, which requires
\begin{equation}
  \CJ^z\dt =\CF_{\IZ_3}(\dt\lambda_1^x),\quad \CF_{\IZ_3}(\dt\CJ^x)=\lambda_1^z\dt
\end{equation} 
Along the quantum self-dual line, $\lambda_1^x =\lambda_1^z$, this gives
\begin{equation}
  \CJ^z\dt =\CF_{\IZ_3}(\dt\lambda_1^x) = \CF_{\IZ_3}(\dt\lambda_1^z) = \CF_{\IZ_3}(\CF_{\IZ_3}(\dt\CJ^x)) = \CJ^x\dt
\end{equation}
Therefore, we conclude the self-dual $\DS$ model can be mapped to the above classical gauge theory with qutrit part being isotropic and along the self-dual line, which can be viewed as the $\IZ_3$ generalization of \cite{Tupitsyn:2008ah}, while qubit part stay anisotropic. If we further take $J^t\rightarrow \infty$, which fixes $\mu_\mathrm{v+l_t}\mu_\mathrm{v} = S_\mathrm{l_t}$ along the temporal direction, then the qutrit plaquette operator along the temporal direction is the same as the one along the spatial direction.

On the other hand, if we require isotropy in $\IZ_2$ gauge part, and let $J^x=J^z$, this will lead to $\coeffb=\coeffc$, however, $\coeffb$, $\coeffc$ can be tuned independently.

\section{VI. Exact diagonalization of the $D$ phase}
\label{SM:DphaseED}

Since the signs $s_\ell=\pm1$ in Eq.~\eqref{eq:csymham} are summed independently on each link, the Hamiltonian can be written as
\begin{align}
H_D={}&-J_p\sum_p\prod_{\ell\in\partial p}Q^Z_\ell
      -J_v\sum_v\prod_{\ell\ni v}Q^X_\ell
      -h_C\sum_\ell\CC_\ell, \label{eq:DphaseH}\\
Q^Z\equiv{}&\CZ+\CZ^\dagger,\qquad
Q^X\equiv\CX+\CX^\dagger .
\end{align}
We take $J_p=J_v=h_C=1$. The plaquette and star terms do not commute and are minimized by incompatible link states with $\CZ=1$ and $\CX=1$, respectively. Therefore, unlike the $\IZ_3$ toric code, $H_D$ is neither a commuting-projector Hamiltonian nor frustration-free.

In the $\CZ$ basis, $H_D$ is stoquastic, and its off-diagonal configuration graph is connected. The Perron-Frobenius theorem therefore guarantees a unique finite-size ground state with strictly positive amplitudes. Since $[\CC_\ell,H_D]=0$ and $\CC_\ell$ permutes the $\CZ$-basis configurations, this ground state satisfies $\CC_\ell=+1$ on every link. In the local basis
\begin{equation}
\ket{0},\qquad
\ket{s}\equiv\frac{\ket{1}+\ket{2}}{\sqrt{2}},
\end{equation}
the qutrit operators reduce to
\begin{equation}
Q^Z=
\begin{pmatrix}
2&0\\
0&-1
\end{pmatrix},
\qquad
Q^X=
\begin{pmatrix}
0&\sqrt{2}\\
\sqrt{2}&1
\end{pmatrix}.
\label{eq:DphaseQ}
\end{equation}
Thus we diagonalize
\begin{equation}
H_D^{\mathrm{eff}}
=-\sum_p\prod_{\ell\in\partial p}Q^Z_\ell
-\sum_v\prod_{\ell\ni v}Q^X_\ell-N_\ell
\end{equation}
with periodic boundary conditions. An $L_x\times L_y$ torus has
$N=L_xL_y$ plaquettes and vertices and $N_\ell=2N$ links. In particular, the $3\times3$ torus has $18$ links and Hilbert-space dimension $2^{18}$. Its four lowest energies in the charge-conjugation-even sector are
\begin{equation}
(E_0,E_1,E_2,E_3)
=(-165.419425,-165.198806,-138.851655,-138.398900),
\end{equation}
giving
\begin{equation}
\Delta E=E_1-E_0=0.220619,\qquad
\Gamma=E_2-E_1=26.347151.
\end{equation}
For the periodic tilted tori with $N=5,\ldots,10$ unit cells, the splitting can be fitted by $\Delta E(N)\simeq12.44e^{-0.447N}$, while $\Gamma$ remains of order one on all geometries we studied. The system sizes are small but the isolated two low energy states and their rapidly decreasing splitting are robust.

The qutrit Hadamard transformation restricts in the even sector to
\begin{equation}
h=\frac{1}{\sqrt{3}}
\begin{pmatrix}
1&\sqrt{2}\\
\sqrt{2}&-1
\end{pmatrix},
\qquad hQ^Zh=Q^X.
\end{equation}
Together with the half translation, it defines the microscopic $\ems$ transformation $\mathcal U_{em}$. The two finite-size states have opposite $\ems$ quantum numbers,
\begin{equation}
\mathcal U_{em}\ket{\psi_\pm}=\pm\ket{\psi_\pm},
\end{equation}
whereas their symmetry-breaking combinations
\begin{equation}
\ket{\Phi_{Z,X}}
=\frac{\ket{\psi_+}\pm\ket{\psi_-}}{\sqrt{2}}
\end{equation}
are exchanged by $\mathcal U_{em}$ and obey
$\CC_\ell\ket{\Phi_{Z,X}}=\ket{\Phi_{Z,X}}$ for every link. 
The numerical spectrum is therefore consistent with $GSD=2$ for thermodynamic limit and the degeneracy arises from spontaneous $\ems$ breaking rather than topological order.

\end{document}